\documentclass[11pt]{article}
\pdfoutput=1
\usepackage{jcappub,amsmath,amssymb}
\usepackage{bm}
\usepackage[table]{xcolor}
\usepackage{graphics}
\usepackage{tikz}
\usetikzlibrary{arrows,positioning,decorations.markings,decorations.pathmorphing,calc}
\usepackage{titlesec}
\titleformat{\paragraph}[runin]{\normalfont\normalsize\bfseries}{\theparagraph}{0pt}{\hspace{0.85cm}}[]

\definecolor{hyperref}{RGB}{026,028,087}

\def\gsim{ \lower .75ex \hbox{$\sim$} \llap{\raise .27ex \hbox{$>$}} }
\def\lsim{ \lower .75ex \hbox{$\sim$} \llap{\raise .27ex \hbox{$<$}} }
\def\be{\begin{equation}}
\def\ee{\end{equation}}
\def\bea{\begin{eqnarray}}
\def\eea{\end{eqnarray}}

\newcommand{\ba}{\begin{array}}
\newcommand{\ea}{\end{array}}
\newcommand{\mn}{\mu\nu}

\newcommand{\commentout}[1]{}

\newcommand{\pa}{\partial}

\newcommand{\cL}{{\cal{L}}}
\newcommand{\cE}{{\cal{E}}}

\newcommand{\comment}[1]{}

\newcommand{\thickhline}{\noalign{\hrule height 2.0pt}}

\newcommand{\bs}{\begin{split}}

\newcommand{\dof}{{\it dof} }

\newcommand{\St}{{St\"uckelberg} }
\newcommand{\Sting}{{St\"uckelberging} }

\def\ba{\begin{eqnarray}}
\def\ea{\end{eqnarray}}

\def\nn{\nonumber}

\def\({\left(}
\def\){\right)}

\definecolor{jn}{RGB}{10, 10, 200} 
\definecolor{js}{RGB}{204, 0, 0} 
\definecolor{pgf}{RGB}{10, 150, 10} 

\newcommand*{\mathcolor}{}
\def\mathcolor#1#{\mathcoloraux{#1}}
\newcommand*{\mathcoloraux}[3]{%
  \protect\leavevmode
  \begingroup
    \color#1{#2}#3%
  \endgroup
}
\newlength{\stheight}
\newcommand\textst[1][fu-grey]{
	\ifmmode\setlength{\stheight}{+1.0ex}
	\else\setlength{\stheight}{+0.5ex}
	\fi
	\bgroup\markoverwith{\textcolor{#1}{\rule[\the\stheight]{2pt}{1.0pt}}}\ULon
} 

\newcommand{\textins}[2][fu-grey]{
	\ifmmode\mathcolor{#1}{#2}
	\else\textcolor{#1}{#2}\@\,
	\fi
}

\newcommand{\jn}[1]{\textins[jn]{#1}} 

\graphicspath{{./}}
\notoc

  \tikzstyle{vecArrow} = [thick, decoration={markings,mark=at position
   1 with {\arrow[semithick]{open triangle 60}}},
   double distance=1.4pt, shorten >= 5.5pt,
   preaction = {decorate},
   postaction = {draw,line width=1.4pt, white,shorten >= 4.5pt}]

\begin{document}

\title{Interacting spin-2 fields \\ in the St\"uckelberg picture}


\author[a]{Johannes Noller}
\author[b]{, James H.C. Scargill}
\author[a]{, Pedro G. Ferreira}

\affiliation[a]{Astrophysics, University of Oxford, DWB, Keble Road, Oxford, OX1 3RH, UK} 
\affiliation[b]{Theoretical Physics, University of Oxford, DWB, Keble Road, Oxford, OX1 3NP, UK} 

\emailAdd{noller@physics.ox.ac.uk}
\emailAdd{james.scargill@physics.ox.ac.uk}
\emailAdd{p.ferreira1@physics.ox.ac.uk}

\abstract{
We revisit and extend the `Effective field theory for massive gravitons' constructed by Arkani-Hamed, Georgi and Schwartz in the light of recent progress in constructing ghost-free theories with multiple interacting spin-2 fields. We show that there exist several dual ways of restoring gauge invariance in such multi-gravity theories, find a generalised Fierz-Pauli tuning condition relevant in this context and highlight subtleties in demixing tensor and scalar modes. The generic multi-gravity feature of scalar mixing and its consequences for higher order interactions are discussed. In particular we show how the decoupling limit is qualitatively changed in theories of interacting spin-2 fields. We relate this to dRGT (de Rham, Gabadadze, Tolley) massive gravity, Hassan-Rosen bigravity and the multi-gravity constructions by Hinterbichler and Rosen. As an additional application we show that EBI (Eddington-Born-Infeld) bigravity and higher order generalisations thereof possess ghost-like instabilities.
}


\maketitle
\newpage


\section{Introduction} \label{sec-intro}

Massive gravity has recently experienced a remarkable renaissance. While at the linearised level the Fierz-Pauli action \cite{Fierz:1939ix} has long been known to consistently describe a single massive spin-2 field, any concrete fully non-linear extension/deformation of general relativity (GR) designed to give the graviton a mass appeared to inevitably run into problems \cite{Aragone:1971kh,Boulware:1973my,Aragone:1979bm,Nappi:1989ny,Duff:1989ea,Creminelli:2005qk,Deffayet:2005ys}.  Generating a mass by adding an `interaction term' (i.e. a potential) for the metric changes the constraint structure of the theory and generically leads to the appearance of a new ghost-like degree of freedom (\dof): the so-called Boulware-Deser ghost \cite{Boulware:1973my}. However, recently it was shown that a consistent and ghost-free theory of a massive spin-2 field can be constructed despite of these obstacles: dRGT (de Rham, Gabadadze, Tolley) gravity \cite{deRham:2010ik,deRham:2010kj,Hassan:2011hr}. 
For some related recent progress see \cite{deRham:2010tw,Hassan:2011vm,deRham:2011qq,deRham:2011rn,Mirbabayi:2011aa,Chamseddine:2011bu,Koyama:2011yg,Golovnev:2011aa,Chamseddine:2011mu,Burrage:2012ja,Hassan:2012qv,deRham:2012kf,Tasinato:2012ze,Tasinato:2013rza,Babichev:2013usa,Fasiello:2012rw,deRham:2013qqa,Gabadadze:2013ria,Ondo:2013wka,Fasiello:2013woa,deRham:2013hsa,deRham:2013awa}.
The particular ghost-free potential that lies at the heart of dRGT gravity is constructed using a fixed Minkowski reference metric $\eta$ in addition to a dynamical metric $g$, so the theory is intrinsically bimetric\footnote{Note that this `bimetricity' is a generic feature of massive gravity proposals, as a tensor other than $g$ is typically needed in order to generate any non-trivial (and non-GR) interaction terms. In dRGT this role is played by a non-dynamical Minkowski metric. However, note another interesting recent proposal for a massive gravity theory: The non-local massive gravity scenario proposed by \cite{Jaccard:2013gla} - also see \cite{Modesto:2013jea}. The proposed fully non-linear action considered there does not rely on an external reference metric at the expense of introducing non-local interactions.}. 
As a result the dRGT interaction terms have been extended to the case of two fully dynamical metrics, culminating in the construction of a consistent and ghost-free theory of bigravity: Hassan-Rosen bigravity \cite{Hassan:2011tf,Hassan:2011zd,Hassan:2011ea}. This theory describes one massless and one massive spin-2 field, consistent with the findings of \cite{Boulanger:2000rq} that any consistent theory of interacting spin-2 fields can contain at most one massless spin-2 field. 
For related cosmological and phenomenological studies see \cite{deRham:2011by,D'Amico:2011jj,Nieuwenhuizen:2011sq,Comelli:2011zm,Comelli:2011wq,Gumrukcuoglu:2011zh,Berezhiani:2011mt,Comelli:2012db,Wyman:2012iw,Gratia:2012wt,DeFelice:2012mx,Vakili:2012tm,Kobayashi:2012fz,D'Amico:2012pi,DeFelice:2013bxa,DeFelice:2013awa,Khosravi:2013axa,Hinterbichler:2013dv,Comelli:2013tja} and \cite{vonStrauss:2011mq,Volkov:2011an,Volkov:2012cf,Akrami:2012vf,Sakakihara:2012iq,Khosravi:2012rk,Berg:2012kn,Akrami:2013ffa,Akrami:2013pna,Maeda:2013bha,Volkov:2013roa,Babichev:2013pfa} for dRGT/massive gravity and Hassan-Rosen bigravity theories respectively. 
Having successfully constructed consistent theories for a single and two dynamical spin-2 fields, a particularly intriguing question arises: What is the picture for N interacting spin-2 fields? Can we construct consistent, ghost-free models of an arbitrary number of spin-2 fields? What is their structure and what are the physical features and associated phenomenology of such theories? In \cite{Hinterbichler:2012cn} the dRGT/Hassan-Rosen interaction terms were generalised by Hinterbichler and Rosen to produce a particular set of consistent and ghost-free interactions for N interacting spin-2 fields, opening the doors for investigating such `multi-gravity' models with a concrete ghost-free proposal. 
For related multi-gravity work also see \cite{
Khosravi:2011zi,Nomura:2012xr,Hassan:2012wr,Hassan:2012wt,Tamanini:2013xia}.

An essential stepping stone in the process leading up to the construction of consistent theories of massive gravity, and their extensions to multiple spin-2 fields, was the systematic development of an `Effective field theory for massive gravitons'\footnote{This is an effective field theory in the sense of constructing a general framework for arbitrary potentials, i.e. non derivative interaction terms between the different spin-2 fields. Derivative interactions other than self-interactions provided by the Ricci scalar are not considered here.}\cite{ArkaniHamed:2002sp} - also see \cite{ArkaniHamed:2001ca,ArkaniHamed:2003vb,Schwartz:2003vj}. This uses the so-called \St trick to project out all relevant \dof in the theory and make the nature of their interaction terms explicit (in this context also see \cite{Siegel:1993sk}). In particular this approach establishes the hierarchy of energy scales in the effective theory, its cutoff scale and how this scale can be altered/raised. Consequently it makes transparent which interactions are most relevant at low energies. Especially interesting is also the relation of these studies to higher dimensional theories of gravity \cite{ArkaniHamed:2001ca,ArkaniHamed:2003vb,Schwartz:2003vj,Deffayet:2005yn,Deffayet:2003zk,deRham:2013awa}.  In the effective field theory context, the particular models investigated by \cite{ArkaniHamed:2002sp} were two site models with Fierz-Pauli interactions as well as some specific generalisations to theories of $N$ spin-2 fields, again with bimetric Fierz-Pauli interactions.

In this paper we endeavour to complete this effective field theory picture by constructing what we dub an `Effective field theory for interacting spin-2 fields'. Specifically this means we extend and generalise the `Effective field theory for massive gravitons' by \cite{ArkaniHamed:2002sp} to deal with generic (i.e. non Fierz-Pauli) N-metric interaction terms -- especially relevant in the light of the afore-mentioned progress in massive/multi-gravity -- and general combinations of interaction terms as discussed by \cite{Hinterbichler:2012cn} rather than restricting ourselves to purely nearest-neighbour interactions as in \cite{ArkaniHamed:2002sp}. 
This allows us to identify several qualitatively new features of generic multi-spin-2 field scenarios. The overarching aim here is to systematically understand the \dof and energy scales involved in such theories,  leading to a better understanding of N spin-2 field theories constructed with known ghost-free interactions of the dRGT/Hassan-Rosen/Hinterbichler-Rosen type as well as potentially paving the way for the discovery of new additional consistent interactions. In the process of constructing an effective field theory view in the \St picture, we discuss some of the new features and subtleties that appear when the approach of \cite{ArkaniHamed:2002sp} is generalised to generic multi-spin-2 field scenarios. In particular we also extend the approach to a detailed analysis of higher order interactions in the so-called decoupling limit as well as uncovering `mixing phenomena' at quadratic order. Throughout we will work in the metric picture - for a discussion of how this relates to interacting spin-2 field theories constructed in the vielbein picture (as in \cite{Hinterbichler:2012cn}) we refer to \cite{vielbein}.\footnote{Note that interaction terms written down in the vielbein picture can not necessarily be represented in the metric picture and vice versa \cite{Hinterbichler:2012cn,Deffayet:2012nr,Deffayet:2012zc}. In order to analyse several generic features of interacting spin-2 field theories, as done in this paper, it is irrelevant which picture is employed. However, in order to analyse concrete theories constructed in one picture or the other, complementing the approach presented here with an explicit mapping to the vielbein case will be very useful. We present this in \cite{vielbein}.}

The outline for this paper is as follows. In section \ref{sec-int} we introduce theories of interacting spin-2 fields, discuss how such theories can be constructed in general and what the dynamical \dof are. In section \ref{sec-link} we review the \St trick in the context of massive gravity and discuss how it can be extended to restore gauge invariance in theories with multiple spin-2 fields. Various subtleties involving this extension are discussed in section \ref{sec-multiSt}, where we also show that there are several dual approaches to introduce \St fields in interacting N spin-2 field theories. We construct a particularly economical approach and also discuss some features of `loops' in theory graphs. In section \ref{sec-gold} we then discuss how the \St scalars -- Goldstone bosons of the $N-1$ broken diffeomorphism invariances in a theory with $N$ spin-2 fields and corresponding to the longitudinal scalar components of the `massive gravitons' -- can be demixed from tensor perturbations and how they acquire a kinetic term in the process. We furthermore derive a generalised Fierz-Pauli tuning condition, which theories of interacting spin-2 fields need to satisfy and discuss gauge-fixing issues relevant to eliminating scalar-tensor mixing at lowest order. In section \ref{sec-can} we show that interacting spin-2 fields generically display scalar mixing, i.e. the \St scalars mix kinetically and the true propagating \dof are in fact linear combinations of the \St scalars. This is a feature absent in bigravity (or single spin-2 field) theories, but generic in the presence of more than two fields. A discussion of higher order interactions and the decoupling limit follows in section \ref{sec-cubic}, where we show that the decoupling limit in multi-gravity theories is qualitatively different from that in bi- and massive gravity. Finally, in sections \ref{sec-EBI},\ref{sec-dRGT} and \ref{sec-EBI4}, we use the machinery developed throughout the paper to analyse three example theories: Eddington-Born-Infeld (EBI) bigravity, Hassan-Rosen bigravity and an EBI bigravity extension to four interacting spin-2 fields respectively. The EBI bigravity case is especially noteworthy, since we show that this theory has ghost-like instabilities. In section \ref{sec-conc} we conclude and summarise our findings. Some further useful results are collected in appendices \ref{appendix-TD}, \ref{appendix-nonlocal} and \ref{appendix-curved}.

\section{Interacting spin-2 fields} \label{sec-int}

{\bf A multi metric theory:} We begin with the following schematic action for N-spin-2 fields $g_{\mu\nu}^{(i)}$  
\be
{\cal S} = {\cal S} _{site} + {\cal S}_{int} + {\cal S}_{matter}.  
\ee
The first term ${\cal S} _{site}$ encompasses all derivative self-interaction terms for the $g_{\mu\nu}^{(i)}$, while ${\cal S}_{int}$ contains (non-derivative) interactions between all $g_{\mu\nu}^{(i)}$ and ${\cal S}_{matter}$ describes the coupling of the spin-2 fields to other matter in the universe. As such we may write
\bea
{\cal S}_{site} &=& \sum_i^N \int d^4 x M_{Pl (i)}^2 \sqrt{g^{(i)}} R\left[g^{(i)}_{\mu\nu}\right] \\
{\cal S}_{int} &=&  {\cal S}_{int}\left[g^{(1)}_{\mu\nu}, \ldots,  g^{(N)}_{\mu\nu} \right]\\
{\cal S}_{matter} &=&  {\cal S}_{matter}\left[\Phi_M, g^{(1)}_{\mu\nu}, \ldots,  g^{(N)}_{\mu\nu}\right],
\eea
where $\Phi_M$ labels all other matter fields ($M$ being the labelling index). Note that we have made two key simplifying assumptions: Firstly, no derivative interactions between different spin-2 fields are considered. Secondly, the derivative self-interactions of each spin-2 field are taken to be the corresponding Ricci scalar (for some recent progress in understanding different `kinetic terms' for spin-2 fields see \cite{Hinterbichler:2013eza,Kimura:2013ika,Folkerts:2011ev}) and the associated mass scale is assumed to be universal, i.e. $M_{Pl (i)}^2 = M_{Pl}^2$. To clarify our notation: Throughout this paper we use bracketed (Latin) indices to label different fields - they are also `site indices' as we will explain shortly. Whether these indices are upper or lower indices has no meaning. Greek indices are space-time indices, which are raised and lowered with the flat Minkowski metric $\eta$.
\\

{\bf The coupling to matter: } By construction ${\cal S}_{site}$ provides consistent and ghost-free self-interactions for all the spin-2 fields\footnote{Note that we are explicitly excluding any auxiliary fields to enter at this stage by making each $g_{\mu\nu}^{(i)}$ dynamical via ${\cal S}_{site}$.}. 
In what follows we will therefore focus on ${\cal S}_{int}$, devising a systematic way to understand the physics coming out of the \dof arising as a result of the symmetries broken by ${\cal S}_{int}$ in theories of multiple interacting spin-2 fields. While we will not have much to say about the matter sector ${\cal S}_{matter}$, it is worth making a few observations before exclusively focusing on ${\cal S}_{int}$ throughout the rest of this paper. Firstly note that we can always choose to minimally couple matter to just one $g_{\mu\nu}^{(i)}$ 
\be
{\cal S}_{matter} =  {\cal S}_{matter}\left[\Phi_M, g^{(i)}_{\mu\nu}\right],
\ee
This ensures that a theory with healthy and consistent ${\cal S}_{site}$ and ${\cal S}_{int}$ remains so after coupling to matter - the constraint structure remains unchanged by the minimal coupling to just one spin-2 field \cite{Hassan:2011zd}. A particular bi-metric coupling has recently been investigated by \cite{Akrami:2013ffa}. Note, however, that multi-metric couplings generically fall into two categories. Either the (weak) equivalence principle is fully respected and consequently matter couples to a single effective metric $\tilde{g}_{\mu\nu}^{(matter)}$ (minimally in the associated Jordan frame). 
\be \label{matteraction}
{\cal S}_{matter} =  {\cal S}_{matter}\left[\Phi_M, \tilde{g}_{\mu\nu}^{(matter)}[g^{(1)}_{\mu\nu}, \ldots,  g^{(N)}_{\mu\nu}]\right].
\ee
$\tilde{g}_{\mu\nu}^{(matter)}$ will in general be a function of all $N$ spin-2 fields $g_{\mu\nu}^{(i)}$ in the theory. We now rewrite the action in terms of  $\tilde{g}_{\mu\nu}^{(i)}$, i.e. we `rotate' our spin-2 field space from $g^{(i)}_{\mu\nu} \to \tilde{g}_{\mu\nu}^{(i)}$, where $\tilde{g}_{\mu\nu}^{(n)} =\tilde{g}_{\mu\nu}^{(matter)}$ for some $n$. Expressed in this basis finding a consistent and ghost-free theory once again becomes equivalent to establishing the correct structure for $\tilde{\cal S}_{site}$ and $\tilde{\cal S}_{int}$ in terms of the $\tilde{g}_{\mu\nu}^{(i)}$.\footnote{In other words, re-writing the matter part of the action as \eqref{matteraction} does not guarantee that the kinetic part of the action can be written $\sum_i R\left[\tilde{g}^{(i)}\right]$ or that interactions between the different $\tilde{g}^{(i)}$ do not introduce ghosts.}
Alternatively, if the matter coupling is `truly bimetric' (as in \cite{Akrami:2013ffa}) different matter species $\Phi_M$ will couple to different effective metrics, leading to weak equivalence principle violations (which it may indeed be interesting to constrain).
\\

\begin{figure}[tp]
\centering
\begin{tikzpicture}[-,>=stealth',shorten >=0pt,auto,node distance=2cm,
  thick,main node/.style={circle,fill=blue!10,draw,font=\sffamily\large\bfseries},arrow line/.style={thick,-},barrow line/.style={thick,->},no node/.style={plain},rect node/.style={rectangle,fill=blue!10,draw,font=\sffamily\large\bfseries},red node/.style={rectangle,fill=red!10,draw,font=\sffamily\large\bfseries},green node/.style={circle,fill=green!20,draw,font=\sffamily\large\bfseries},yellow node/.style={rectangle,fill=yellow!20,draw,font=\sffamily\large\bfseries}]

 \node[main node](100){};
  \node[main node] (101) [right of=100] {};
 
  \node[main node,fill=black!100,scale=0.7] (2)  [right=3cm of 101]{};
   \node[main node] (1) [above=1cm of 2] {};
  \node[main node] (3) [below left=1cm of 2] {};
   \node[main node] (4) [below right=1cm of 2] {};
   
    \node[main node,fill=black!100,scale=0.7] (5)  [right=4cm of 2]{};
  \node[main node] (7) [below left=1cm of 5] {};
   \node[main node] (8) [below right=1cm of 5] {};
   
     \node[draw=none,fill=none](80)[above left=1.5cm of 5]{};
       \node[draw=none,fill=none](81)[above right=1.5cm of 5]{};
        \node[draw=none,fill=none](82)[above=1.5cm of 5]{};
         \node[draw=none,fill=none](83)[right=0.7cm of 100]{};

  \path[every node/.style={font=\sffamily\small}]
  (100) edge node {} (101)
    (1) edge node {} (2)
     (4) edge node {} (2)
     (3) edge node  {} (2)
      (7) edge node [above left] {} (5)
       (8) edge node [above right] {} (5);


\draw[-,dashed] (5) to (80);
\draw[-,dashed] (5) to (81);
\draw[-,dashed] (5) to (82);
    
\draw [<->] ($(8)+(0.2,0.2)$) 
    arc (-45:225:1.65);

   \node[draw=none,fill=none](92)[below of=83]{(a) Bimetric};
   
   \node[draw=none,fill=none](93)[below of=2]{(b) Trimetric};
   
   \node[draw=none,fill=none](94)[below of=5]{(c) $N$-metric};
\end{tikzpicture}
\caption{Different types of multi-metric interactions and their graphical representation. Small black nodes denote interaction vertices (except for bimetric interactions where we drop the black node) and large light nodes are the sites corresponding to  spin-2 fields participating in the interaction (and their associated copy of $GC_{(i)}$).} \label{fig-interactions}
\end{figure}
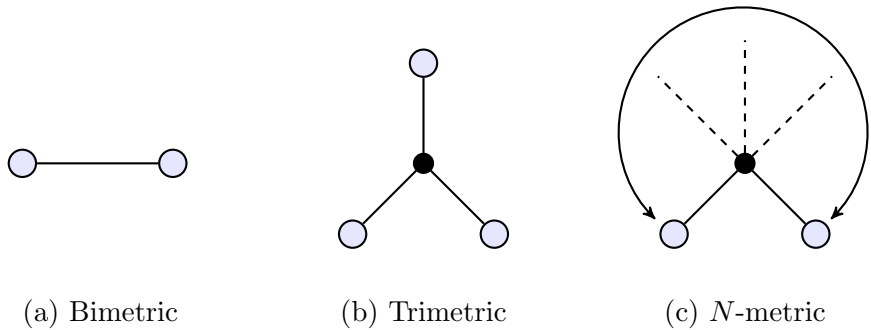

{\bf A single interaction term:} The kinetic terms in ${\cal S}_{site}$ manifestly respect $N$ general coordinate invariances - one for each spin-2 field. Following the convention of  \cite{ArkaniHamed:2002sp,ArkaniHamed:2003vb,Schwartz:2003vj} we denote these by $GC_{(i)}$, where the label $i$ indicates with which site/spin-2 field a given symmetry group is to be associated. As a result of the presence of $N$ $GC_{(i)}$ there are $N$ diffeomorphism invariances
\be
g_{\mu\nu}^{(i)}(x) \to \pa_\mu d_{(i)}^\alpha  \pa_\nu d_{(i)}^\beta  g_{\alpha\beta}^{(i)}(d(x)),
\ee
where $d_{(i)}$ is a diffeomorphism. However, ${\cal S}_{int}$ will generically break several $GC_{(i)}$. Let us begin by considering individual interaction terms. Following in the footsteps of \cite{ArkaniHamed:2002sp,ArkaniHamed:2003vb,Schwartz:2003vj,Hinterbichler:2012cn}, it is helpful to represent these terms graphically with the ultimate aim of painting a theory graph representing a full multi-metric theory. Figure \ref{fig-interactions} shows the possible interaction terms ordered by the number of participating spin-2 fields (we choose the same `theory graph conventions' as \cite{Hinterbichler:2012cn}). Every site/node corresponds to a spin-2 field, a line connecting two sites denotes a bimetric interaction term and $n$ sites connected via lines to a single small black node denotes an $n$-metric interaction vertex. 

Consider one such vertex for $n$ fields in isolation as in figure \ref{fig-interactions} (c). Before the interaction term is introduced there are $n$ non-interacting spin-2 fields, so $n$ unbroken copies of $GC_{(i)}$ which result in 2 degrees of freedom (\dof) on each site corresponding to those of a massless graviton. Consequently there are $2n$ degrees of freedom in total. The interaction term results in the breaking of $n-1$ copies of this gauge invariance, leaving only one copy intact (this copy being the diagonal subgroup of all $n$ previously unbroken $GC_{(i)}$). As a consequence of the broken symmetry we have $n-1$ (propagating) Goldstone modes which get eaten by $n-1$ spin-2 fields. These fields become massive as a result. The full theory therefore contains one massless spin-2 field (2 \dof) and $n-1$ massive spin-2 fields (with 5 \dof each). These take up a total of $2 + 5(n-1)$ \dof.
However, a priori each spin-2 field contributes 6 \dof, resulting in $6n$ \dof for the case we are considering here. The surviving copy of $GC_{(i)}$ removes 4 \dof and, as we just worked out, the single massless and $n-1$ massive fields take up another $2 + 5(n-1)$ \dof. This leaves us with $6n - 4 - 2 - 5 (n-1) = n-1$ \dof. These are $n-1$ potential Boulware-Deser ghosts and only if ${\cal S}_{int}$ has the right structure to enforce constraints, eliminating the \dof associated to these would-be ghosts, is the theory healthy.  In the following sections we will discuss the St\"uckelberg trick, which will allow us to explicitly understand the physics associated to all these \dof, in particular the nature of their interactions. 
\\

\comment{Our aim in this paper is much more modest, however. Performing the full constraint analysis for arbitrary interaction terms is very challenging indeed, albeit ultimately necessary to find the most general `healthy' interaction terms. In the following sections we will discuss the St\"uckelberg trick and a generalised decoupling limit ultimately giving a very straightforward recipe for checking whether a given theory has a ghost in this decoupling limit. Showing that a given theory contains no ghost in this limit is no guarantee that the full theory is healthy, but it is a very useful tool in order to quickly check whether any candidate theory is worth investigating further. If an instability is present at the decoupling level, then the full theory has a ghost. }

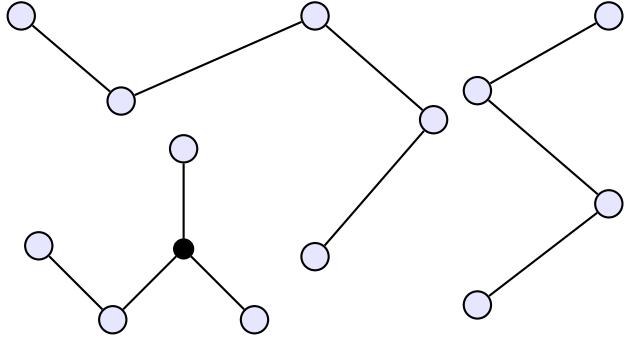
\begin{figure}[tp]
\centering
\begin{tikzpicture}[-,>=stealth',shorten >=0pt,auto,node distance=1cm,
  thick,main node/.style={circle,fill=blue!10,draw,font=\sffamily\small\bfseries}]

  \node[main node] (1) {};
 \node[main node, right=100pt,at=(1.east)](2) {};
  \node[main node, right=100pt,at=(2.east)](3) {};
   \node[main node, below right=50pt,at=(2.east)](4) {};
     \node[main node,above right=10pt,at=(4.east)](5){};
     \node[main node,below right=40pt,at=(1.east)](6){};
     \node[main node,below right=20pt,at=(6.east)](7){};
              \node[main node, below=80pt,at=(2.south)](10){};
                  \node[main node, below=60pt,at=(3.south)](11){};
                      \node[main node, below=70pt,at=(5.south)](12){};
                      
   \node[main node,fill=black!100,minimum size=0.01cm,scale=0.7] (20)  [below=1cm of 7]{};
  \node[main node] (30) [below left=1cm of 20] {};
   \node[main node] (40) [below right=1cm of 20] {};
   \node[main node] (50) [above left=1cm of 30] {};
                      
                       \path[every node/.style={font=\sffamily\small}]
                   (20) edge node {} (30)
                     (30) edge node {} (50)
                    (20) edge node {} (40)
                     (20) edge node {} (7)     
    (1) edge node {} (6)
      (6) edge node {} (2)
      (2) edge node {} (4)
       (4) edge node {} (10)
            (3) edge node {} (5)
      (5) edge node {} (11)
       (11) edge node {} (12);

\end{tikzpicture}
\caption{Here we picture a system of 13 interacting spin-2 fields with interaction terms involving two or three fields. Three graphs internally connected by interactions, but disconnected from all the other fields, remain. As a result we have three surviving copies of $GC$ and $13-3=10$ propagating modes/Goldstone bosons for this setup. Each line for bimetric interactions breaks one copy of $GC_{(i)}$ and hence contributes one Goldstone boson $\pi^{(i)}$, whereas $N$-metric interactions break $N-1$ copies of $GC_{(i)}$ contributing $N-1$ Goldstone bosons $\pi^{(i)}$.} \label{fig-Nfields}
\end{figure}

{\bf Multiple interaction terms: } But before moving on to the St\"uckelberg trick, let us briefly think about how to put the above reasoning for a single interaction vertex together in order to produce a theory graph for an arbitrary ${\cal S}_{int}$. We can write a general interaction term with $n$ participating spin-2 fields
\bea
{\cal S}_{int,1} &=&  \sum_i \int d^4 x \sqrt{g^{(i)}} \Lambda_i \\
{\cal S}_{int,n} &=&  \sum_{i_1, \ldots, i_n}^{N} c_{i_1 \ldots i_n} \int d^4 x f^{(i_1 \ldots i_n)}_n \left[g^{(i_1)}_{\mu\nu}, \ldots, g^{(i_n)}_{\mu\nu} \right],
\eea
where the $c_{i_1 \ldots i_n} $ are constant coefficients. Note that the $n=1$ case is not fundamentally different from the other $n$, but since the only Lorentz scalar we can form with just one spin-2 field is a constant, the $n=1$ term is simply a cosmological constant. Now consider a generic multi-metric theory with interaction terms as depicted in figure \ref{fig-Nfields}\footnote{Note we have restricted ourselves to bimetric and trimetric interaction terms in figure \ref{fig-Nfields} to provide a simple example, but there is of course no fundamental reason why more spin-2 fields cannot participate.}.
If the theory of interest is described by a single fully connected theory graph, then ${\cal S}_{int}$ breaks all but one copy of $GC$, reducing the $N$ general coordinate invariances $GC_{(i)}$ down to a single diagonal subgroup. As pointed out in  \cite{Hinterbichler:2012cn} the situation is somewhat different for more general ${\cal S}_{int}$ as  depicted in figure \ref{fig-Nfields}. The number of unbroken copies of $GC_{(i)}$ corresponds to the number of disconnected `islands' in the theory graph. As a result, in figure \ref{fig-Nfields}, 3 copies of $GC_{(i)}$ remain intact, so that the theory contains 3 massless and 10 massive spin-2 fields, which have eaten the 10 Goldstone bosons arising from the broken copies of $GC_{(i)}$. Establishing whether any of the potential 10 Boulware-Deser ghosts are present requires knowledge of the exact nature of the interaction terms.

A variety of interaction terms have been considered in the context of multiple interacting spin-2 fields: nearest neighbour Fierz-Pauli interactions as well as more general Fierz-Pauli terms, e.g. for truncated Kaluza-Klein models \cite{Schwartz:2003vj}, are still plagued by ghosts. dRGT massive gravity  \cite{deRham:2010ik,deRham:2010kj,Hassan:2011hr} was the first successful (ghost-free) model of a single interacting massive graviton - its action is that of a bimetric theory with one non-dynamical metric (the Minkowski $\eta_{\mu\nu}$) and interaction terms formed out of elementary symmetric polynomials of $\sqrt{g^{\mu\nu}\eta_{\nu\alpha}}$. The same type of interaction term has been generalised to the case of a fully dynamical bimetric theory and shown to be ghost-free by Hassan and Rosen \cite{Hassan:2011tf,Hassan:2011zd,Hassan:2011ea} (see example II in section \ref{sec-dRGT}) and subsequently for $N$-metric interactions by Hinterbichler and Rosen  \cite{Hinterbichler:2012cn}. Whether other types of ghost-free interaction terms exist for such theories remains to be seen. Consequently our philosophy throughout this paper (until we consider some concrete examples at the end, at least) will be to remain agnostic about the particular form of any interaction term and analyse N spin-2 field theories without imposing any particular interaction structure.

\section{Link fields and the St\"uckelberg trick} \label{sec-link}

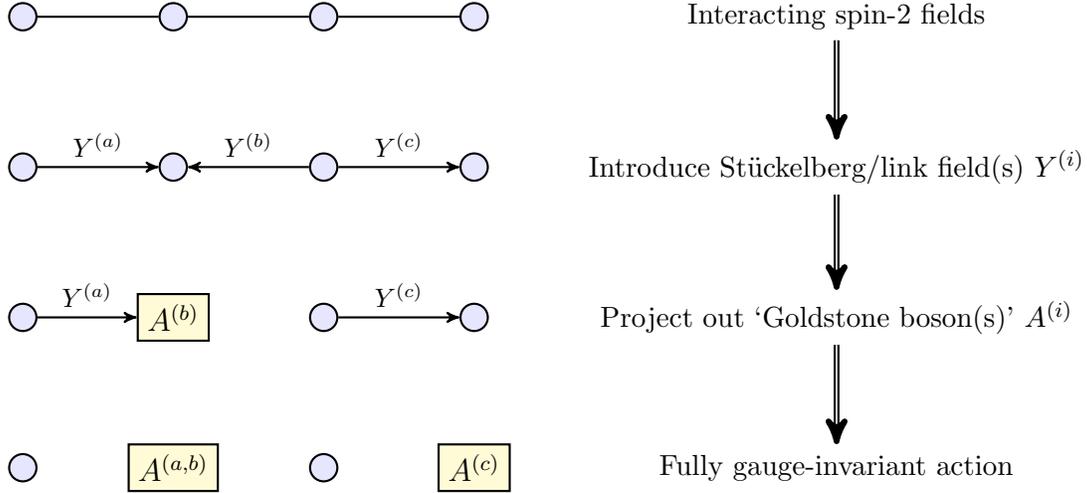
\begin{figure}[tp]
\centering
\begin{tikzpicture}[->,>=stealth',shorten >=0pt,auto,node distance=2cm,
  thick,main node/.style={circle,fill=blue!10,draw,font=\sffamily\large\bfseries},arrow line/.style={thick,-},barrow line/.style={thick,->},no node/.style={plain},rect node/.style={rectangle,fill=blue!10,draw,font=\sffamily\large\bfseries},red node/.style={rectangle,fill=red!10,draw,font=\sffamily\large\bfseries},green node/.style={circle,fill=green!20,draw,font=\sffamily\large\bfseries},yellow node/.style={rectangle,fill=yellow!20,draw,font=\sffamily\large\bfseries}]

  \node[main node] (1) {};
  \node[main node] (2) [right of=1] {};
    \node[main node] (3) [right of=2] {};
      \node[main node] (4) [right of=3] {};
  \node[main node] (5) [below of=1] {};
  \node[main node] (6) [below of=2] {};
    \node[main node] (7) [below of=3] {};
  \node[main node] (8) [below of=4] {};
    \node[main node] (9) [below of=5] {};
  \node[yellow node] (10) [below of=6] {$A^{(b)}$};
    \node[main node] (11) [below of=7] {};
  \node[main node] (12) [below of=8] {};
      \node[main node] (13) [below of=9] {};
  \node[yellow node] (14) [below of=10] {$A^{(a,b)}$};
    \node[main node] (15) [below of=11] {};
  \node[yellow node] (16) [below of=12] {$A^{(c)}$};
  
  \node[draw=none,fill=none](80)[below right of=1]{};
   \node[draw=none,fill=none](81)[above right of=5]{};
   \node[draw=none,fill=none](82)[right=2.5cm of 4]{Interacting spin-2 fields};
   
   \node[draw=none,fill=none](83)[below of=82]{Introduce St\"uckelberg/link field(s) $Y^{(i)}$};
   
   \node[draw=none,fill=none](84)[below of=83]{Project out `Goldstone boson(s)' $A^{(i)}$};
   
   \node[draw=none,fill=none](85)[below of=84]{Fully gauge-invariant action};

  \path[every node/.style={font=\sffamily\small}]
    (5) edge node [above] {$Y^{(a)}$} (6)
    (7) edge node [above] {$Y^{(b)}$} (6)
    (7) edge node {$Y^{(c)}$} (8)
       (9) edge node [above] {$Y^{(a)}$} (10)
          (11) edge node [above] {$Y^{(c)}$} (12);
            
            \draw[arrow line] (1) -- (2);
              \draw[arrow line] (2) -- (3);
               \draw[arrow line] (3) -- (4);
               
               \draw[barrow line, double] (82) -- (83);
              \draw[barrow line, double] (83) -- (84);
               \draw[barrow line, double] (84) -- (85);
            
\end{tikzpicture}
\caption{The St\"uckelberg trick: We start with a set of sites/spin-2 fields connected by (in the case here bimetric) interaction terms. St\"uckelberg/link fields are introduced, mapping interaction terms onto individual sites. Note that it is an arbitrary choice to which site (that participates in the interaction) the term is mapped and that there is no directedness to the lines before introducing a link field. We then expand the \St fields around their new site basis, projecting out a Goldstone boson associated with the symmetry broken by the original interaction term. Doing this to all the interaction terms we end up with a theory in which all the gauge-invariances broken by interaction terms in the original theory are restored at the expense of introducing the $A^{(i)}$ as gauge degrees of freedom. This way of applying the \St trick corresponds to {\it approach I} as shown in equation \eqref{bimetricExampleI}. Dual approaches are discussed below. Note that each link leads to interaction terms solely depending on one Goldstone boson $A^{(i)}$, i.e. there are no cross-terms between different Goldstone bosons in the approach shown here (there is scalar-tensor mixing, however, which we will later see reintroduces Goldstone mixing). } \label{fig-Stueck}
\end{figure}

{\bf The \St field:} A generic multi-metric action such as
\be \label{fixedSint}
{\cal S} = \sum_i^N \int d^4 x M_{Pl}^2 \sqrt{g^{(i)}} R\left[g^{(i)}_{\mu\nu}\right] + {\cal S}_{int}\left[g^{(1)}_{\mu\nu}, \ldots,  g^{(N)}_{\mu\nu} \right] 
\ee
is effectively a gauge-fixed action\footnote{This is partially gauge-fixed in the sense that the gauge-invariance for the surviving single copy of $GC_{(i)}$ is still fully intact.}, as indicated by the fact that $N-1$ copies of $GC_{(i)}$ are broken by ${\cal S}_{int}$. We can restore the broken gauge invariances by employing the so-called St\"uckelberg trick. Consider a site $i$ in our theory graph with associated spin-2 field $g_{\mu\nu}^{(i)}$ and general coordinate invariance $GC_{(i)}$ (parametrised by diffeomorphisms $d^\mu_{(i)} (x)$), which has been broken by a single interaction term. ${\cal S}_{int}$ is consequently {\it not} invariant under the gauge transformation
\be
g_{\mu\nu}^{(i)}(x) \overset{d_{(i)}}{\longrightarrow} \pa_\mu d_{(i)}^\alpha  \pa_\nu d_{(i)}^\beta  g_{\alpha\beta}^{(i)}(d(x)).
\ee
We now introduce the St\"uckelberg field $Y^\mu_{(i)}$ and use it to replace all occurrences of $g_{\mu\nu}^{(i)}$ in the interaction term via the following substitution, mimicking the diffeomorphism invariance symmetry
\be
g_{\mu\nu}^{(i)}(x) \to G_{\mu\nu}^{(i)} \equiv \pa_\mu Y^\alpha_{(i)} \pa_\nu Y^\beta_{(i)} g_{\alpha\beta}^{(i)}(Y(x)).
\ee
Note that, as far as the $R^{(i)}$ term is concerned, this is a gauge transformation with gauge function $Y^\mu_{(i)}$, which leaves $R^{(i)}$ invariant. $G_{\mu\nu}^{(i)}$ is now also invariant under gauge transformation parametrised by $d^\mu_{(i)} (x)$, where $Y^\mu_{(i)}$ transforms as
\be
Y^\mu_{(i)}(x) \overset{d_{(i)}}{\longrightarrow} d^{-1}_{(i)} Y^\mu_{(i)}(x).
\ee 
Having replaced $g_{\mu\nu}^{(i)}$ with $G_{\mu\nu}^{(i)}$, ${\cal S}_{int}$ is therefore now invariant under gauge transformations generated by $d^\mu_{(i)} (x)$.\footnote{Note that we will discuss how $Y^\mu_{(i)}$ transforms under diffeomorphisms associated with different sites $d^\mu_{(j)}$ with $i \neq j$ further below.} We will see shortly that the \St replacement corresponds to introducing extra gauge degrees of freedom into the action.
\\

{\bf Functional composition: }  Taking a lead from \cite{ArkaniHamed:2002sp}, it will be useful to introduce some notation at this point, writing gauge transformations in terms of functional composition $\circ$. Consider three objects $\phi, a_\mu,g_{\mu\nu}$ transforming as a scalar, one-form and metric under $GC_{(i)}$ respectively. $d^\mu_{(i)} (x)$ is the gauge function generating these transformations. Establishing a dictionary for writing these transformations we have
\begin{align}
\phi(x) &\overset{d_{(i)}}{\longrightarrow}  \phi(d(x))      &\phi &\overset{d_{(i)}}{\longrightarrow}  \phi \circ d_{(i)}
\\ a_\mu(x) &\overset{d_{(i)}}{\longrightarrow}  \pa_\mu d^\alpha_{(i)} a_\alpha (d(x))       &{\bf a} &\overset{d_{(i)}}{\longrightarrow}  {\bf a} \circ d_{(i)} \\
g_{\mu\nu}^{(i)}(x) &\overset{d_{(i)}}{\longrightarrow} \pa_\mu d_{(i)}^\alpha  \pa_\nu d_{(i)}^\beta  g_{\alpha\beta}^{(i)}(d(x))     &{\bf g} &\overset{d_{(i)}}{\longrightarrow} {\bf g} \circ d_{(i)}.
\end{align}
Clearly gauge transformations are much simpler in the functional composition language as we no longer need to worry about the valence of the object in question. For an arbitrary tensor ${\bf p}_{(i)}$, which transforms under $GC_{(i)}$, but remains invariant under 
other copies of general coordinate invariance $GC_{(j)}$ (where $i \neq j$) we therefore have
\begin{align}
{\bf p}_{(i)} &\overset{d_{(i)}}{\longrightarrow} {\bf p}_{(i)} \circ d_{(i)}       &{\bf p}_{(i)} &\overset{d_{(j)}}{\longrightarrow} {\bf p}_{(i)}.
\end{align}
The \St transformation that makes an action, which is built out of ${\bf p}_{(i)}$ and other fields not transforming under $GC_{(i)}$, invariant under $GC_{(i)}$ is therefore 
\be
{\bf p}_{(i)} \overset{d_{(i)}}{\longrightarrow} {\bf p}_{(i)} \circ Y_{(i)} 
\ee
where ${\bf p}$ and $Y$ transform as
\begin{align}
{\bf p}_{(i)} &\overset{d_{(i)}}{\longrightarrow} {\bf p}_{(i)} \circ d_{(i)}    &Y_{(i)} &\overset{d_{(i)}}{\longrightarrow} d^{-1}_{(i)} \circ Y_{(i)}.
\end{align}
\\

{\bf Link fields: } So far we have only thought about making an interaction term, which is initially not invariant under a particular $GC_{(i)}$, invariant under the same group of transformations.  In a setting such as dRGT massive gravity, where only one metric is dynamical, this is all that is required. However, in the context of $N$ dynamical spin-2 fields, the \St trick can be made much more powerful than this. By specifying a more complex transformation law for the \St field $Y^{\mu}$, this field can be viewed as a `link field' mapping an object living on site $i$ (i.e. transforming under $GC_{(i)}$ and invariant under $GC_{(j)}$, where $i \neq j$) into an object living on site $j$ (i.e. transforming under $GC_{(j)}$ and invariant under $GC_{(k)}$, where $k \neq j$, but e.g. $k=i$ is allowed). This means that $Y^{\mu}$ has to transform under $GC_{(i)} \times GC_{(j)}$ as
\be
Y_{(ij)} \overset{d_{(i)},d_{(j)}}{\longrightarrow} d^{-1}_{(i)} \circ Y_{(ij)} \circ d_{(j)}.
\ee
Let us check that ${\bf g}_{(i)} \circ Y_{(ij)}$ now has the correct transformation properties under both $GC_{(i)}$ and $GC_{(j)}$
\be
{\bf G}_{(j)} \equiv {\bf g}_{(i)} \circ Y_{(ij)} \overset{d_{(i)},d_{(j)}}{\longrightarrow} {\bf g}_{(i)} \circ d_{(i)} \circ d^{-1}_{(i)} \circ Y_{(ij)} \circ d_{(j)} = {\bf G}_{(j)} \circ d_{(j)}. 
\ee
It is instructive to think of the link field $Y_{(ij)}$ as generating the pull-back of ${\bf g}_{(i)}$ from site $i$ onto site $j$. The dRGT case, with only one dynamical spin-2  field, in this language corresponds to mapping onto the site of a non-dynamical metric, such as $\eta_{\mu\nu}$, which does not transform under any $GC_{(j)}$. Note that, alternatively, one could also map the non-dynamical metric, which does not transform under any $GC_{(j)}$, to the site of the dynamical metric. What about more general interactions?

For a bimetric interaction between two spin-2 fields ${\bf g}_{(i)}$ and ${\bf g}_{(j)}$, as already considered in \cite{ArkaniHamed:2002sp}, we can use the generalised \St field $Y^{\mu}_{(ij)}$ to map the whole interaction onto one site, say $j$
\be
{\cal S}_{int} = \int d^4 x \sqrt{g_{(j)}} f({\bf g}_{(i)},{\bf g}_{(j)}) \to \int d^4 x \sqrt{g_{(j)}} f({\bf g}_{(i)} \circ Y_{(ij)},{\bf g}_{(j)},) = \int d^4 x \sqrt{g_{(j)}} f({\bf G}_{(j)},{\bf g}_{(j)}). 
\ee 
The resulting ${\cal S}_{int}$ is invariant under $GC_{(i)}$ by construction (nothing transforms under $GC_{(i)}$) and as long as $f(\ldots)$ describes a Lorentz scalar, it will also be invariant under $GC_{(j)}$. Note that the $\sqrt{g_{(j)}}$ term in the initial ${\cal S}_{int}$ could be absorbed into $f$, but we have kept it here to indicate that the form of the interaction term quite often suggests a natural site to which to map the term.

Extending this formalism to an $n$ spin-2 field interaction vertex is now straightforward. Consider a single such vertex 
\be
{\cal S}_{int} = \int d^4 x f({\bf g}_{(i_1)},\ldots,{\bf g}_{(i_n)}),
\ee
which transforms under $GC_{(i_1)} \ldots GC_{(i_n)}$ and hence breaks $n-1$ gauge symmetries. We choose an arbitrary site $i_k$ where $k \in \{ 1,n\}$ and introduce $n-1$ link fields $Y^\mu_{(i_j i_k)}$ for each $j \in \{1,n\} \neq k$ (we are of course free to also define $Y^\mu_{(i_k i_k)}$, which is the identity). Note we have not projected out the determinant of any of the participating spin-2 fields to stay fully general here. Applying the \St fields as before and defining $j_1, \ldots j_{n-1}$ to label all sites except for $i_k$ we have
\be \label{NmetricLink}
{\cal S}_{int} \to \int d^4 x  f\left({\bf g}_{(i_k)},  {\bf g}_{(j_1)} \circ Y_{(j_1 i_k)}, \ldots,  {\bf g}_{(j_{n-1})} \circ Y_{(j_{n-1} i_k)} \right).
\ee
This action is now invariant under all $GC_{(j_1)}, \ldots, GC_{(j_{n-1})}$ by construction, with all spin-2 fields being mapped onto the $i_k$ site, i.e.  transforming under $GC_{(i_k)}$. In this way one can restore the full $n$ gauge invariances for an $n$-vertex interaction term. Throughout the rest of this paper we will call the act of introducing \St fields and restoring gauge invariances in the process ``to \St '' - for ease of terminology we will somewhat loosely use this as a verb. It is also worth emphasizing that there are important subtleties when applying this iteratively to several interactions in a theory graph, which we will discuss below in section \ref{sec-multiSt}.
\\

{\bf Projecting out Goldstone modes: } In understanding the nature of the {\St}/link fields $Y^{\mu}_{(ij)}$ it is useful to expand them about the identity
\be \label{goldstone1}
Y^{\mu}_{(ij)} \to x^\mu_{(i)} + A^\mu_{(ij)}.
\ee
Written in this way we can see that the \St trick fundamentally is nothing more than a coordinate transformation, which takes coordinate $x^\mu_{(i)} \to Y^{\mu}_{(ij)} = x^\mu_{(i)} + A^\mu_{(ij)}$. To see this explicitly, consider an interaction vertex and notice that for a St\"uckelberged spin-2 field we can write
\bea \label{goldstone2}
G_{\mu\nu}(Y) &\equiv & \pa_\mu Y^\alpha \pa_\nu Y^\beta g_{\alpha\beta}(Y) \to  \pa_\mu (x^\alpha + A^\alpha) \pa_\nu (x^\beta + A^\beta) g_{\alpha\beta}(x+A) \\
&=& \left( \delta_\mu^\alpha + \pa_\mu A^\alpha \right)  \left( \delta_\nu^\beta + \pa_\nu A^\beta \right) \left( g_{\alpha\beta} + A^\sigma \pa_\sigma g_{\alpha \beta} + \frac{1}{2}A^{\sigma}A^{\lambda} \pa_\sigma \pa_\lambda g_{\alpha\beta} \ldots   \right), \nn
\eea
where we have dropped explicit indices $i,j$. The $A^\mu$ field then is the Goldstone boson of the broken $GC_{(i)}$ invariance associated with $g_{\mu\nu}$ and encodes all the effects of symmetry breaking. The gauge-fixed nature of ${\cal S}_{int}$ as written in equation \eqref{fixedSint} now becomes apparent. The fully gauge-invariant ${\cal S}_{int}$ with expanded \St fields is given by combining equations \eqref{NmetricLink},\eqref{goldstone1} and \eqref{goldstone2}. This is a theory with $N$ massless spin-2 fields and $N-1$ gauge bosons. But we can gauge-fix this action, removing all the gauge bosons $A^\mu_{(ij)}$ by moving to unitary gauge in which $Y^\mu_{(ij)} = x^\mu_{(i)}$. In this gauge we recover \eqref{fixedSint} and all gauge bosons have been eaten to produce $N-1$ massive spin-2 fields (a single massless spin-2 field remains due to the single surviving copy of $GC$).

Having projected out the Goldstone boson of a broken $GC_{(i)}$, we can further decompose $A^\mu$ as
\be
A^\mu \to A^\mu + \pa^\mu \pi,
\ee
i.e. into a `vector'/spin one and a `scalar' mode, which correspond in the decoupling limit to the longitudinal scalar and vector components of the `massive graviton'\footnote{We insist on inverted commas for the term `graviton' in this context, since the statement about $A^\mu$ of course applies to all of the spin-2 fields which are subjected to the \St treatment in an interacting theory with $N$ such fields. Prior to coupling this to matter via ${\cal S}_{matter}$, which will decide what other matter in the universe sees as gravity, there is consequently no way of discriminating one such field as `the graviton' and we resist calling the excitations of all spin-2 fields `gravitons', since as far as other matter in the universe is concerned, these extra fields have very little to do with the geometry perceived by matter.} associated with $g_{\mu\nu}^{(i)}$ in the gauge-fixed action \eqref{fixedSint}. Note that following \cite{ArkaniHamed:2002sp,Schwartz:2003vj} we have introduced a fake U(1) symmetry, with $\pi$ as its Goldstone boson. $\pi$ is of particular interest, since the longitudinal scalar mode of the `graviton' is the one that typically threatens to become the Boulware-Deser ghost. We also expect it to yield the most dangerous interaction terms (i.e. the ones growing most quickly with energy \cite{ArkaniHamed:2003vb}, or equivalently, the ones suppressed by the smallest scale, which means they will remain relevant in the decoupling limit, as we shall see below in section \ref{sec-cubic}). Finally note that introducing the fake U(1) symmetry should be supplemented by adding appropriate gauge-fixing terms. However, their exact form doesn't matter here \cite{ArkaniHamed:2002sp, Hinterbichler:2011tt}. 
\\

{\bf Summary so far: }
Until now we have reviewed the construction of \cite{ArkaniHamed:2002sp} to introduce and use \St fields in order to restore gauge invariances broken by bimetric interaction terms. This effectively projects out the $n-1$ Goldstone modes of the $n-1$ $GC_{(i)}$ symmetries broken by the interaction terms and which got eaten to produce $n-1$ massive `gravitons' in the initial gauge-fixed action.  We found it straightforward to extend this formalism to the case of a single $n$ spin-2 field interaction vertex. We will now move on to consider more complicated theories of $N$ interacting spin-2 fields and find that various subtleties not present in the simple setups considered so far appear and of which care must be taken.

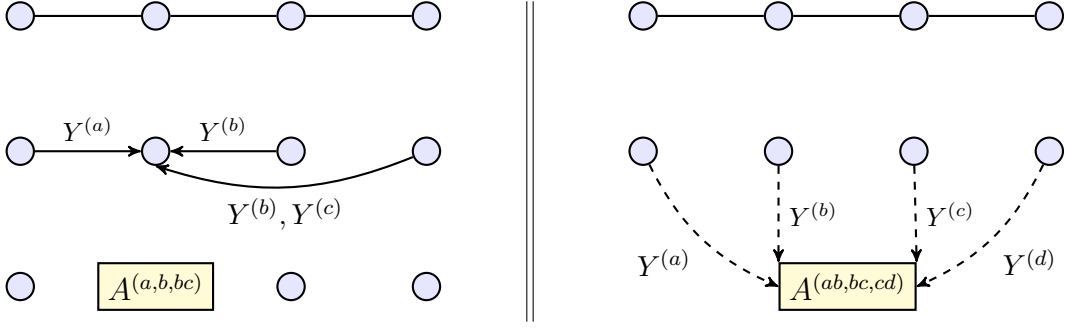
\begin{figure}[tp]
\centering
    \begin{tabular}{ c || c }
{  \begin{tikzpicture}[->,>=stealth',shorten >=0pt,auto,node distance=1.8cm,
  thick,main node/.style={circle,fill=blue!10,draw,font=\sffamily\large\bfseries},arrow line/.style={thick,-},barrow line/.style={thick,->},no node/.style={plain},rect node/.style={rectangle,fill=blue!10,draw,font=\sffamily\large\bfseries},sarrow line/.style={thick,->,shorten >=1pt},green node/.style={circle,fill=green!20,draw,font=\sffamily\large\bfseries},yellow node/.style={rectangle,fill=yellow!20,draw,font=\sffamily\large\bfseries}]

  \node[main node] (1) {};
  \node[main node] (2) [right of=1] {};
    \node[main node] (3) [right of=2] {};
      \node[main node] (4) [right of=3] {};
      \node[main node] (9) [below of=1] {};
  \node[main node] (10) [below of=2] {};
    \node[main node] (11) [below of=3] {};
  \node[main node] (12) [below of=4] {};
      \node[main node] (13) [below of=9] {};
  \node[yellow node] (14) [below of=10] {$A^{(a,b,bc)}$};
    \node[main node] (15) [below of=11] {};
  \node[main node] (16) [below of=12] {};

 \node[draw=none,fill=none] (22) [right=0.65cm of 4] {};

  \path[every node/.style={font=\sffamily\small}]
       (9) edge node [above] {$Y^{(a)}$} (10.west)
          (11) edge node [above] {$Y^{(b)}$} (10);

            \draw[arrow line] (1) -- (2);
              \draw[arrow line] (2) -- (3);
               \draw[arrow line] (3) -- (4);
              
              \draw[barrow line] (12) to[bend left=20] node[below,rotate=0] {$Y^{(b)},Y^{(c)}$} (10.south);
            
\end{tikzpicture}} &
{
\begin{tikzpicture}[->,>=stealth',shorten >=0pt,auto,node distance=1.8cm,
  thick,main node/.style={circle,fill=blue!10,draw,font=\sffamily\large\bfseries},arrow line/.style={thick,-},barrow line/.style={thick,->},no node/.style={plain},rect node/.style={rectangle,fill=blue!10,draw,font=\sffamily\large\bfseries},sarrow line/.style={thick,->,shorten >=1pt},yellow node/.style={rectangle,fill=yellow!20,draw,font=\sffamily\large\bfseries}]

  \node[main node] (1) {};
  \node[main node] (2) [right of=1] {};
    \node[main node] (3) [right of=2] {};
      \node[main node] (4) [right of=3] {};
      \node[main node] (9) [below of=1] {};
  \node[main node] (10) [below of=2] {};
    \node[main node] (11) [below of=3] {};
  \node[main node] (12) [below of=4] {};
   \node[draw=none,fill=none] (14) [right=0.58cm of 10] {};
 \node[yellow node] (15) [below of=14] {$A^{(ab,bc,cd)}$};
 
 \node[draw=none,fill=none] (22) [left=0.65cm of 1] {};

  \path[dashed, every node/.style={font=\sffamily\small,dashed}]
          (10) edge node [right,dashed] {$Y^{(b)}$} (15.north west)
         (11) edge node [right,dashed] {$Y^{(c)}$} (15.north east);
            
            
            \draw[arrow line] (1) -- (2);
              \draw[arrow line] (2) -- (3);
               \draw[arrow line] (3) -- (4);

              \draw[barrow line,dashed] (12) to[bend left=20] node[below right,rotate=0] {$Y^{(d)}$} (15.east);
              
              \draw[barrow line,dashed] (9) to[bend right=20] node[below left,rotate=0] {$Y^{(a)}$} (15.west);
            
\end{tikzpicture}}
    \end{tabular}
\caption[text]{Different ways of St\"uckelberging: Here we show graphs corresponding to {\it approach II} analogous to the case shown in equation \eqref{bimetricExampleII} (left) and {\it approach III} analogous to the case shown in equation \eqref{bimetricExampleIII} (right). Note that, on the second line in {\it approach II} (left), even though the right-most node is only mapped via $Y^{(c)}$, the interaction term connecting it to its neighbouring node is mapped via $Y^{(c)}$ {\it and} $Y^{(b)}$, which is why we label the `mapping' in the second line with both of these link fields. As a result there will be an interaction term for Goldstone bosons associated to the $b$ and $c$ index in the final St\"uckelberged action, whereas the other links give contributions depending on Goldstone bosons associated to indices $a$ and $b$ only, i.e. they do not introduce extra cross-interaction terms. We denote this by $A^{(a,b,bc)}$ in the final line. In {\it approach III} link fields are introduced for all sites, so that link produces an interaction term involving two Goldstone bosons in the final action. Hence we have  $A^{(ab,bc,cd)}$.
} 
\label{fig-DiffStueck}
\end{figure}
\section{St\"uckelberging interacting spin-2 fields} \label{sec-multiSt}

{\bf Different ways to \St{}: } In the previous section we discussed how, for an $n$ spin-2 field interaction term, one can introduce $n-1$ \St fields $Y^\mu$ to map all the participating nodes/spin-2 fields to the same site. How does this prescription generalise to the case of a full theory graph, i.e. when one has multiple interaction terms as in figure \ref{fig-Nfields}? Below we describe three approaches to this.

{\it Approach I:} We treat each interaction term in isolation and apply the \St trick as in equation \ref{NmetricLink}. This guarantees that every \St field we introduce corresponds to a propagating mode (with the caveat discussed in the `too many links' paragraph below). However, this also means that the same metric $g_{\mu\nu}^{(i)}$, if participating in multiple interaction terms, may be mapped to different sites when St\"uckelberging different interaction terms. For example consider a three-site model with two bimetric interaction terms. Then we may choose to \St as follows
\begin{align} 
\nn {\cal S}_{int} &= \int d^4 x \left(   \sqrt{g_{(j)}} f_1({\bf g}_{(i)},{\bf g}_{(j)}) +  \sqrt{g_{(k)}} f_2({\bf g}_{(i)},{\bf g}_{(k)})  \right) \\
&\nn \to \int d^4 x \left(  \sqrt{g_{(j)}} f_1({\bf g}_{(i)} \circ Y_{(ij)},{\bf g}_{(j)}) + \sqrt{g_{(k)}} f_2({\bf g}_{(i)} \circ Y_{(ik)},{\bf g}_{(k)})  \right) \\
&= \int d^4 x \left(  \sqrt{g_{(j)}} f_1({\bf G}_{(j)},{\bf g}_{(j)}) +  \sqrt{g_{(k)}} f_2({\bf G}_{(k)},{\bf g}_{(k)})     \right).  \label{bimetricExampleI}
\end{align}
This is consistent - after all nothing forces us to map the same spin-2 field to the same site throughout the action - but by definition means we don't have a simple global prescription for mapping a particular spin-2 field onto a given site. This is also illustrated in figure \ref{fig-Stueck}.

 {\it Approach II:} Consider a theory graph as in figure \ref{fig-Nfields}. We know each separate `island' corresponds to one surviving copy of $GC_{(i)}$. To make this explicit it may be desirable to map all the fields participating in interaction terms in that island to the same site. Consider the three-site model from above again\footnote{Note that we are, in an abuse of notation, denoting the mapping of a metric determinant $g_{(i)}$ from site $i$ to site $j$ by $g_i \circ Y_{(ij)}$.}
 \begin{align}
\nn {\cal S}_{int} &= \int d^4 x \left(   \sqrt{g_{(j)}} f_1({\bf g}_{(i)},{\bf g}_{(j)}) +  \sqrt{g_{(k)}} f_2({\bf g}_{(i)},{\bf g}_{(k)})  \right) \\
&\nn \to \int d^4 x \left(  \sqrt{g_{(j)}} f_1({\bf g}_{(i)} \circ Y_{(ij)},{\bf g}_{(j)}) + \sqrt{g_{(k)}\circ Y_{(kj)}} f_2({\bf g}_{(i)} \circ Y_{(ij)},{\bf g}_{(k)}  \circ Y_{(kj)})  \right) \\
&= \int d^4 x  \sqrt{g_{(j)}} \left( f_1({\bf G}_{(j)},{\bf g}_{(j)}) +  f_2({\bf G}_{(j)},\tilde{\bf G}_{(j)})     \right). \label{bimetricExampleII}
\end{align}
This can clearly become somewhat cumbersome, since e.g. a spin-2 field which is $n$ `steps' removed from the site to which it will be mapped, will have to be St\"uckelberged as 
\be
{\bf g}_{(n)} \to {\bf g}_{(n)} \circ Y_{(n,n-1)} \circ Y_{(n-1,n-2)}  \circ \ldots \circ Y_{(1,0)},
\ee
but does have the advantage of making the surviving copy of $GC_{(i)}$ explicit.\footnote{The term $Y_{(n,n-1)} \circ Y_{(n-1,n-2)}  \circ \ldots \circ Y_{(1,0)}$ is what \cite{ArkaniHamed:2002sp} refer to as a plaquette.} A similar, slightly more complex, example is shown in the left graph in figure \ref{fig-DiffStueck}.  The number of \St fields introduced still corresponds to the propagating \dof. However, the major drawback of this approach is that the interaction terms become much more complex since multiple \St fields interact at each vertex now, whereas before only $n-1$ fields interacted for an $n$-field vertex. As a concrete example compare equations \eqref{bimetricExampleI} and \eqref{bimetricExampleII}. Whereas in the first approach both vertices only has a single Goldstone boson $A^\mu$ not directly interacting with any other bosons, in the second approach the second vertex will have interactions between $A^\mu_{(ij)}$ and  $A^\mu_{(ik)}$. This significantly complicates interaction terms and can make working with this approach very unwieldy.
 
\begin{figure}[tp]
\centering
\begin{tikzpicture}[->,>=stealth',shorten >=0pt,auto,node distance=2cm,
  thick,main node/.style={circle,fill=blue!10,draw,font=\sffamily\large\bfseries},arrow line/.style={thick,-},barrow line/.style={thick,->},no node/.style={plain},rect node/.style={rectangle,fill=blue!10,draw,font=\sffamily\large\bfseries},red node/.style={rectangle,fill=red!10,draw,font=\sffamily\large\bfseries},green node/.style={circle,fill=green!20,draw,font=\sffamily\large\bfseries},yellow node/.style={rectangle,fill=yellow!20,draw,font=\sffamily\large\bfseries},blue node/.style={rectangle,fill=blue!10,draw,font=\sffamily\large\bfseries}]

  \node[main node] (1) {};
  \node[main node] (2) [right of=1] {};
    \node[main node] (3) [right of=2] {};
    
      \node[main node] (111) [above of=1]{};
  \node[main node] (222) [above of=2] {};
    \node[main node] (333) [above of=3] {};
  
  \node[draw=none,fill=none](80)[below right of=1]{};
   \node[draw=none,fill=none](82)[right=2.5cm of 3]{\it Approach I};
   
   \node[draw=none,fill=none](83)[below of=82]{\it Approach II};
   
   \node[draw=none,fill=none](84)[below of=83]{\it Approach III};
   
   \node[draw=none,fill=none](85)[above of=82]{Gauge-breaking interactions};

      \node[main node] (9) [below of=1] {};
  \node[main node] (10) [below of=2] {};
    \node[main node] (11) [below of=3] {};

 \node[draw=none,fill=none] (22) [right=0.65cm of 4] {};

      \node[main node] (90) [below of=9] {};
  \node[main node] (100) [below of=10] {};
    \node[main node] (110) [below of=11] {};
 \node[rect node] (150) [below of=100] {};

   \path[every node/.style={font=\sffamily\small}]
       (2) edge node [above] {$Y^{(21)}$} (1.east)
          (2) edge node [above] {$Y^{(23)}$} (3);
  
  \path[every node/.style={font=\sffamily\small}]
       (10) edge node [above] {$W^{(21)}$} (9.east);
       
              \draw[barrow line] (11) to[bend left=20] node[below,rotate=0] {$W^{(21)},W^{(31)}$} (9.south);

  \path[dashed, every node/.style={font=\sffamily\small,dashed}]
          (100) edge node [right,dashed] {$V^{(2x)}$} (150.north);
            

            \draw[arrow line] (111) -- (222);
              \draw[arrow line] (222) -- (333);

              \draw[barrow line,dashed] (110) to[bend left=20] node[below right,rotate=0] {$V^{(3x)}$} (150.east);
              
              \draw[barrow line,dashed] (90) to[bend right=20] node[below left,rotate=0] {$V^{(1x)}$} (150.west);

\end{tikzpicture}
\caption[text]{Another example comparing {\it approaches I-III} for a simple three spin-2 field system with two bimetric interaction terms. $Y^{(ij)},W^{(ij)},V^{(ij)}$ are the \St fields used in {\it approaches I-III} respectively. Note that the choice of the direction of each $Y^{ij}$ is arbitrary (we can always reverse link fields by inverting , since $Y_{(ij)}^{-1} = Y_{(ji)}$). Also the choice of target node in {\it approach II} is arbitrary. Again note that, while the right-most node in {\it approach II} is mapped with $W^{(31)}$ only, the interaction term connecting it to the central node picks up a dependence on both $W^{(31)}$ and $W^{(21)}$, which is why we label the `mapping' with both of these fields. The choices made here are intended to bring out the differences between all approaches for the simplest possible system. We discuss these differences and demonstrate the equivalence of all approaches below.
}
\label{fig-StueckEq}
\end{figure}
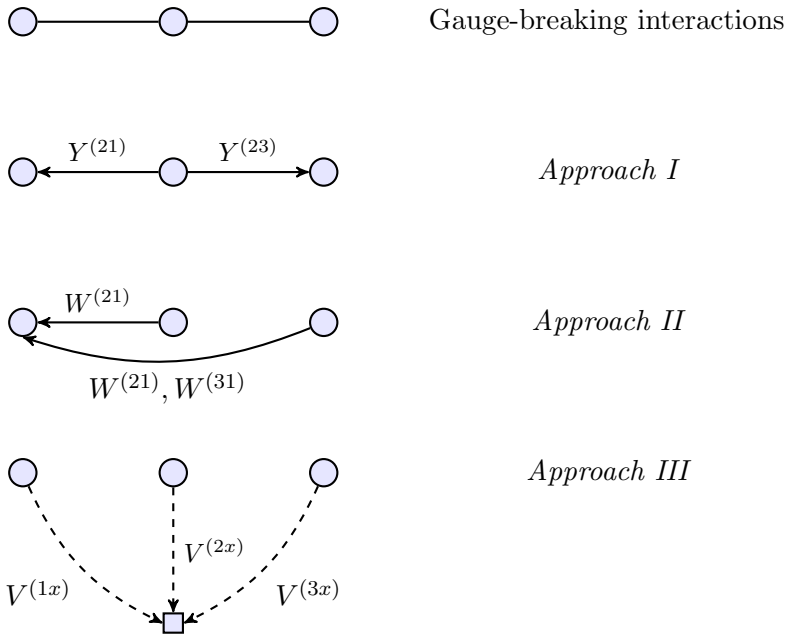

 {\it Approach III:} Finally one may hope to restore gauge-invariance with an algorithmic prescription that does not rely on inspection of the interaction terms. As such we might want to \St all spin-2 fields in the theory. There are of course different ways of doing this, but a natural choice is to combine this with the advantage of {\it approach II} and send all fields to some site $l$. This site can be one of the existing ones (on the same island or not), in which case the approach is equivalent to {\it approach II} in the case of a single connected island, or an extra `imaginary' site. 
\begin{align}
\nn {\cal S}_{int} &= \int d^4 x \left(   \sqrt{g_{(j)}} f_1({\bf g}_{(i)},{\bf g}_{(j)}) +  \sqrt{g_{(k)}} f_2({\bf g}_{(i)},{\bf g}_{(k)})  \right) \\
&\nn \to \int d^4 x \left(  \sqrt{g_{(j)} \circ Y_{(jl)}} f_1({\bf g}_{(i)} \circ Y_{(il)},{\bf g}_{(j)}  \circ Y_{(jl)}) + \sqrt{g_{(k)}\circ Y_{(kl)}} f_2({\bf g}_{(i)} \circ Y_{(il)},{\bf g}_{(k)}  \circ Y_{(kl)})  \right) \\
&= \int d^4 x  \sqrt{g_{(l)}}  \left( f_1({\bf G}_{(l)}^{(1)},{\bf G}_{(l)}^{(2)}) +  f_2({\bf G}_{(l)}^{(3)},{\bf G}_{(l)}^{(4)})     \right). \label{bimetricExampleIII}
\end{align}
A similar example is shown in the right graph in figure \ref{fig-DiffStueck}. While affording us with a simple global prescription for performing the \St trick and making the surviving copy of $GC_{(i)}$ explicit, this approach has an obvious drawback. We have introduced too many would-be Goldstone bosons. Namely we have introduced $N$ $Y_{(il)}$, but we know there are only $N-m$ actual Goldstone bosons, where $m$ is the number of disconnected islands in the theory. As a result there are $m$ non-linear combinations of the would-be Goldstone bosons that actually do not propagate. In other words, when transforming into the right basis these combinations have no kinetic terms and act as auxiliary fields that can be integrated out, leaving us with the correct number of propagating degrees of freedom. 

An important insight from considering the three different approaches, is that there is no unique way of applying the \St trick to $N$ interacting spin-2 fields and that both the interaction terms between different Goldstone bosons and the apparent number of degrees of freedom depends on the choice of approach. An analogous conclusion was reached in \cite{Schwartz:2003vj} in the context of a particular `circle Lagrangian' ($n$ spin-2 fields interacting bimetrically with a theory graph that forms a single closed loop - see the `too many links' section below). In what follows we will follow {\it approach I}, since it generates the correct number of degrees of freedom by default and produces the most economical interaction terms (i.e. the ones with the fewest interacting Goldstone bosons). 
\\

{\bf The equivalence of approaches I-III: } The existence of the different possible approaches discussed above is due to the fact that there is no unique prescription for introducing gauge degrees of freedom in the process of St\"uckelberging. Here we wish to show that all approaches presented so far are ultimately dual to one another (as they should be - after all they describe the same physics).  Using one approach over another may be far more economical in any given case, but a well-defined mapping between different ways of introducing \St fields always exists. This can most easily be seen with a concrete example. Consider three spin-2 fields connected by two bimetric interactions as depicted in figure \ref{fig-StueckEq}. The interaction terms post-\Sting in {\it approaches I-III}\footnote{Note that, even inside any given approach, there are arbitrary choices of target nodes to be made - cf. the caption of figure \ref{fig-StueckEq}} are given by
\begin{align}
\label{L1} {\cal L}_I &= f_1\left({\bf g}_{(1)}, {\bf g}_{(2)} \circ {Y}_{(21)}\right)  +  f_2\left({\bf g}_{(2)} \circ {Y}_{(23)},{\bf g}_{(3)}\right) \\
{\cal L}_{II} &= f_1\left({\bf g}_{(1)}, {\bf g}_{(2)} \circ {W}_{(21)}\right)  +  f_2\left({\bf g}_{(2)} \circ {W}_{(21)},{\bf g}_{(3)} \circ {W}_{(31)}\right)  \label{L2} \\
{\cal L}_{III} &= f_1\left({\bf g}_{(1)}\circ {V}_{(1x)}, {\bf g}_{(2)} \circ {V}_{(2x)}\right)  +  f_2\left({\bf g}_{(2)} \circ {V}_{(2x)},{\bf g}_{(3)}\circ {Y}_{(3x)}\right) \label{L3}
\end{align}
It is now trivial to relate the \St (or link) fields used to one another. $Y_{(ij)}$'s and $W_{(ij)}$'s are related via
\be \label{YsWs}
Y_{(21)} = W_{(21)}, \quad\quad W_{(31)} = Y_{(23)}^{-1} \circ Y_{(21)}
\ee
whereas  $Y_{(ij)}$'s and $V_{(ij)}$'s satisfy 
\begin{align} \label{VsYs}
\nn V_{(2x)} \circ V_{(1x)}^{-1} &= Y_{(21)}, \\
\nn V_{(2x)} \circ V_{(3x)}^{-1} &= Y_{(23)}, \\
V_{(3x)} \circ V_{(1x)}^{-1} &= Y_{(23)}^{-1} \circ  Y_{(21)}.
\end{align}
That all three $V$'s can be expressed in terms of the two $Y$'s is a direct manifestation of the fact that there are really only two \St field degrees of freedom here (i.e. in {\it approach III} we introduced one too many would-be Goldstone bosons $V$). In this sense equation \eqref{VsYs} can be seen as providing the constraint eliminating one of the $V$'s. We can now explicitly relate interaction terms \eqref{L1}-\eqref{L3} to each other. Consider \eqref{L2}
\begin{align}
\nn {\cal L}_{II} &= f_1\left({\bf g}_{(1)}, {\bf g}_{(2)} \circ {W}_{(21)}\right)  +  f_2\left({\bf g}_{(2)} \circ {W}_{(21)},{\bf g}_{(3)} \circ {W}_{(31)}\right) \\
\nn &= f_1\left({\bf g}_{(1)}, {\bf g}_{(2)} \circ {Y}_{(21)}\right)  +  f_2\left({\bf g}_{(2)} \circ {Y}_{(21)},{\bf g}_{(3)} \circ Y_{(23)}^{-1} \circ Y_{(21)} \right) \\
\nn &\to f_1\left({\bf g}_{(1)}, {\bf g}_{(2)} \circ {Y}_{(21)}\right)  +  f_2\left({\bf g}_{(2)},{\bf g}_{(3)} \circ Y_{(23)}^{-1} \right) \\
&\to f_1\left({\bf g}_{(1)}, {\bf g}_{(2)} \circ {Y}_{(21)}\right)  +  f_2\left({\bf g}_{(2)} \circ Y_{(23)},{\bf g}_{(3)} \right) = {\cal L}_I \label{2to1}
\end{align}
In the second line we have used \eqref{YsWs} to substitute for the $W$'s. In the third and fourth lines we have acted on all fields in the interaction terms with $Y_{(21)}^{-1}$ and $Y_{(23)}$ respectively. Note that both $f_1(\ldots)$ and $f_2(\ldots)$ are gauge-invariant under diffeomorphisms in the action post-\Sting (just as $R_{(1)}$ and $R_{(2)}$ already were before). This e.g. means that 
\be
f_1\left({\bf g}_{(1)}, {\bf g}_{(2)} \circ {Y}_{(21)}\right)  = f_1\left({\bf g}_{(1)} \circ {Y}_{(21)}^{-1}, {\bf g}_{(2)}\right) 
\ee
allowing us to explicitly apply a `diffeomorphism' $Y_{(ij)}$ to $f_2(\ldots)$ in \eqref{2to1} while choosing to leave $f_1(\ldots)$ unchanged. Similar manipulations, showing the equivalence of ${\cal L}_{III}$ to ${\cal L}_{II}$ and ${\cal L}_{I}$, can also be carried out. The gauge-invariant structure of interaction terms post-\Sting is therefore what allows us to treat each interaction term in isolation in {\it approach I} instead of having a global prescription always sending each ${\bf g}_{(i)}$ to the same site as in {\it approaches II-III}.
\\

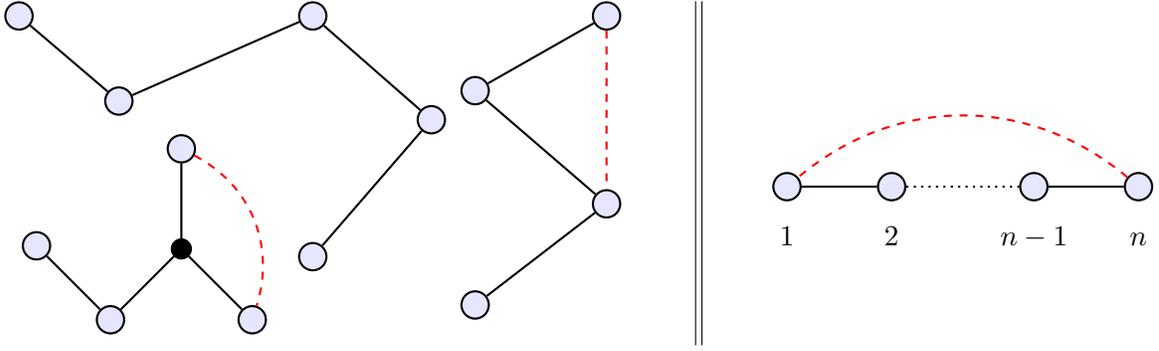
\begin{figure}[tp]
\centering
    \begin{tabular}{ c || c }
{\begin{tikzpicture}[-,>=stealth',shorten >=0pt,auto,node distance=1cm,
  thick,main node/.style={circle,fill=blue!10,draw,font=\sffamily\small\bfseries},barrow line/.style={thick,red,dashed,-}]

  \node[main node] (1) {};
 \node[main node, right=100pt,at=(1.east)](2) {};
  \node[main node, right=100pt,at=(2.east)](3) {};
   \node[draw=none,fill=none] (222) [right=0.5cm of 3] {}; 
   \node[main node, below right=50pt,at=(2.east)](4) {};
     \node[main node,above right=10pt,at=(4.east)](5){};
     \node[main node,below right=40pt,at=(1.east)](6){};
     \node[main node,below right=20pt,at=(6.east)](7){};
              \node[main node, below=80pt,at=(2.south)](10){};
                  \node[main node, below=60pt,at=(3.south)](11){};
                      \node[main node, below=70pt,at=(5.south)](12){};
                      
   \node[main node,fill=black!100,minimum size=0.01cm,scale=0.7] (20)  [below=1cm of 7]{};
  \node[main node] (30) [below left=1cm of 20] {};
   \node[main node] (40) [below right=1cm of 20] {};
   \node[main node] (50) [above left=1cm of 30] {};
                      
    \draw[barrow line] (7) to[bend left=40]  (40);                
     \draw[barrow line] (3) to[bend left=0]  (11);         
                      
                       \path[every node/.style={font=\sffamily\small}]
                   (20) edge node {} (30)
                     (30) edge node {} (50)
                    (20) edge node {} (40)
                     (20) edge node {} (7)     
    (1) edge node {} (6)
      (6) edge node {} (2)
      (2) edge node {} (4)
       (4) edge node {} (10)
            (3) edge node {} (5)
      (5) edge node {} (11)
       (11) edge node {} (12);

\end{tikzpicture}}&
{\begin{tikzpicture}[-,>=stealth',shorten >=0pt,auto,node distance=1cm,
  thick,main node/.style={circle,fill=blue!10,draw,font=\sffamily\small\bfseries},barrow line/.style={thick,dotted,-},rarrow line/.style={thick,red,dashed,-}]

 \node[draw=none,fill=none] (222) {}; 
  \node[draw=none,fill=none] (2222) [left=0.5cm of 222] {}; 
  \node[main node,above=1.5cm of 222](1) {};
 \node[main node, right=1cm,at=(1.east)](2) {};
  \node[main node, right=1.5cm,at=(2.east)](3) {};
   \node[main node, right=1cm,at=(3.east)](4) {};

  \node[draw=none,fill=none] (11) [below=0.2cm of 1] {$1$}; 
    \node[draw=none,fill=none] (12) [below=0.2cm of 2] {$2$}; 
      \node[draw=none,fill=none] (13) [below=0.2cm of 3] {$n-1$}; 
        \node[draw=none,fill=none] (14) [below=0.28cm of 4] {$n$}; 
                      
    \draw[barrow line] (2) to (3);   
      \draw[rarrow line] (4) to [bend right=40] (1);   
                      
                       \path[every node/.style={font=\sffamily\small}]  
    (1) edge node {} (2)
      (3) edge node {} (4);

\end{tikzpicture}}
\end{tabular}
\caption{Loops in theory graphs as discussed in the `Too many links' section. Potentially dangerous interactions added to previous figure, and leading to the creation of loops, are dashed lines (left). These will introduce interaction terms such as \eqref{Sloop}, coupling together all the Goldstone bosons in the loop. A `circle Lagrangian', corresponding to a straight line with the first and final loop connected and especially relevant in the context of `deconstructing dimensions', is also shown (right).} \label{fig-Danger}
\end{figure}
{\bf A background metric: }
Finally we have to address the important issue of which metric we choose to raise and lower indices in the St\"uckelberged theory. A natural choice would seem to be the metric of the site onto which we are mapping spin-2 fields. However, this is not an optimal choice, because different interaction terms will map onto different sites with the \St convention we have chosen here ({\it approach I}). So when combining parts of the scalar field action for all the $\pi^{(i)}$ it would be desirable to have indices (especially on partial derivatives) throughout raised with the same metric (at lowest order). In other words we would like to expand all metrics about the same background metric. For the sake of simplicity we will choose the non-dynamical (flat) Minkowski metric here, but in principle any other background metric (that allows a perturbative treatment of $\delta g_{\mu\nu}^{(i)}$) could be chosen too (see appendix \ref{appendix-curved} for more on what changes). As such we expand 
\begin{align} \label{backgroundmetric}
{\bf g}_{(i)} &\to {\bm \eta} + {\bf h}_{(i)}&    \Longleftrightarrow&   &g_{\mu\nu}^{(i)} &\to \eta_{\mu\nu} + h_{\mu\nu}^{(i)}
\end{align}
Again we iterate that this is a particular choice of background metric where we have chosen ${\bf g}_0 = {\bf \eta}$. Indices will be raised and lowered with the background metric and we denote partial derivatives acting on the scalar mode $\pi$ by indices too, i.e.
\be
\pa_{\mu_1} \ldots \pa_{\mu_i} \pa^{\nu_1} \ldots \pa^{\nu_j} \pi \equiv \pi^{\nu_1 \ldots \nu_j}_{\mu_1 \ldots \mu_i}.
\ee 
Since we have chosen a flat metric to raise and lower indices, all indices on $\pi$ always commute. It is perhaps worth emphasizing at this point that, even though all interaction terms `live' on individual sites following the \St trick, information about the interactions of different sites in the gauge-fixed theory is not lost. It is still encoded in the Goldstone bosons $A^\mu_{(ij)}$ and its scalar mode $\pi^{(ij)}$ (in an abuse of notation we have only labelled these with he site they are on so far, but in cases where it is not obvious which scalar mode is considered, we will label them explicitly) as well as in the metric perturbations $h_{\mu\nu}^{(i)}$.
\\

{\bf `Too many' links: } So far we have only considered theories with the `right' number of interaction vertices, so that each \St field corresponds to a propagating mode. In other words it is a true Goldstone boson (compare our discussion between {\it approaches I-III} above). Specifically for $N$ interacting spin-2 fields, which form $m$ `islands', i.e. where the interaction term preserves $m$ copies of $GC_{(i)}$, the number of interaction vertices satisfied
\be \label{loopsCon}
n_2 + 2 n_3 + 3 n_4 + \ldots = \sum_i (i-1) n_i = N-m,
\ee
where $n_i$ counts the number of interaction vertices with $i$ participating spin-2 fields. What happens if we violate this constraint? In terms of the theory graph this corresponds to introducing loops. Figure \ref{fig-Danger} shows examples of such theories: the graph from figure \ref{fig-Nfields} with added interactions that violate \eqref{loopsCon} and introduce loops and another graph corresponding to the `circle Lagrangian' (i.e. one big loop) as considered by \cite{ArkaniHamed:2001ca,ArkaniHamed:2002sp,ArkaniHamed:2003vb,Schwartz:2003vj}. Loops in theory graphs (in the metric picture used throughout this paper) modify the constraint structure of the theory \cite{ASM-thesis}\footnote{We thank Angnis Schmidt-May for bringing this and reference \cite{ASM-thesis} to our attention.} and have been conjectured to introduce ghosts into the theory \cite{Nomura:2012xr,ASM-thesis} - also see related discussions in \cite{Khosravi:2011zi,Hinterbichler:2012cn}.\footnote{Note that theories with loops are non-equivalent in the metric and vielbein pictures. In fact theories with loops in the vielbein picture are known to be ghost-free for certain types of interaction terms \cite{Hinterbichler:2012cn}.} 
While we won't attempt to prove or disprove this conjecture here, we can observe that this adds a Goldstone interaction term to the St\"uckelberged theory that is qualitatively different from the others.

Consider a single loop made up of bimetric interactions only (such as the `circle Lagrangian' in figure \ref{fig-Danger}) with $n$ sites and, since it is a loop, $n$ links. We now \St $n-1$ links. It doesn't matter which, but for concreteness suppose we only leave out the node $n$ to node $1$ link, denoting its interaction by 
 \begin{align}
{\cal S}_{loop} = \int d^4 x   \sqrt{g_{(n)}} f({\bf g}_{(1)},{\bf g}_{(n)})
\end{align}
Now the degrees of freedom projected out from $n-1$ are sufficient to describe all Goldstone bosons arising from symmetry breaking. What about the final link? We know it does not break any more symmetries, yet if we \St it as all the other links we seemingly add an extra would be Goldstone boson. However, note that the link fields introduced already are sufficient to make the interaction term corresponding to the final link gauge-invariant. We denote the \St field connecting node $j$ and $j-1$ with $Y_{(j-i,j)}$ (if the original link was chosen the other way, note that we can always use $Y_{(j,j-1)}^{-1} = Y_{(j-i,j)}$ to find a link field mapping from site $j-1$ to $j$). We can now make ${\cal S}_{loop}$ gauge-invariant via
\begin{align}
\nn {\cal S}_{loop} &= \int d^4 x   \sqrt{g_{(n)}} f({\bf g}_{(1)},{\bf g}_{(n)})\\
&\nn \to \int d^4 x   \sqrt{g_{(n)}} f({\bf g}_{(1)} \circ Y_{(1,2)} \circ Y_{(2,3)} \circ \ldots \circ Y_{(n-1,n)},{\bf g}_{(n)}) \\
&= \int d^4 x  \sqrt{g_{(n)}} f({\bf G}_{(n)},{\bf g}_{(n)})). \label{Sloop}
\end{align}
This now restores gauge-invariance to the full action and introduces the correct number of Goldstone bosons. Closing the loop has had the effect of adding an extra interaction term, ${\cal S}_{loop}$, which couples all the \St fields together. We investigate whether this can be done in a ghost-free way in \cite{loops}, but can already note here that the pattern shown in \eqref{Sloop} is general. Whenever we encounter a loop, this means that the fully gauge-invariant action resulting from applying the \St trick will contain an interaction term coupling all the Goldstone bosons from the loop together.

\section{From Goldstone modes to demixed scalar-tensor theories} \label{sec-gold}

{\bf The pure scalar action at quadratic order: } Consider the interaction term post-St\"uckelberging. We project out the Goldstone bosons $A^\mu_{(ij)}$ and the associated longitudinal scalar modes $\pi_{(ij)}$ as in equation \eqref{goldstone2}. Finally metric perturbations $h_{\mu\nu}^{(i)}$ are expanded around a flat background as in equation \eqref{backgroundmetric}. This will result in an interaction term mixing scalar, vector and tensor modes. Let us briefly focus on its pure scalar part, effectively dropping all occurrences of $A^\mu_{(ij)}$ and setting all spin-2 fields to be the Minkowski $\eta$. At linear order all terms must be total derivatives, since at the level of the pure-scalar action here every $\pi_{(ij)}$ is acted on by two derivatives. Things become interesting at quadratic order.

In a theory with only bimetric interactions the different $\pi_{(ij)}$ don't interact and at quadratic order for each Goldstone boson we have
\be
{\cal S}_{int, pre-mixing}^{1-scalar} \sim \int d^4 x \left( a \pi_{\mu\nu} \pi^{\mu\nu} + b \pi_{\mu}^{\mu} \pi_\nu^\nu \right) \to \int d^4 x (a + b) \pi_{\mu\nu} \pi^{\mu\nu}. 
\ee
In the last step we have performed two integration by parts. If the interaction term is not chosen so that $a+b=0$, this term leads to a fourth order (in derivatives) equation of motion for $\pi$, signalling the presence of an Ostrogradski ghost. Setting $a+b=0$ is the equivalent of the Fierz-Pauli tuning for bimetric interactions. Even though the full effective scalar action will pick up extra contributions from mixing with metric perturbations, we will see that these cannot cure an Ostrogradski instability already present at the level of the pure scalar action pre-mixing. Also notice that, pre-mixing, the scalar action has no kinetic term for $\pi$.

In a theory with higher order spin-2 field interactions (trimetric and above), interaction terms mix different Goldstone bosons together. Since we are only considering terms up to quadratic order in the fields, this will show up via cross-terms between different $\pi_{\mu,(ij)}^\mu$. As such the quadratic interaction terms look like
\begin{align}
\nn {\cal S}_{int, pre-mixing}^{N-scalar} &\sim \sum_{i,j}^{i,j \neq k} \int d^4 x \left( a_{(ij)} \pi^{(ik)}_{\mu\nu} \pi^{\mu\nu}_{(jk)} + b_{(ij)} \pi_{\mu,(ik)}^{\mu} \pi_{\nu,(jk)}^\nu \right) \\
&\to \sum_{i,j}^{i,j \neq k} \int d^4 x \left( a_{(ij)} + b_{(ij)} \right) \pi^{(ik)}_{\mu\nu} \pi^{\mu\nu}_{(jk)},  \label{GenFPA}
\end{align}
where $k$ labels the site the interaction term is mapped to and $i,j$ run over all other sites participating in the interaction. All terms $\pi^{(ik)}_{\mu\nu} \pi^{\mu\nu}_{(jk)}$ introduce ghosts (both for $i=j$ and $i \neq j$) and we therefore find the $N$ spin-2 field interaction generalisation of Fierz-Pauli tuning: the requirement that 
\be
a_{(ij)} + b_{(ij)} = 0 \label{GenFPC}
\ee
for all $i,j$. Once again these instabilities will not be removed by mixing with the tensor modes, so it is imperative to check whether there are ghosts at the quadratic level in the pure scalar action pre-mixing to ensure stability. Also note that the presence of a generalised Fierz-Pauli coupling essentially amounts to a vanishing of the quadratic pure scalar self-interaction term up to total derivatives - if this also takes place at higher orders this is directly connected to the raising of the effective cutoff scale of the theory and potential ghost-freedom of the theory \cite{deRham:2010ik,deRham:2010kj,Hassan:2011hr,Hinterbichler:2011tt}. We will discuss this further in sections \ref{sec-cubic} and \ref{sec-dRGT}. As a final comment, it is worth mentioning that requiring the generalised Fierz-Pauli condition \eqref{GenFPC} to hold is equivalent to the following: Firstly we diagonalise the interactions of tensor-perturbations $h$ prior to performing the \St trick and then --- this is the point equivalent to enforcing \eqref{GenFPC} --- require that the resulting massive eigen-fields $\tilde h$ have Fierz-Pauli mass terms at quadratic order in the fields. \Sting these eigen-fields then automatically leads to an action for the \St scalars that separately satisfy the Fierz-Pauli condition (there is no mixing at quadratic order anymore, since one works with the eigen-fields $\tilde h$).
\\

{\bf Conformal transformation(s) :} As mentioned above, the scalar Goldstone bosons $\pi_{ij}$ have no proper kinetic terms before mixing with tensor modes. Upon considering the full scalar tensor theory the fields $\pi_{(ij)}$ and $h_{\mu\nu}^{(i)}$ become kinetically mixed. We can already see this from St\"uckelberging one spin-2 field, projecting out its Goldstone bosons $A^\mu$ and $\pi$ and expanding around the non-dynamical background metric
\begin{align} \label{goldstone3}
G_{\mu\nu}(Y) &\equiv  \pa_\mu Y^\alpha \pa_\nu Y^\beta g_{\alpha\beta}(Y) =  \pa_\mu (x^\alpha + A^\alpha) \pa_\nu (x^\beta + A^\beta) g_{\alpha\beta}(x+A) \\
&\to  \left( \delta_\mu^\alpha + \pa_\mu A^\alpha \right)  \left( \delta_\nu^\beta + \pa_\nu A^\beta \right) \left( g_{\alpha\beta} + A^\sigma \pa_\sigma g_{\alpha \beta} + \frac{1}{2}A^{\sigma}A^{\lambda} \pa_\sigma \pa_\lambda g_{\alpha\beta} \ldots   \right)\nn \\
&\to  \left( \delta_\mu^\alpha + \pa_\mu A^\alpha + \pi_\mu^\alpha \right)  \left( \delta_\nu^\beta + \pa_\nu A^\beta + \pi_\nu^\beta \right) \left( \eta_{\alpha\beta} + h_{\alpha\beta} + \left( A^\sigma + \pi^\sigma \right) \pa_\sigma h_{\alpha \beta} + \ldots   \right).\nn
\end{align}
This introduces kinetic mixing between $h_{\alpha\beta}^{(i)}$ and $\pi_{(ik)}$, which we can remove at the linear level by a Weyl rescaling of the spin-2 fields, i.e. a (linearised) conformal transformation of the metric.

In a theory with only one dynamical metric, like in dRGT, it is sufficient to transform the dynamical metric like ${\bf h}_{(i)}$ as
\begin{align}
{\bf h}_{(i)} &\to {\bf \bar{h}}_{(i)} + c_{(i)} \pi_{(ik)} {\bm \eta}. 
\end{align}
This already introduces a proper kinetic term for $\pi$. However, even in cases where there are only bimetric interactions, several dynamical spin-2 fields require a more complex transformation. This is because fluctuations ${\bf h}_{(k)}$ of the target site metric around the common background ${\bf \eta}$ will also pick up an interaction with Goldstone bosons $\pi_{(ik)}$\footnote{The common background has been chosen to be Minkowski here, but we again emphasize that any non-dynamical metric may be chosen without alterations to the formalism.}. This mixing is not removed by the above transformation. 

For a general theory with $N$ interacting spin-2 fields and higher order interactions (trimetric and above) there can be mixings between tensor perturbations from each spin-2 field and all of the Goldstone bosons. As an example consider a tetra-metric vertex upon St\"uckelberging and expanding around some common background metric. Ignoring vector modes we have
\begin{align}
\nn {g}_{(0)}^{\mn} {g}^{(1)}_{\nu\alpha}  {g}_{(2)}^{\alpha\beta} {g}^{(3)}_{\beta\mu}  \to &\; 
 ({\eta}^{\mn} + {h}_{(0)}^{\mn})     ({\eta}_{\nu\alpha} + {h}^{(1)}_{\nu\alpha} + 2\pi^{(10)}_{\nu\alpha} + \hdots)     \\  &\cdot  ({\eta}^{\alpha\beta} + {h}_{(2)}^{\alpha\beta} + 2\pi_{(20)}^{\alpha\beta} + \hdots)  ({\eta}_{\beta\mu} + {h}^{(3)}_{\beta\mu} + 2\pi^{(30)}_{\beta\mu} + \hdots)
\end{align}
which will mix all participating ${\bf h}_{(i)}$ and $\pi_{(jk)}$. 
The general linearised conformal transformation demixing a theory with arbitrary interaction vertices at the linear level will therefore be
\be \label{conformalgeneral}
{\bf h}_{(i)} \to {\bf \bar{h}}_{(i)} + \sum_j c_{(j)} \pi_{(ji)} {\bm \eta} \qquad \forall i.
\ee
This can simplify greatly depending on the nature of the interaction terms - for each index $i$ it is sufficient to let the index $j$ only run over all sites that ${\bf h}_i$ interacts with in the original gauge-fixed theory (i.e. the theory pre-St\"uckelberging). For a `line theory' (as depicted in figure \ref{fig-Danger} without the `closing link' connecting the initial and final site) with only bimetric interactions between neighbouring sites and no loops, for example, \eqref{conformalgeneral} reduces to
\be
{\bf h}_{(i)} \to {\bf \bar{h}}_{(i)} + c_{(i-1)} \pi_{(i-1,i)} {\bm \eta} + c_{(i+1)} \pi_{(i+1,i)} {\bm \eta}\qquad \forall i.
\ee 

Note that generically, in order to explicitly eliminate any mixing in the action at linear order, one needs to fix residual gauge symmetries associated with each $\bf h$ via the addition of appropriate gauge fixing terms. We discuss this below and provide an explicit example in \ref{sec-EBI}. Finally it is worth pointing out that mixing will generically persist at higher orders, see e.g. \cite{Hinterbichler:2011tt}.
\\

{\bf Scalar-tensor mixing: } What form do the terms mixing scalar and tensor modes take? And does the conformal transformation proposed above allow us to remove this mixing at the lowest order? Let us begin somewhat backward by looking at the kinetic term, i.e. the Ricci scalar, for a given spin-2 field ${\bf g}_{(i)} = {\bm \eta} + {\bf h}_{(i)}$. The linearisation of the Einstein-Hilbert Lagrangian $\sqrt{-g} R$ is
\be
{\cal L}_{\text{EH}}^{\text{lin}} = h_{\mn} \hat{\cE}^{\mn \alpha \beta} h_{\alpha \beta},
\ee
where the so-called massless kinetic operator $\hat{\cE}^{\mn \alpha \beta}$ is 
\be
 \hat{\cE}^{\mn \alpha \beta} = -\tfrac{1}{2}(\eta^{\mu (\alpha} \eta^{\beta) \nu} \Box - \eta^{\nu (\beta}\partial^{\alpha) \mu} - \eta^{\mu (\beta} \partial^{\alpha) \nu} +\eta^{\alpha \beta} \partial^{\mn} - \eta^{\mn} \eta^{\alpha \beta} \Box + \eta^{\mn} \partial^{\alpha \beta} ),
\ee
where we are symmetrising with unit weight. Now consider the conformal transformation we use to demix scalar and tensor modes. We restrict ourselves to the simple case ${\bf h}_{(i)} \to {\bf \bar{h}}_{(i)} + c_{(i)} \pi_{(i)} {\bm \eta}$ for now and find that the Einstein-Hilbert term $\sqrt{-g_{(i)}}{R}_{(i)}$ will, upon conformally transforming, generate a mixing of the form 
\begin{align} \label{mixing1}
\hat{\cE}^{\mn \alpha \beta}(\phi \eta_{\alpha \beta}) &=  X_{(1)}^{\mn}(\phi), \\
\hat{\cE}^{\mn \alpha \beta}(\pa_\alpha \phi \pa_\beta \pi) &= -\tfrac{1}{2}X_{(2)}^{\mn}(\phi,\pi),
\end{align}
where
\be \label{X-def}
X_{(n)}^{\mn}(\pi_1,\hdots,\pi_n) = \delta^{\mu \mu_1 \hdots \mu_n}_{[\nu \nu_1 \hdots \nu_n]} \pi^{\nu_1}_{\mu_1,(1)} \hdots \pi^{\nu_n}_{\mu_n,(n)}
\ee
and the antisymmetric tensor $\delta^{\mu \mu_1 \hdots \mu_n}_{[\nu \nu_1 \hdots \nu_n]}$ is defined in \eqref{delta-def} and $X_{(n)}^{\mn}$ thus satisfies $\partial_\mu X_{(n)}^{\mn} = 0$. Before the conformal transformation a number of mixing terms may be present, depending on the nature of the interaction term ${\cal S}_{\text{int}}$. In addition terms quadratic in the tensor perturbations ${\bf h}_{(i)}$ will introduce mixing terms via the conformal transformation. Overall, at lowest order in the mixing, i.e. at second order in the fields, we can list all possible contributing terms
\be \label{mixing2}
{\cal L}_{\text{EH}}^{\text{lin},(i)}, \quad\quad h_{\mu\nu}^{(i)} \pi^{\mu\nu}_{(j)}, \quad\quad h_{\mu}^{\mu,(i)} \pi^{\nu}_{\nu,(j)}, \quad\quad h_{\mn}^{(i)} h^{\mn}_{(j)}, \quad\quad h_{\mu}^{\mu,(i)} h_{\nu}^{\nu,(j)}.
\ee
Again we emphasize that, apart from  ${\cal L}_{\text{EH}}^{\text{lin}}$, which of these is present is determined by the form of ${\cal S}_{\text{int}}$. Using equation \eqref{mixing1} for linearised Einstein-Hilbert terms and computing the corresponding behaviour under conformal transformations for the other terms, we find that the `mixing terms' \eqref{mixing2} are mapped to
\begin{eqnarray}
{\cal L}_{\text{EH}}^{\text{lin},(i)} &\to& \bar{\cal L}_{\text{EH}}^{\text{lin},(i)} + 2 c_{(i)} \bar{h}_{\mu\nu}^{(i)} X^{\mu\nu}_{(1)} (\pi_{(i)}) + 3 c_{(i)}^2 \pi_{(i)} \Box \pi_{(i)}, \label{Rlinconf} \\
{h}_{\mu\nu}^{(i)} X^{\mu\nu}_{(1)} (\pi_{(j)}) &=& h_{\mu\nu}^{(i)} \pi^{\mu\nu}_{(j)} - h_{\mu}^{\mu,(i)} \Box \pi_{(j)} \to  \bar{h}_{\mu\nu}^{(i)} X^{\mu\nu}_{(1)} (\pi_{(j)})
- 3 c_{(i)} \pi_{(i)} \Box \pi_{(j)}, \label{dRGTlikeconf} \\
h_{\mu}^{\mu,(i)} \Box \pi_{(j)} &\to & \bar{h}_{\mu}^{\mu,(i)} \Box \pi_{(j)} + 4 c_{(i)} \pi_{(i)}   \Box \pi_{(j)}, \label{traceboxconf} \\
\nn h_{\mn}^{(i)}h^{\mn}_{(j)} &\to& \bar{h}_{\mn}^{(i)}\bar{h}^{\mn}_{(j)} + c_{(i)} \pi_{(i)}  \bar{h}_{\mu}^{\mu,(j)} + c_{(j)} \pi_{(j)}  \bar{h}_{\mu}^{\mu,(i)} \\ &&+ 4 c_{(i)} c_{(j)} \pi_{(i)} \pi_{(j)}, \label{h^2conf} \\
\nn h_{\mu}^{\mu,(i)} h_{\nu}^{\nu,(j)} &\to & \bar{h}_{\mu}^{\mu,(i)} \bar{h}_{\nu}^{\nu,(j)} + 4 c_{(i)} \pi_{(i)}  \bar{h}_{\mu}^{\mu,(j)} + 4 c_{(j)} \pi_{(j)}  \bar{h}_{\mu}^{\mu,(i)} \\ &&+ 16 c_{(i)} c_{(j)} \pi_{(i)} \pi_{(j)}, \label{hhconf}
\end{eqnarray}
where $\to$ denotes a conformal transformation for all $h_{(i)}$ as discussed in the previous section and we have assumed a simple form for the conformal transformation $h_{(i)} \to \bar{h}{(i)} + c_{(i)}\pi_{(i)}$.  If a more complicated form of the conformal transformation is appropriate, i.e. several $\pi$ partake in the transformation for $h_{(i)}$, additional cross-terms will appear. We discuss this below and in section \ref{sec-can}.  Equations \eqref{Rlinconf}-\eqref{hhconf} now allow us to draw a number of conclusions for generic interaction terms ${\cal S}_{\text{int}}$. 

\begin{enumerate}
\item dRGT-like couplings of the form $h_{\mu\nu}^{(i)} X_{(1)}^{\mu\nu} (\pi_{(i)})$\footnote{In general the dRGT-like mixing terms are of the form $\sum_j h_{\mu\nu}^{(i)} X_{(j)}^{\mu\nu} (\pi_{(i)})$ - the conformal transformation will only remove the lowest order mixing $h_{\mu\nu}^{(i)} X_{(1)}^{\mu\nu} (\pi_{(i)})$.} can manifestly be removed by cancelling off contributions from \eqref{Rlinconf} and \eqref{dRGTlikeconf}. In other words, by tuning the $c_{(i)}$ these mixing terms can be removed. Note that, at higher orders in the mixing, no local field redefinition which can removing scalar-tensor mixings is guaranteed to exist (although non-local redefinitions may achieve this \cite{deRham:2010ik,Hinterbichler:2011tt}).

\item In general ${\cal S}_{\text{int}}$ will  generate mixing terms of a similar form, but also involving cross couplings $h_{\mu\nu}^{(i)} X_{(1)}^{\mu\nu} (\pi_{(j)})$ with $i \neq j$. This happens when a site $i$ has several interaction terms linking to it and as a result the conformal transformation for $h_{(i)}$ required to demix scalar and tensor modes picks up a a dependence on several $\pi$. 
Suppose the original theory has some such cross-coupling $h_{\mu\nu}^{(0)} X_{(1)}^{\mu\nu} (\pi)$ and $h_{\mu\nu}^{(0)} X_{(1)}^{\mu\nu} (\phi)$ for some $h_{\mu\nu}^{(0)}$. Then the appropriate conformal transformation in order to achieve scalar-tensor demixing will be $h_{(i)} \to \bar h_{(i)} + c_1 \pi \eta + c_2 \phi \eta$, resulting in \eqref{Rlinconf} also giving rise to $h_{\mu\nu}^{(0)} X_{(1)}^{\mu\nu} (\pi)$ and $h_{\mu\nu}^{(0)} X_{(1)}^{\mu\nu} (\phi)$ terms. These can be cancelled off against each other by choosing $c_1$ and $c_2$ appropriately and we will give an explicit example in section \ref{sec-can}.


\item Suppose that, in addition to potential contributions like $h_{\mu\nu}^{(i)} X_{(1)}^{\mu\nu} (\pi_{(i)})$, ${\cal S}_{\text{int}}$ also generates `mixing terms' \eqref{traceboxconf} and \eqref{hhconf}. Now, even after the conformal transformation, mixing terms involving $\bar{h}^{\mu,(i)}_\mu$ will persist. We can remove these by appropriately gauge-fixing. We will discuss this in more detail below, but for now it is sufficient to stress that there is always enough gauge-freedom to move to a traceless gauge where $\bar{h}^{\mu,(i)}_\mu = 0$.

\item If ${\cal S}_{\text{int}}$ introduces tensor-tensor interactions at quadratic order, i.e. $h_{\mn}^{(i)} h^{\mn}_{(j)}$ and $h_{\mu}^{\mu,(i)} h_{\nu}^{\nu,(j)}$, these have two interesting effects. Upon conformally transforming these firstly lead to   $\pi_{(j)} \bar{h}_{\mu}^{\mu,(i)}$ mixing terms, which can be removed by working in a traceless gauge. Secondly, they will generically give rise to $\pi_{(i)} \pi_{(j)} $ cross-terms for the pure-scalar action and consequently also to $\pi_{(i)}^2$ mass terms for the Goldstone bosons $\pi_{(i)}$.
\end{enumerate}

To summarise the last two sections: The conformal transformations proposed generate kinetic terms for all `Goldstone bosons' $\pi_{(i)}$ (which is essential in order to proceed with canonically normalising all fields) and also allow us to eliminate scalar-tensor mixing at quadratic order in the fields. For further discussion of the nature of scalar-tensor mixing terms see appendix \ref{appendix-nonlocal}. However, as we shall see in section \ref{sec-can}, the different $\pi_{(i)}$ generically remain kinetically mixed, so we cannot at this stage identify them as separate propagating \dof. In section \ref{sec-can} we discuss how this scalar mixing can be demixed and  the appropriate propagating \dof can be identified.
\\

{\bf Gauge transformations:  }
Consider a single interaction vertex in a multi-metric theory and the effect of the \St trick on it. If everything is brought to site $j$, the (infinitesimal) gauge transformation properties of the fields are
\begin{align}
\delta x^{(i) \mu} &= \xi^{(i) \mu} \qquad \forall i \\
\delta h^{(i)}_{\mn} &= \pa_\mu \xi^{(i)}_\nu + \pa_\nu \xi^{(i)}_\mu + \cL_\xi h_{\mn} \qquad \forall i \\
\delta A^{(i)}_\mu &= - \xi^{(i)}_\mu - A^{(i) \alpha} \pa_\alpha \xi^{(i)}_\mu + \hdots + \xi^{(j)}_\mu + \xi^{(j) \alpha} \pa_\alpha A^{(i)}_\mu + \pa_\mu \Lambda^{(i)} \qquad i \neq j \\
\delta \pi^{(i)} &= - \Lambda^{(i)} \qquad i \neq j,
\end{align}
where the ellipsis denotes terms containing higher powers of $A^{(i)}$, and the Lie derivative is $\cL_\xi h_{\mn} = \xi^\alpha \pa_\alpha h_{\mn} + \pa_\mu \xi^\alpha h_{\alpha \nu} + \pa_\nu \xi^\alpha h_{\mu \alpha}$. Taking the decoupling limit the gauge transformations are reduced to
\begin{align}
\delta h^{(i)}_{\mn} &= \pa_\mu \xi^{(i)}_\nu + \pa_\nu \xi^{(i)}_\mu \qquad \forall i \\
\delta A^{(i)}_\mu &= \pa_\mu \Lambda^{(i)} \qquad i \neq j \\
\delta \pi^{(i)} &= 0 \qquad i \neq j,
\end{align}
where $\xi$, and $\Lambda$ have been suitably rescaled. The gauge freedom associated with $\xi$ should now be fixed by imposing suitable gauge conditions (we will discuss the Lorentz gauge below).
This is particularly relevant when eliminating scalar-tensor mixing at quadratic level in fields. As we saw above, quadratic scalar-tensor mixings involving the trace $h_{\mu,(i)}^{\mu}$ generically remain after the conformal transformation designed to demix the action (cf. equations \eqref{dRGTlikeconf} and \eqref{hhconf}). Here we will show that this mixing is eliminated once appropriate gauge fixing terms are added to the action. 

Let us begin by recalling the Lorentz and transverse traceless gauges commonly used in standard General Relativity ($m^2 = 0$), where we generalise to $N$ non-interacting spin-2 fields labelled by an index $i$. Now the Lorentz gauge condition corresponds to $\pa_\mu h_{\nu, (i)}^\mu = 0$ and fixes the gauge freedom up to residual gauge transformations $\Box \xi_\mu^{(i)} = 0$. One possible gauge choice to fix the remaining gauge freedom is the transverse, traceless gauge (essentially the gravitational equivalent of the Coulomb gauge), which results in the complete set of gauge conditions 
\begin{align}
\pa_\mu h_{\nu, (i)}^\mu &= 0,  &h_{0\mu,(i)} &= 0,  &h_{\mu,(i)}^\mu &= 0. 
\end{align}
The requirement $h_{\mu,(i)}^\mu = 0$ is one of the gauge fixing conditions, i.e. it does not fix all the gauge degrees of freedom, but only a subset of the conditions associated with the transverse traceless gauge. In fact it fixes only a single real space degree of freedom \cite{Hinterbichler:2011tt}. In order to eliminate the scalar-tensor mixing at quadratic level in the interacting spin-2 field theories considered above $h_{\mu,(i)}^\mu = 0$ is a sufficient requirement, so we do not need to fully specify the gauge fixing conditions in order to ensure demixing. 

So let us see how this can be implemented in the massive gravity/interacting spin-2 field case. The analogous gauge fixing conditions in the Lorentz gauge are
\begin{align}
\pa^\nu h_{\mn}^{(i)} - \frac{1}{2} \pa_\mu h^{\nu,(i)}_\nu + m A_\mu^{(i)} &= 0 \\
\pa_\mu A^\mu_{(i)} + m \left( \frac{1}{2} h_{\nu,(i)}^\nu + 3 \pi_{(i)}   \right) &= 0.
\end{align}
This gauge still leaves a residual gauge freedom $(\Box - m^2)\xi_\mu^{(i)} = 0$. And this residual gauge freedom is enough so that we can always consistently set the trace of $h_{(i)}$ to zero and consequently eliminate scalar-tensor mixing at the quadratic level. 

Finally we may wonder what the precise form of $\xi_\mu^{(i)}$ enforcing tracelessness is. Consider the decoupling limit gauge transformations for tensor modes
\be
h_{\mu\nu}^{' (i)} = h_{\mn}^{(i)} - \pa_\nu \xi_{\mu}^{(i)} - \pa_\mu \xi_{\nu}^{(i)}
\ee
We now take the trace and require $h_{\mu}^{' \mu} $ to vanish. In other words we want to find out whether we can always consistently set the trace to zero with some gauge transformation, regardless of the initial form of $h_\mu^{' \mu}$. The result is
\be
 h_{\mu,(i)}^{\mu} = \pa_\mu \xi^{\mu}_{(i)} + \pa^\mu \xi_{\mu}^{(i)} = 2\pa_\mu \xi^\mu_{(i)},
\ee
where we recall that are we are raising and lowering indices with the flat Minkowski metric $\eta_{\mn}$. This is the condition on $\xi^\mu_{(i)}$ in order to eliminate $h_\mu^{\mu, (i)}$ . 
\\

\section{Scalar mixing and propagating modes}\label{sec-can}

\begin{figure}[tp]
\centering
\begin{tikzpicture}[-,>=stealth',shorten >=0pt,auto,node distance=2cm,
  thick,main node/.style={circle,fill=blue!10,draw,font=\sffamily\large\bfseries},arrow line/.style={thick,-},barrow line/.style={thick,->},no node/.style={plain},rect node/.style={rectangle,fill=blue!10,draw,font=\sffamily\large\bfseries},red node/.style={rectangle,fill=red!10,draw,font=\sffamily\large\bfseries},green node/.style={circle,fill=green!20,draw,font=\sffamily\large\bfseries},yellow node/.style={rectangle,fill=yellow!20,draw,font=\sffamily\large\bfseries}]

 \node[main node](100){};
  \node[main node] (101) [right of=100] {};
  
    \node[main node] (1000)  [right=1.5cm of 101]{};
   \node[main node] (1001) [right=0.5cm of 1000] {};
   \node[draw=none,fill=none] (10000) [right=0.5cm of 1001] {};
  \node[main node] (1002) [right=0.5cm of 10000] {};
   \node[main node] (1003) [right=0.5cm of 1002] {};
 
  \node[main node] (2)  [right=2cm of 1003]{};
   \node[main node] (1) [above=1cm of 2] {};
  \node[main node] (3) [below left=1cm of 2] {};
   \node[main node] (4) [below right=1cm of 2] {};
   
    \node[main node,fill=black!100,scale=0.7] (5)  [right=4cm of 2]{};
  \node[main node] (7) [below left=1cm of 5] {};
   \node[main node] (8) [below right=1cm of 5] {};
   
     \node[draw=none,fill=none](80)[above left=1.5cm of 5]{};
       \node[draw=none,fill=none](81)[above right=1.5cm of 5]{};
        \node[draw=none,fill=none](82)[above=1.5cm of 5]{};
         \node[draw=none,fill=none](83)[right=0.7cm of 100]{};

  \path[every node/.style={font=\sffamily\small}]
  (100) edge node {} (101)
    (1) edge node {} (2)
     (4) edge node {} (2)
     (3) edge node  {} (2)
         (1000) edge node {} (1001)
     (1002) edge node {} (1003)
      (7) edge node [above left] {} (5)
       (8) edge node [above right] {} (5);


\draw[-,dashed] (1001) to (1002);

\draw[-,dashed] (5) to (80);
\draw[-,dashed] (5) to (81);
\draw[-,dashed] (5) to (82);
    
\draw [<->] ($(8)+(0.2,0.2)$) 
    arc (-45:225:1.65);

   \node[draw=none,fill=none](92)[below of=83]{(a) Bimetric};
   
   \node[draw=none,fill=none](95)[below of=10000]{(b) Line theory};
   
   \node[draw=none,fill=none](93)[below of=2]{(c) Branching vertex};
   
   \node[draw=none,fill=none](94)[below of=5]{(d) N-metric};
\end{tikzpicture}
\caption{Different types of theories and interaction vertices discussed in section \ref{sec-can}. From left to right we consider: (a) Isolated bimetric interactions simply connecting two sites not connected to any other sites. (b) A `line theory' made up of $N$ sites with nearest neighbour interactions only and hence forming a line with $N-1$ links. Note that connectig the initial and final site with a link would reproduce a `circle theory` which forms one large loop as discussed above. (c) Branching vertices where one `central' node is conected to several outer nodes with bimetric interactions. The constructions discussed in the line theory and branching vertices examples together are sufficient to build an arbitrarily complex, loop-free theory with $N$ sites only containing bimetric interactions. (d) N-metric interactions coupling together $N-1$ Goldstone bosons.} \label{fig-scalarmixing}
\end{figure}
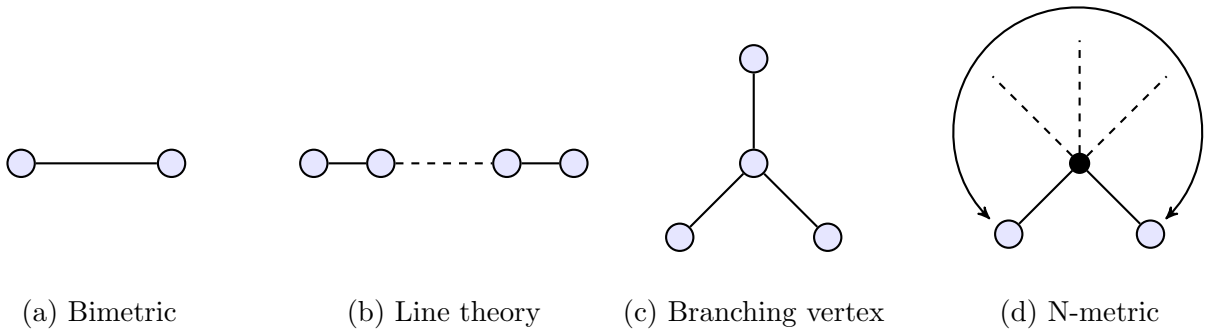

Having Weyl transformed all the relevant spin-2 fields for each interaction term in ${\cal S}_{int}$, all the $\pi_{(i)}$ modes have proper kinetic terms now and the scalar-tensor mixing has been eliminated at the lowest order. However, in a generic theory the Weyl transformation will have induced kinetic mixing between the different $\pi$ modes. In addition, tensor-tensor terms such as $h_{\mn}^{(i)} h^{\mn}_{(j)}$ will generically generate `mass terms' $\pi_{(i)} \pi_{(j)}$ via the conformal transformation. In other words, we still have not completely diagonalised the action (again, up to quadratic order in the fields) and scalar mixing (both kinetic- and mass-mixing generically persists). Here we show why this mixing arises and how to demix the scalar action allowing us to identify what the actual propagating modes are. Note that, for the purposes of this section, Latin indices are site labels (unless otherwise stated and even when not bracketed) and are not summed over, whereas Greek indices are space-time indices and the Einstein summation convention applies as usual.
 \\

{\bf No scalar mixing in bimetric theories: }
In a bimetric theory no scalar mixing arises as there is only a single would-be Goldstone boson $\pi$. Let us show this explicitly. The interaction term is of the form
\be \label{toybimaction}
{\cal S}_{\text{int}} = f \left( {{\bf g}_{(0)}} , {{\bf g}_{(1)}} \right),
\ee
where $f$ is some function of the two spin-2 fields $ {{\bf g}_{(0)}}$ and ${{\bf g}_{(1)}}$. This corresponds to graph (a) in figure \ref{fig-scalarmixing}. Performing the \St trick, expanding around a background and dropping the \St vector $A^\mu_{(0 1)}$ this results in\footnote{Note that we have defined the perturbations ${h}^{\mu\nu}_{(0)}$ relative to the inverse metric ${\bf g}_{(0)}^{-1}$, i.e. ${g}_{(0)}^{\mn} = \eta^{\mn} + {h}^{\mu\nu}_{(0)}$. If instead we define a perturbation $\tilde{h}^{\mu\nu}_{(0)}$ with respect to the metric  via ${g}^{(0)}_{\mn} = \eta_{\mn} + \tilde{h}_{\mu\nu}^{(0)}$, then a relative sign change takes place (to lowest order), i.e. $\tilde{h}^{\mu\nu}_{(0)} = - {h}^{\mu\nu}_{(0)} + {\cal O}(h^2) $. For comparison see \eqref{linetheorySt} where this becomes relevant explicitly, since ${\bf g}_{(i)}$ and ${\bf g}_{(i)}^{-1}$ both enter.}
\begin{eqnarray}
{\cal S}_{\text{int}}  \to f \left( \eta^{\mu\nu} + {h}^{\mu\nu}_{(0)} \; , \;  (\delta^\alpha_\mu + \pi^\alpha_{\mu}) (\delta^\beta_\nu + \pi^\beta_\nu) (\eta_{\alpha\beta} + h_{\alpha\beta}^{(1)} + \pi^\sigma \pa_\sigma h^{(1)}_{\alpha \beta} + \ldots ) \right). 
\end{eqnarray}
To go further we must know something about the nature of $f$. Here we choose a symmetric interaction term, i.e. we consider terms $ f \left( {{\bf g}_{(0)}} , {{\bf g}_{(1)}} \right)$, which do not distinguish between $ {{\bf g}_{(0)}}$ and ${{\bf g}_{(1)}}$. For large parts of our argument the only important feature of this choice will be its symmetric nature. However, for concreteness we take it to be of the dRGT/Hassan-Rosen bigravity form \cite{Hassan:2011tf,Hassan:2011zd,Hassan:2011ea} as discussed in more detail in section \ref{sec-dRGT} below. The scalar-tensor mixing terms pre-conformal-transformation now are
\be \label{bim-line-mixing-1}
m^2 h_{\mn}^{(i)} X^{\mn}_{(1)}(\pi).
\ee
In order to remove the scalar-tensor mixing we perform the conformal transformation
\be
h_{\mn}^{(i)} \to \bar{h}_{\mn}^{(i)} + c_1 \pi \eta_{\mn}.
\ee
Under this transformation equation \eqref{bim-line-mixing-1} is mapped to
\begin{eqnarray}
m^2  \left[   \bar{h}_{\mn}^{(i)}  X^{\mn}_{(1)}(\pi)  +  c_1 \pi X^\mu_{\mu,(1)} (\pi) \right] 
=  m^2 \left[ \bar{h}_{\mn}^{(i)}  X^\mu_{\mu,(1)}(\pi)  +  3 c_1 \pi \Box \pi \right].  \label{bim-m2-mixing}
\end{eqnarray} 
As we saw above, additional mixing terms are generated from ${\cal L}_{\text{EH}}^{\text{lin}}$, namely we have
\begin{eqnarray}
\nn h_{\mu\nu} \hat{\cal E}\epsilon^{\mu \nu \alpha \beta} h_{\alpha\beta} &\to & \bar{h}_{\mu\nu} \hat{\cal E}^{\mu \nu \alpha \beta} \bar{h}_{\alpha\beta} + 2 c_1 \bar{h}_{\mn}  X^{\mn}_{(1)}(\pi) +  c_1^2 \pi X^\mu_{\mu,(1)} (\pi) \\
&=& \bar{h}_{\mu\nu} \hat{\cal E}^{\mu \nu \alpha \beta} \bar{h}_{\alpha\beta} + 2 c_1 \bar{h}_{\mn}  X^{\mn}_{(1)}(\pi)  + 3 c_1^2 \pi \Box \pi . \label{bim-Rlin-mixing}
\end{eqnarray}
In order to eliminate the scalar-tensor mixing we require $m^2 = - 2 c_1$. This means the pure scalar action that remains once equations \eqref{bim-m2-mixing} and \eqref{bim-Rlin-mixing} are combined is
\begin{eqnarray}
-\frac{3}{4} m^4 \pi \Box \pi.
\end{eqnarray}
In other words, $\pi$ is a propagating Goldstone mode and no further demixing needs to be performed after the scalar-tensor mixing has been eliminated.
\\

{\bf Scalar mixing - an example: } The situation is different if more than two spin-2 fields interact with bimetric interactions. We start with a trimetric theory with two bimetric interaction terms. 
\be \label{toylineaction}
{\cal S}_{\text{int}} = m_I^2 f \left( {{\bf g}_{(i-1)}} , {{\bf g}_{(i)}} \right) + m_{II}^2 f \left( {{\bf g}_{(i)}} , {{\bf g}_{(i+1)}} \right).
\ee
 This corresponds to graph (b) in figure \ref{fig-scalarmixing}. Performing the \St trick and expanding around the background this now results in
\begin{eqnarray}
\nn && m_{I}^2 f \left( \eta^{\mu\nu} + {h}^{\mu\nu}_{(i-1)} \; , \;  (\delta^\alpha_\mu + \pi^\alpha_{(1),\mu}) (\delta^\beta_\nu + \pi_{(1),\nu}^\beta) (\eta_{\alpha\beta} - h_{\alpha\beta}^{(i)} + \ldots ) \right) \\
&+ & m_{II}^2 f \left( \eta^{\mu\nu} + h^{\mn}_{(i)} \; , \; (\delta^\alpha_\mu + \pi^\alpha_{(2),\mu}) (\delta^\beta_\nu + \pi^\beta_{(2),\nu}) (\eta_{\alpha\beta} + h_{\alpha\beta}^{(i+1)} + \ldots ) \right), \label{linetheorySt}
\end{eqnarray}
where $\pi_{(1)}$ and $\pi_{(2)}$ are the \St scalars from the first and second bimetric interaction term respectively. 
Moving to the scalar-tensor mixing terms for the `central' $i$ node pre-conformal-transformation we now have
\be 
m^2 \left( \alpha_{1}^2 h_{\mn}^{(i)} X^{\mn}_{(1)}(\pi_{(1)})  \pm \alpha^2_2 h_{\mn}^{(i)} X^{\mn}_{(1)}(\pi_{(2)}) \right). \label{line-STmixing}
\ee
The $\pm$ sign and the parameters $\alpha^2$ are of some importance here. The sign of a $h_{\mn}^{(i)} X^{\mn}_{(1)}(\pi_{(j)})$ mixing term depends on how we define and introduce the \St fields $Y_{ij}$ as well as $h_{\mn}^{(i)}$. Depending on what signs are chosen for $h_{\mn}^{(i)}$ and $\pi_{(ij)}$ and whether we use $Y_{ji}$ or $Y_{ij}$ for two particular neighbouring sites $i$ and $j$, `incoming' links $Y_{(ji)}$ will generate a positive $+ h_{\mn}^{(i)} X^{\mn}_{(1)}(\pi_{(ji)})$ coupling, whereas `outgoing' links $Y_{(ij)}$ will generate a negative $- h_{\mn}^{(i)} X^{\mn}_{(1)}(\pi_{(ij)})$ coupling or vice versa. What is important here is not the convention that is chosen, but that incoming and outgoing links generate mixing terms of opposite signs. Depending on how we \St the interaction terms there can therefore be a relative sign difference in the couplings in equation \eqref{line-STmixing} encoded by the $\pm$ sign. Secondly, $\alpha^2$ encodes the difference between coupling strengths $m_{I}$ and $m_{II}$ for the two bimetric interactions. In other words we have $m_{I}^2 = \alpha_1^2 m^2$ and $m_{II}^2 = \alpha^2_2 m^2$, where the scale $m^2$, which is factored out, is arbitrary. 

Let us now return to \eqref{line-STmixing} and demix the fields. For now we solely focus on the quadratic interaction terms (both self-interaction as well as mixing terms) from the $h_{\mn}^{(i)}$ couplings. After all, we already know what the $h_{\mn}^{(i-1)}$ and $h_{\mn}^{(i+1)}$ couplings will do (they are analogous to the simple bimetric examples considered above). The conformal transformation for $h_{\mn}^{(i)}$ will be
\be
h_{\mn}^{(i)} \to \bar{h}_{\mn}^{(i)} + c_1 \pi_{(1)} \eta_{\mn} +  c_2 \pi_{(2)} \eta_{\mn}.
\ee
Under this transformation, dropping all $h_{\mn}^{(i-1)}$- and $h_{\mn}^{(i+1)}$-dependent terms and choosing $m^2$ such that $\alpha_1^2 = 1$, from equation \eqref{line-STmixing} we have
\begin{eqnarray}
 &m^2  &\left[   \bar{h}_{\mn}^{(i)}  \left( X^{\mn}_{(1)}(\pi_{(1)}) \pm \alpha^2_2 X^{\mn}_{(1)}(\pi_{(2)}) \right) \right. \label{line-m2-mixing}\\
\nn &&+ \left. c_1 \pi_{(2)} X^\mu_{\mu,(1)} (\pi_{(1)}) + c_2 \pi_{(2)} X^\mu_{\mu,(1)}(\pi_{(1)}) \pm c_1 \alpha^2_2 \pi_{(1)} X^\mu_{\mu,(1)} (\pi_{(2)}) \pm c_2 \alpha_2^2 \pi_{(2)} X^\mu_{\mu,(1)}(\pi_{(1)}) \right] \\
\nn &=&  m^2 (\bar{h}_{\mn}^{(i)}   (X^\mu_{\mu,(1)}(\pi_{(1)}) \pm \alpha_2^2 X^\mu_{\mu,(1)}(\pi_{(2)}) ) +  3 c_1 \pi_{(1)} \Box \pi_{(2)} + 3 (c_2 \pm \alpha_2^2 c_1) \pi_{(2)} \Box \pi_{(1)} \pm 3 c_2 \alpha^2_2 \pi_{(2)} \Box \pi_{(2)}). 
\end{eqnarray} 
Now additional mixing terms are generated from ${\cal L}_{\text{EH}}^{\text{lin}}$ under the conformal transformation and we have
\begin{eqnarray}
\nn h_{\mu\nu} \hat{\cal E}^{\mu \nu \alpha \beta} h_{\alpha\beta} &\to & \bar{h}_{\mu\nu} \hat{\cal E}^{\mu \nu \alpha \beta} \bar{h}_{\alpha\beta} + 2 c_1 \bar{h}_{\mn}  X^{\mn}_{(1)}(\pi_{(1)}) + 2 c_2 \bar{h}_{\mn}  X^{\mn}_{(1)}(\pi_{(2)}) \\
\nn &+& c_1^2 \pi_{(1)} X^\mu_{\mu,(1)} (\pi_{(1)}) + c_1 c_2 \pi_{(2)} X^\mu_{\mu,(1)}(\pi_{(1)}) + c_1 c_2 \pi_{(1)} X^\mu_{\mu,(1)} (\pi_{(2)}) + c_2^2 \pi_{(2)} X^\mu_{\mu,(1)}(\pi_{(2)}) \\
\nn &=& \bar{h}_{\mu\nu} \hat{\cal E}^{\mu \nu \alpha \beta} \bar{h}_{\alpha\beta} + 2 c_1 \bar{h}_{\mn}  X^{\mn}_{(1)}(\pi_{(1)}) + 2 c_2 \bar{h}_{\mn}  X^{\mn}_{(1)}(\pi_{(2)}) \\
&+& 3 c_1^2 \pi_{(1)} \Box \pi_{(1)} + 6 c_1 c_2 \pi_{(2)} \Box \pi_{(1)} + 3 c_2^2 \pi_{(2)} \Box \pi_{(2)}. \label{line-Rlin-mixing}
\end{eqnarray}
In order to eliminate the scalar-tensor mixing we require $\pm \alpha_2^2 m^2 = - 2 c_2 = \mp 2 \alpha_2^2 c_1$. This means the pure scalar action that remains once equations \eqref{line-m2-mixing} and \eqref{line-Rlin-mixing} are combined is
\begin{eqnarray}
\nn &-&\frac{3}{4} m^4 \left(  \pi_{(1)} \Box \pi_{(1)} + 2 \alpha_2^2 \pi_{(2)} \Box \pi_{(1)} + \alpha_2^4 \pi_{(2)} \Box \pi_{(2)}   \right) \\
= &-& \frac{3}{4} m^4 (\pi_{(1)} + \alpha_2^2 \pi_{(2)}) \Box (\pi_{(1)} + \alpha_2^2 \pi_{(2)}), \label{line-central-mixing}
\end{eqnarray}
where we have worked with $(X^{\mn}_{(1)}(\pi_{(1)}) + \alpha^2_2 X^{\mn}_{(1)}(\pi_{(2)}))$, i.e. a positive relative sign, in order to avoid clutter from now on - it is straightforward to modify \eqref{line-central-mixing} and the below in the case of a negative relative sign. Now the $h_{(i-1)}$ and $h_{(i+1)}$ mixings will generate additional kinetic terms 
\be
-  \frac{3}{4} m^4 (\pi_{(1)} \Box \pi_{(1)}  + \alpha_2^4 \pi_{(2)} \Box \pi_{(2)} ) = -  \frac{3}{8} m^4 \left[ (\pi_{(1)} + \alpha_2^2 \pi_{(2)}) \Box (\pi_{(1)} + \alpha_2^2 \pi_{(2)})  + (\pi_{(1)} - \alpha_2^2 \pi_{(2)}) \Box (\pi_{(1)} - \alpha_2^2 \pi_{(2)}) \right].
\ee
This form now makes it obvious that we can eliminate the kinetic scalar mixing by moving to
\be
\{ \pi_{(1)}, \pi_{(2)}\}  \to \{\phi_{(1)} = \frac{1}{\sqrt{2}} \left(\pi_{(1)} + \alpha_2^2 \pi_{(2)} \right) \; , \; \phi_{(2)}  = \frac{1}{\sqrt{2}} \left( \pi_{(1)} - \alpha_2^2 \pi_{(2)} \right) \},
\ee
i.e. we essentially map $\{ \pi_{(1)}, \pi_{(2)}\} $ to the normalised eigenvectors $\phi_{(1)} ,\phi_{(2)} $ of the matrix encoding the kinetic mixing between $\{ \pi_{(1)}, \pi_{(2)}\}$. We can now finally write the full kinetic scalar interaction term as
\be
-  \frac{3}{4} m^4 \left[ 3 \phi_{(1)} \Box \phi_{(1)}  +  \phi_{(2)} \Box \phi_{(2)}   \right].
\ee
Note that we have so far focussed on diagonalising the kinetic terms of our trimetric theory here. That is of course consistent, but will identify \dof which are no longer kinetically mixed, yet in principle still interact through potential terms such as $\phi_{(1)} \phi_{(2)}$. If one wants identify the proper eigenmodes of the theory, i.e. independently evolving \dof (at least up to quadratic order in the fields), one should first canonically normalise the fields $\phi$ and then also diagonalise the mass terms for the \St scalars. These will come from tensor-tensor terms such as $h_{\mn} h^{\mn}$ after the conformal transformation has come into effect. This would result in identifying the independently propagating and canonically normalised \dof $\hat\chi$. We describe how this procedure works in detail below. Already at this point, however, we emphasize that neither the kinetic eigenmodes $\phi$ nor the $\hat \chi$ we will describe below are the original \St scalars $\pi_{(1)}$ and $\pi_{(2)}$, but instead are linear combinations thereof. Also we stress that we have not canonically normalised the $\phi$ fields at this point yet.  
\\

\comment{
We can therefore identify the demixed propagating modes $\chi_{(i)}$  which are $\chi_{(i-2)} = \alpha^2_{(i)} \pi_{(i-1,i)} - \alpha_{(i-1)}^2 \pi_{(i-2,i-1)} $ for $i = 3,\ldots,N $ as well as $\chi_{(N-1)} = \alpha^2_{(2)} \pi_{(1,2)} + \alpha^2_{(N)} \pi_{(N-1,N)}$. In terms of the $\chi_{(i)}$ the scalar action finally takes the form
\be
\frac{3}{4} m^4 \left[\sum_{i=1}^{N-2} 3 \chi_{(i)} \Box \chi_{(i)}  +  \chi_{(N-1)} \Box \chi_{(N-1)}   \right].
\ee
again it is the $N-1$ $\chi_{(i)}$, linear combinations of the $\pi_{(ij)}$, that are the propagating Goldstone modes.
\\}

{\bf The general demixing procedure and mass-mixing: }
What procedure should one follow if we aim to fully eliminate the kinetic- and mass-mixing between the \St scalars and establish the independently propagating scalar \dof of an interacting spin-2 field theory up to quadratic order in the fields? Having established the independently propagating modes will be of importance once we consider higher order interactions and their scale in the next section. Taking into account both kinetic and mass mixing, the pure scalar quadratic part of the Lagrangian is
\be
\pi^\text{T} \left( K \Box + M \right) \pi,
\ee
where we have arranged the \St fields into a column vector $\pi$ and encoded kinetic- and mass-mixing in the matrices $K$ and $M$ respectively. De-mixing at the quadratic level then amounts to finding a basis in which the matrices $K$ and $M$ are diagonal; if and only if they commute then they can be simultaneously diagonalised by a unitary transformation acting on $\pi$. A good example for this case is the `loop' theory considered in \cite{deRham:2013awa}.\footnote{There a \St field is introduced for each link (including the loop-closing one) and performing the kinetic demixing one finds that one linear combination of all these fields does not propagate. In other words the right number of \dof remains and 
one recovers a form physically equivalent to the `plaquette' procedure described in section \ref{sec-multiSt}.} There each of $K$ and $M$ is a circulant matrix, which can be diagonalised by a discrete Fourier transform.

However for a more general theory (a good specific example being `line' theories, which we will discuss in some detail below) this is not the case: $K$ and $M$ do not commute and hence cannot be simultaneously diagonalised by one unitary transformation. Instead one must first diagonalise $K$ and then perform a rescaling of the new fields (a non-unitary transformation), which essentially amounts to their canonical normalisation. Let us call the vector of fields this results in $\phi$ and $\hat \phi$ (before and after canonical normalisation, i.e. applying $N$, respectively). As a result the kinetic matrix for $\hat\phi$ is proportional to the identity and the mass matrix for $\hat \phi$ is some $\tilde M$, which is real and symmetric - so it is diagonalizable by an orthogonal matrix $U^{M}$. Since $\tilde M$ and $I$ commute, the kinetic and mass terms for $\hat\phi$ can be diagonalised simultaneously - in fact by $U^{M}$. This is because ${U^{M}}^T U^{M} = U^{M} {U^{M}}^T = I$, so when the mass matrix $\tilde M$ is diagonalised by applying $U^{M}$ to $\hat\phi$, resulting in a new field vector $\hat \chi$, the resulting kinetic term for $\hat \chi$ is still diagonal and the fields are still canonically normalised. As a result we have successfully diagonalised both the mass and kinetic matrices. More formally one has\footnote{We here normalise scalar kinetic terms to $- \frac{1}{2} \hat\phi \Box \hat\phi$ and hence $- \frac{1}{2} \hat\chi \Box \hat\chi$. Note the choice of sign (especially since we have not specified the signature of the spin-2 fields involved anywhere yet).}
\be
\pi_{i} \left( K_{i j} \Box + M_{i j} \right) \pi_{j} = - \frac{1}{2} \sum_{i} \hat{\chi}_{i} \left( \Box +  \mu^2_{(i)} \right) \hat{\chi}_{i},
\ee
where $\mu^2_{(i)}$ are the diagonalised mass-terms and 
\be
\pi_i = U^K_{i j} N_{j k} U^M_{k l} \hat{\chi}_l, \label{diagonalisation transformation}
\ee
where $U^K$ is the matrix used to diagonalise $K$ (e.g. the matrix of eigenvectors of $K$) and $N$ is the diagonal matrix used to canonically normalise the fields. This means that $N = \text{diag}(\{\sqrt{2 \lambda_i}^{-1}\})$, where $\lambda_i$ are the eigenvalues of $K$. As discussed above $U^M$ is the orthogonal matrix of eigenvectors of $\tilde M = N^\text{T} {U^K}^\text{T} M U^K N$. The fields $\hat{\chi}$ are now completely de-mixed up to quadratic order in the fields and are consequently the independently propagating scalar \dof of the theory. As we will see, working in terms of the $\hat \chi$ makes analysing the suppression scale of higher order terms straightforward.
\\

{\bf Scalar mixing with generic bimetric interactions - a `line theory': }
Let us now perform the first step of the general procedure --- de-mixing the kinetic scalar terms --- for a more complicated theory. Namely for a `line theory' with $N$ sites as depicted in graph (b) of figure \ref{fig-scalarmixing}
\be \label{toylineNaction}
{\cal S}_{\text{int}} = \sum_{i=1}^{N-1} m_{(i+1)}^2 f \left( {{\bf g}_{(i)}} , {{\bf g}_{(i+1)}} \right).
\ee
Suppose we choose to \St this interaction term by introducing $N-1$ \St fields $Y_{(i,i+1)}$, i.e. we always map from site $i$ to $i+1$, never the reverse. Except for the initial and final site, each site therefore has one incoming and one outgoing link, resulting in a sign difference between $h_{\mn}^{(i)} X^{\mn}_{(1)}(\pi_{i-1,i})$ and $h_{\mn}^{(i)} X^{\mn}_{(1)}(\pi_{i,i+1})$. Each $Y_{(i,i+1)}$ will give rise to a \St scalar $\pi_{(i,i+1)}$. In terms of these \St scalars the kinetic scalar interactions post-conformal-transformation will schematically be
\begin{eqnarray}
\nn &-&\frac{3}{4} m^4 \Bigg[ \alpha_{(2)}^4 \pi_{(1,2)} \Box \pi_{(1,2)} + \alpha_{(N)}^4 \pi_{(N-1,N)} \Box \pi_{(n-1,n)}  \\
&+& \left. \sum_{i=3}^{N} (\alpha_{(i)}^2 \pi_{(i-1,i)} - \alpha_{(i-1)}^2 \pi_{(i-2,i-1)}) \Box (\alpha_{(i)}^2 \pi_{(i-1,i)} - \alpha_{(i-1)}^2  \pi_{(i-2,i-1)}) \right], \label{linemixing}
\end{eqnarray}
where $m_{i}^2 \alpha_{(i)}^2 = m^2$. As discussed above equation \eqref{linemixing} can be re-written in matrix format as
\be
-\frac{3}{4}m^4 \pi^T
\comment{ \left( \begin{smallmatrix}
   2 & 1 &  &   &  &  &  \\
 1 &  2 & 1 &    &  & &  \\
    & 1 &  &    &  &  &  \\
    &  &  & \ddots  &  &   &   \\
   & & &  &  & 1 & \\
   &  &  &    & 1 & 2 & 1\\
   &  &  &    &  & 1 & 2
 \end{smallmatrix} \right)}
 K \Box \pi,
\ee 
where $K$ here is a tri-diagonal matrix with $2$ on the diagonal and $-1$ on the super- and sub-diagonals. $K$ is real and symmetric, so we can identify its eigenvectors and eigenvalues to construct the orthogonal matrices $U^K$ that diagonalise the kinetic terms by mapping the $\pi_{(i,i+1)}$ to linear combinations as identified by the eigenvectors.
\\

{\bf Scalar mixing with generic bimetric interactions - branching vertices: }
Having considered a single bimetric interaction and its generalisation to a `line theory', we are only missing one piece to build a generic theory only containing bimetric interaction terms. To build such a theory we also need to know how to handle branching vertices like
\be \label{toybranchaction}
{\cal S}_{\text{int}} = m_{I}^2 f \left( {{\bf g}_{(0)}} , {{\bf g}_{(1)}} \right) +m_{II}^2   f \left( {{\bf g}_{(0)}} , {{\bf g}_{(2)}} \right) + m_{III}^2  f \left( {{\bf g}_{(0)}} , {{\bf g}_{(3)}} \right).
\ee
Such a branching vertex is depicted in graph (c) in figure \ref{fig-scalarmixing}. By combining such branching vertices (with a `central' node and several `outer' ones) with the line-type interactions discussed above, we can build a generic theory with only bimetric interaction terms. For concreteness we will only discuss a 4 spin-2 field branching vertex here, but generalisation to higher order branching vertices is straightforward. Returning to \eqref{toybranchaction}, we again \St interaction terms, expand around the background and focus on the mixing terms between $h_{(i)}$ and $\pi_{(j,0)}$ (though from now on the second index will be omitted)
\begin{eqnarray}
\nn && m^2 \sum_{i=1}^3 \alpha^2_{(i)} \left( h^{(0)}_{\mn}-h^{(i)}_{\mn} \right) X^{\mn}_{(1)}(\pi_{(i)}) \\ &=& m^2 \left[ h^{(0)}_{\mn} X^{\mn}_{(1)} \left( \sum_{i=1}^3 \alpha^2_{(i)} \pi_{(i)} \right) - \sum_{i=1}^3 \alpha^2_{(i)} h^{(i)}_{\mn} X^{\mn}_{(1)}(\pi_{(i)}) \right].
\end{eqnarray}
The first term corresponds to the scalar-tensor interactions at the `central' node ${\bf g}_{(0)}$, whereas the second term sums up contributions from all `outer' nodes. Note that we have chosen \St fields $Y$ such that all relative signs for the `central' interaction term are positive. The conformal transformations for $h_{(i)}$ will now be
\begin{align}
h_{\mn}^{(0)} \to& \;\bar{h}_{\mn}^{(0)} + c_{(1)} \pi_{(1)} \eta_{\mn} +  c_{(2)} \pi_{(2)} \eta_{\mn} +  c_{(3)} \pi_{(3)} \eta_{\mn} \\
h_{\mn}^{(i > 0)} \to& \;\bar{h}_{\mn}^{(i)} - c_{(i)} \pi_{(i)} \eta_{\mn},
\end{align}
with, as above, $c_{(i)} = -\frac{1}{2} m^2 \alpha_{(i)}^2$.  This results in a mixed kinetic scalar action
\begin{align}
&-\frac{3}{4}m^4 \left[ \left(\sum_{i=1}^{3} \alpha^2_{(i)} \pi_{(i)} \right) \Box \left(\sum_{j=1}^{3} \alpha^2_{(j)} \pi_{(j)} \right) + \sum_{i=1}^{3} \alpha^4_{(i)} \pi_{(i)} \Box \pi_{(i)} \right],
\end{align}
where again the first term corresponds to contributions from the `central' node ${\bf g}_{(0)}$ and mimics the structure of `central' nodes in the line theory example discussed above. Demixing the kinetic scalar action (again we focus on the first step in the demixing procedure and ignore any scalar mass terms for the time being) we obtain
\begin{align}
& -\frac{3}{4}m^4 \bigg[ \frac{4}{3} \Sigma_3 \Box \Sigma_3 + \frac{2}{3} \left( \alpha^2_{(3)} \pi_{(3)} - \frac{1}{2} \Sigma_2 \right) \Box \left( \alpha^2_{(3)} \pi_{(3)} - \frac{1}{2} \Sigma_2 \right) \nn \\
&\qquad\qquad+\:\frac{1}{2} \left( \alpha^2_{(2)} \pi_{(2)} - \alpha^2_{(1)} \pi_{(1)} \right) \Box \left( \alpha^2_{(2)} \pi_{(2)} - \alpha^2_{(1)} \pi_{(1)} \right) \bigg] \nn \\
\equiv& -\frac{3}{4}m^4 \bigg[ \frac{4}{3} \phi_{(1)} \Box \phi_{(1)} + \frac{2}{3} \phi_{(2)}  \Box \phi_{(2)}   + \frac{1}{2} \phi_{(3)}  \Box \phi_{(3)}  \bigg],
\end{align}
where $\Sigma_n = \sum_{i=1}^n \alpha^2_{(i)} \pi_{(i)}$. The modes $\phi_{(i)}$ that diagonalise the kinetic scalar action are (note we have not normalised these yet)
\begin{eqnarray}
\nn \phi_{(1)} &=& \Sigma_3, \\
\nn \phi_{(2)} &=& \alpha_{(3)}^2 \pi_{(3)} - \frac{1}{2} \Sigma_2, \\
\phi_{(3)} &=&  \alpha^2_{(2)} \pi_{(2)} - \alpha^2_{(1)} \pi_{(1)}.  
\end{eqnarray}
Once again the demixed, propagating scalar modes are linear combinations of the original \St scalars $\pi_{(ij)}$. Note, however, that at this stage the $\phi_{(i)}$ derived here are interacting propagating modes, since without having diagonalised the mass terms there will still be mass mixing terms $\phi_{(i)} \phi_{(j)}$ at this stage.
\\

{\bf Scalar mixing for higher order interaction vertices: }
Finally we wish to consider interaction vertices which couple together more than two spin-2 fields\footnote{Ghost-free higher order vertices have been constructed in the vielbein language in \cite{Hinterbichler:2012cn}. These vielbein terms cannot always be mapped to the metric picture used throughout this paper \cite{Hinterbichler:2012cn,Deffayet:2012nr,Deffayet:2012zc}. We provide the mapping between our approach in the metric and vielbein picture in \cite{vielbein}, but here we focus on the generic behaviour of any higher order interaction vertex in the metric picture.}, cf. graph (d) in figure \ref{fig-scalarmixing}. As discussed above, an $N$-metric interaction will break $N-1$ copies of diffeomorphism invariance and hence we introduce $N-1$ \St fields. We remind ourselves that this means such an interaction term is mapped to  
\be
{\cal S}_{\text{int}} = f \left( {{\bf g}_{(1)}} , {{\bf g}_{(2)}} , \dots , {{\bf g}_{(N)}} \right) \to f \left( {{\bf g}_{(1)}} \circ Y_1 , \dots , {{\bf g}_{(N-1)}} \circ Y_{N-1} , {{\bf g}_{(N)}} \right).
\ee
(Again, for brevity the second index on the $Y$'s has been omitted.) Note that, as discussed before, cross-terms and scalar self-interactions in the quadratic pure-scalar action pre-conformal transformation have to be eliminated to satisfy the generalised Fierz-Pauli condition \eqref{GenFPC}. Turning to the mixing between $h_{(i)}$ and $\pi_{(j)}$, at quadratic order this can be written
\be
\sum_{i=1}^N \sum_{j=1}^{N-1} h^{(i)}_{\mn} \left( k_{(i j)} X^{\mn}_{(1)}(\pi_{(j)}) + \tilde{k}_{(i j)} \eta^{\mn} \pi^{\rho}_{(j) \rho} \right), \label{N-metric-quadratic}
\ee
where $k_{(ij)}$ and $k_{(ij)}$ are constant coefficients. The conformal transformation employed to eliminate the scalar-tensor mixing is consequently
\be
h_{\mn}^{(i)} \to \;\bar{h}_{\mn}^{(i)} - \frac{1}{2} \sum_{j=1}^{N-1} k_{(i j)} \pi_{(j)} \eta_{\mn},
\ee
where any surviving $\bar{h}_\alpha^\alpha$ terms are removed via gauge-fixing. With this conformal transformation the demixed kinetic scalar part of  \eqref{N-metric-quadratic} (demixed in terms of having eliminated scalar-tensor mixing) becomes
\be
- \frac{1}{4} \sum_{i=1}^N \sum_{j,k = 1}^{N-1} k_{(i j)} \left( 3 k_{(i k)} + 8 \tilde{k}_{(i k)} \right) \pi_{(j)} \Box \pi_{(k)}, \label{N-metric-quadratic-2}
\ee
where we are again ignoring any mass terms for the time being. The different $\pi_{(i)}$ are mixed here, which is a general feature of $N$ spin-2 field interactions.\footnote{In other words, such interactions kinetically mix all the participating $\pi_{(i)}$, in contrast to the line theory considered above, which only kinetically mixed `neighbouring' $\pi_{(i)}$.} This can be demixed to (in principle) $N-1$ different propagating modes, corresponding to the $N-1$ broken copies of $GC_{(i)}$ and which are linear combinations of the different $\pi_{(i)}$ as expected. For example, for a dRGT-like scalar-tensor interaction where $\tilde k_{(i j)} = 0$, demixing the scalar \dof corresponds to diagonalising the (in general up to rank $N-1$) matrix $\sum_i k_{(ij)} k_{(ik)}$. 

If one imposes additional symmetry requirements on the nature of the interactions, one can, however, also reduce the effective number of propagating modes (equivalently: the extra symmetry can lead to some of the $N-1$ modes becoming identical). We now illustrate this by showing that \eqref{N-metric-quadratic-2} simplifies remarkably upon imposing some prima facie rather modest assumptions, highlighting two particular cases:

\begin{enumerate}

\item  Consider the case when the interactions of a given scalar $\pi_{(j)}$ to two metrics are totally proportional, that is the `interaction strengths' factorise as $k_{(i j)} = k_{(i)} \gamma_{(j)},\ \tilde{k}_{(i j)} = k_{(i)} \tilde{\gamma}_{(j)}$. In this case \eqref{N-metric-quadratic} can be written
\be
\sum_{i=1}^N k_{(i)} h^{(i)}_{\mn} \sum_{j=1}^{N-1}  \left( \frac{\gamma_{(j)}}{2} \left( \eta^{\mn} \Box - \partial^{\mn} \right) + \tilde{\gamma}_{(j)} \eta^{\mn} \Box \right) \pi_{(j)},
\ee
i.e. everything depending on the $i$ site index factors out. The demixed scalar action \eqref{N-metric-quadratic-2} therefore becomes 
\begin{align}
 \nn & -\frac{1}{4} \left( \sum_{i=1}^N k_{(i)}^2 \right) \sum_{j,k = 1}^{N-1} \gamma_{(j)} \left( 3 \gamma_{(k)} + 8 \tilde{\gamma}_{(k)} \right) \pi_{(j)} \Box \pi_{(k)} \\
 =& -\frac{1}{4} \left( \sum_{i=1}^N k_{(i)}^2 \right) \left[ 3 \Sigma_{\gamma} \Box \Sigma_{\gamma} + 8 \Sigma_\gamma \Box \tilde{\Sigma}_{\tilde{\gamma}} \right] \nn \\
 \nn =& -\frac{1}{4} \left( \sum_{i=1}^N k_{(i)}^2 \right) \left[ 3 \Sigma^{'}_{\gamma} \Box \Sigma^{'}_{\gamma} - \frac{16}{3} \tilde{\Sigma}_{\tilde{\gamma}} \Box \tilde{\Sigma}_{\tilde{\gamma}} \right] \\
 \equiv& -\frac{1}{4} \left( \sum_{i=1}^N k_{(i)}^2 \right) \left[ 3 \phi_{(1)} \Box \phi_{(1)} - \frac{16}{3} \phi_{(2)} \Box \phi_{(2)} \right],
\end{align}
where $\Sigma_\gamma = \sum_{i=1}^{N-1} \gamma_{(i)} \pi_{(i)}$ and $\tilde\Sigma_{\tilde\gamma} = \sum_{i=1}^{N-1} \tilde\gamma_{(i)} \pi_{(i)}$. In the third line we have de-mixed the scalar action (at quadratic order in the fields) and introduced $\Sigma^{'}_\gamma = \Sigma_\gamma + \frac{4}{3} \tilde{\Sigma}_{\tilde{\gamma}}$ in the process, so that the de-mixed modes are $\phi_{(1)} = \Sigma^{'}_\gamma$ and $\phi_{(2)} = \tilde\Sigma_{\tilde\gamma}$. We emphasize that, as a result of our `factorisation assumption', there are only two distinct propagating modes left over now. 

There is an interesting observation that can be made for this case. Namely that the $\tilde{\Sigma}_{\tilde{\gamma}}$ field is ghost-like (it has the wrong sign kinetic term). There are at least two ways in which the introduction of such a ghost can be avoided. Firstly one of the hallmarks of dRGT-type \emph{bimetric} interactions is that $\tilde{\gamma} = 0$. If this is also true for a given higher order interaction here, this eliminates this particular ghost, leaving us with a single, healthy propagating mode. Secondly, if $\tilde{\gamma}_i \propto \gamma_i$, and hence $\tilde{\Sigma}_{\tilde{\gamma}}$ is not independent of $\Sigma_\gamma$, then only a single $\Sigma_{\gamma}$ field remains again and whether or not this field is ghost-like depends on the exact proportionality constant between $\tilde{\gamma}_{(i)}$ and $\gamma_{(i)}$.

\item Now consider the case where $\tilde{k}_{(i j)}$ instead factorises as $\tilde{k}_{(i j)} = \tilde{k}_{(i)} \gamma_{(j)}$. This corresponds to the interactions of a given \emph{tensor} $h_{(i)}$ to two different \emph{scalars} being totally proportional. Now \eqref{N-metric-quadratic} becomes
\be
\sum_{i=1}^N h^{(i)}_{\mn} \left( \frac{k_{(i)}}{2}\left(\eta^{\mn} \Box - \partial^{\mn} \right) + \eta^{\mn} \tilde{k}_{(i)} \Box \right) \sum_{j=1}^{N-1}  \gamma_j \pi_{(j)},
\ee
which is similar to the expression in case 1 above, except that the `flavour' index on the differential operator is now shared with the tensor and not the scalar. This has important consequences, as \eqref{N-metric-quadratic-2} now turns into
\be
-\frac{1}{4} \left( \sum_{i=1}^N k_{(i)} \left( 3k_{(i)} + 8 \tilde{k}_{(i)} \right)\right) \left( \sum_{j=1}^{N-1} \gamma_{(j)} \pi_{(j)} \right) \Box \left( \sum_{k=1}^{N-1} \gamma_k \pi_{(k)} \right),
\ee
and we immediately see that there is only a single propagating mode $\phi_{(1)} = \Sigma_\gamma = \sum_{i=1}^{N-1} \gamma_{(i)} \pi_{(i)}$, highly reminiscent of the way in which the `central' node in our bimetric interaction branching vertex above only contributed to the $\sum \pi_{(i)}$ mode. It is worth emphasizing that, for example by adding different bimetric interactions connecting the outer nodes of this particular $N$ spin-2 field interaction to other nodes, the symmetry of the interaction term can be broken and the full $N-1$ degrees of freedom contributed by the $N$ spin-2 field interaction vertex can be brought out again.

\end{enumerate}

{\bf Canonical normalisation: }
By now we have eliminated both scalar-tensor as well as kinetic scalar-scalar mixing at quadratic order.
Step 2 in the procedure summarised in equation \eqref{diagonalisation transformation} above is to canonically normalise the eigenmodes of the kinetic matrix $K$: the $\phi$ fields. In fact we can canonically normalise all the propagating \dof of the theory at this point --- they are all $h^{\mn}_{(i)}, A^{\mu}_{(ik)},\phi_{(l)}$, where the $\phi_{(l)}$ may still interact via their mass terms. Note that, since the scalar modes $\phi$ generically combine contributions from different vertices, $l$ here is no longer a site label but just an ordering index for the $\phi$ fields as determined via the eigenvectors of $K$. Canonical normalisation is straightforward now, especially since we have related all $\phi_{(l)}$ to the same mass scale $m^2$ by absorbing appropriate powers of $\alpha^2_{(ik)}$ into the definition of each $\phi_{(l)}$. Note that we have labelled the different masses $m$ and corresponding $\alpha$'s with only one index (whenever it is unambiguous to do so) so far, but just to be explicit we here label them with the full two indices. Canonical normalisation\footnote{We recall that having a universal normalising scale $M_{Pl}$ for all ${\bf h}_{(i)}$ is an assumption we have imposed. If different $M_{Pl}^{(i)}$ are chosen, it is straightforward to generalise the normalisation procedure outlined here.} 
then instructs us to normalise as follows
\begin{align}
h_{\mu\nu}^{(i)} &\to \frac{1}{M_{Pl}} \hat h_{\mu\nu}^{(i)}     &A^\mu_{(ik)} &\to \frac{1}{m_{(ik)} M_{Pl}} \hat A^\mu_{(ik)} &\phi_{(l)} &\to \frac{1}{m^2 M_{Pl} \sqrt{\lambda_{(l)}}} \hat \phi_{(l)} , \label{cannorm}
\end{align}
ignoring numerical pre-factors, where a hat denotes a canonically normalised field and where $\lambda_{(l)}$ is the eigenvalue of the eigenmode $\phi_{(l)}$ of the kinetic matrix $K$. Vector perturbations $A^{\mu}_{(i k)}$ are consequently normalised by the scale $m_{(i k)}$ associated with the interaction vertex for which $A^{\mu}_{(i k)}$ is a \St vector. Since the $\phi_{(l)}$ are linear combinations of the \St scalars $\pi_{(i k)}$, they all get normalised by the same mass scale $m^2$, which we recall is related to the strength of interaction terms by $m^2 \alpha_{(i k)}^2 = m_{(i k)}^2 $. Again we emphasize that this means the dependence on all other mass scales has been absorbed into our definition of the $\phi_{(l)}$ via their dependence on the $\alpha_{(i k)}^2$. Re-phrasing equation \eqref{cannorm} in the language of \eqref{diagonalisation transformation}, for the scalar modes the canonical normalisation amounts to
\be
N_{jk} = \frac{1}{m^2 M_{Pl} \sqrt{\lambda_{(j)}}} \delta_{jk}
\ee
again up to numerical pre-factors and where the index $j$ is not summed over here.

\section{Higher order interactions and the decoupling limit } \label{sec-cubic}

\begin{figure}[tp]
\centering
\begin{tikzpicture}[-,>=stealth',shorten >=0pt,auto,node distance=2cm,
  thick,main node/.style={circle,fill=blue!10,draw,font=\sffamily\large\bfseries},arrow line/.style={thick,-},barrow line/.style={thick,->},no node/.style={plain},rect node/.style={rectangle,fill=blue!10,draw,font=\sffamily\large\bfseries},red node/.style={rectangle,fill=red!10,draw,font=\sffamily\large\bfseries},green node/.style={circle,fill=green!20,draw,font=\sffamily\large\bfseries},yellow node/.style={rectangle,fill=yellow!20,draw,font=\sffamily\large\bfseries}]

   \node[draw=none,fill=none](1)[]{};     
   \node[draw=none,fill=none](2)[below=5cm of 1]{}; 

 \node[draw=none,fill=none](22)[below=2.1cm of 1]{}; 
  \node[draw=none,fill=none](222)[below=2.6cm of 1]{};    
    \node[draw=none,fill=none](2222)[below=0.8cm of 222]{};  
        \node[draw=none,fill=none](22222)[below=0.2cm of 2222]{};  
            \node[draw=none,fill=none](222222)[below=0.6cm of 22222]{};  
   
   \draw[->,thick,black] (22) to (1);
   \draw[-,thick,black,dotted] (22) to (222);
     \draw[-,thick,black] (222) to (2222);
      \draw[-,thick,black,dotted] (2222) to (22222);
     \draw[-,thick,black] (22222) to (222222);
   
   \node[draw=none,fill=none](5)[below=2cm of 1]{}; 
   \node[draw=none,fill=none](501)[left=0.5cm of 5]{${\Lambda_3^{(1)}}$}; 
   \node[draw=none,fill=none](502)[right=0.5cm of 5]{}; 
   \node[draw=none,fill=none](6)[below=0.1cm of 5]{}; 
   \node[draw=none,fill=none](601)[left=0.5cm of 6]{}; 
   \node[draw=none,fill=none](602)[right=0.5cm of 6]{}; 
   \node[draw=none,fill=none](7)[below=0.1cm of 6]{}; 
   \node[draw=none,fill=none](701)[left=0.5cm of 7]{${\Lambda_5^{(1)}}$}; 
   \node[draw=none,fill=none](702)[right=0.5cm of 7]{}; 
   
    \node[draw=none,fill=none](92)[below=0.1cm of 2]{(a) Two bimetric links};
   
     \draw[-,ultra thick,red] (501) to (502);
       \draw[-,ultra thick,red] (701) to (702);


    \node[draw=none,fill=none](8)[below=3.5cm of 1]{}; 
   \node[draw=none,fill=none](801)[left=0.5cm of 8]{${\Lambda_3^{(2)}}$}; 
   \node[draw=none,fill=none](802)[right=0.5cm of 8]{}; 
   \node[draw=none,fill=none](9)[below=0.1cm of 8]{}; 
   \node[draw=none,fill=none](901)[left=0.5cm of 9]{}; 
   \node[draw=none,fill=none](902)[right=0.5cm of 9]{}; 
   \node[draw=none,fill=none](10)[below=0.1cm of 9]{}; 
   \node[draw=none,fill=none](1001)[left=0.5cm of 10]{$\Lambda_5^{(2)}$}; 
   \node[draw=none,fill=none](1002)[right=0.5cm of 10]{}; 
   
     \draw[-,ultra thick,blue,dotted] (801) to (802);
       \draw[-,ultra thick,blue,dotted] (1001) to (1002);


 
    \node[draw=none,fill=none](1a)[right=4cm of 1]{};     
   \node[draw=none,fill=none](2a)[below=5cm of 1a]{}; 
   \draw[->,thick,black] (2a) to (1a);
   
   \node[draw=none,fill=none](5a)[below=2cm of 1a]{}; 
   \node[draw=none,fill=none](501a)[left=0.5cm of 5a]{${\Lambda_3^{(1)}}$}; 
   \node[draw=none,fill=none](502a)[right=0.5cm of 5a]{}; 
   
    \node[draw=none,fill=none](92a)[below=0.1cm of 2a]{(b) Their `decoupling limit'};
   
     \draw[-,ultra thick,red] (501a) to (502a);


    \node[draw=none,fill=none](8a)[below=3.5cm of 1a]{}; 
   \node[draw=none,fill=none](801a)[left=0.5cm of 8a]{${\Lambda_3^{(2)}}$}; 
   \node[draw=none,fill=none](802a)[right=0.5cm of 8a]{}; 
   
     \draw[-,ultra thick,blue,dotted] (801a) to (802a);


 
    \comment{\node[draw=none,fill=none](1b)[right=4cm of 1a]{};     
   \node[draw=none,fill=none](2b)[below=5cm of 1b]{}; 
   \draw[->,thick,black] (2b) to (1b);}

        \node[draw=none,fill=none](1b)[right=4cm of 1a]{};     
   \node[draw=none,fill=none](2b)[below=5cm of 1b]{}; 
 \node[draw=none,fill=none](22b)[below=2.5cm of 1b]{}; 
  \node[draw=none,fill=none](222b)[below=3.2cm of 1b]{};

   \draw[->,thick,black] (22b) to (1b);
   \draw[-,thick,black,dotted] (22b) to (222b);
     \draw[-,thick,black] (222b) to (2b);
   
   \node[draw=none,fill=none](5b)[below=2cm of 1b]{}; 
   \node[draw=none,fill=none](501b)[left=0.5cm of 5b]{$\Lambda_3^{(1)}$}; 
   \node[draw=none,fill=none](502b)[right=0.5cm of 5b]{}; 
   \node[draw=none,fill=none](6b)[above=0.4cm of 5b]{}; 
   \node[draw=none,fill=none](601b)[left=0.5cm of 6b]{$\Lambda_3^{(-)}$}; 
   \node[draw=none,fill=none](602b)[right=0.5cm of 6b]{}; 

    \node[draw=none,fill=none,align=center](92b)[below=0.1cm of 2b]{(c) `Decoupling limit' \\ of a trimetric line};
   
     \draw[-,ultra thick,red] (501b) to (502b);
      \draw[-,ultra thick,blue,dotted] (601b) to (602b);


    \node[draw=none,fill=none](8b)[below=3.5cm of 1b]{}; 
   \node[draw=none,fill=none](801b)[left=0.5cm of 8b]{${\Lambda_3^{(2)}}$}; 
   \node[draw=none,fill=none](802b)[right=0.5cm of 8b]{}; 
   \node[draw=none,fill=none](10b)[below=0.3cm of 8b]{}; 
   \node[draw=none,fill=none](1001b)[left=0.5cm of 10b]{$\Lambda_3^{(+)}$}; 
   \node[draw=none,fill=none](1002b)[right=0.5cm of 10b]{}; 
   
    \node[draw=none,fill=none](11b)[below=0.3cm of 10b]{}; 
   \node[draw=none,fill=none](1101b)[left=0.2cm of 11b]{}; 
   \node[draw=none,fill=none](1102b)[right=0.2cm of 11b]{}; 
   
     \draw[-,ultra thick,red] (801b) to (802b);
       \draw[-,ultra thick,blue,dotted] (1001b) to (1002b);
        \draw[-,ultra thick,black] (1101b) to (1102b);

\draw[->,thick,black,dashed] (602b) to (802b);  
    \node[draw=none,fill=none](1bb)[right=0.5cm of 1b]{}; 
    \draw[->,thick,black,dashed] (602b) to (1bb);  
     \node[draw=none,fill=none](1102bb)[right=0.05cm of 1102b]{}; 
      \draw[->,thick,black,dashed] (1002b) to (802b);  
       \draw[->,thick,black,dashed] (1002b) to (1102bb);  
        \node[draw=none,fill=none](1101bb)[left=0.0cm of 1101b]{$ \Lambda_{3,min}^{(+)}$};

 
       \node[draw=none,fill=none](1c)[right=4cm of 1b]{};     
   \node[draw=none,fill=none](2c)[below=5cm of 1c]{}; 
 \node[draw=none,fill=none](22c)[below=2.5cm of 1c]{}; 
  \node[draw=none,fill=none](222c)[below=3.2cm of 1c]{};

   \draw[->,thick,black] (22c) to (1c);
   \draw[-,thick,black,dotted] (22c) to (222c);
     \draw[-,thick,black] (222c) to (2c);
   
   \node[draw=none,fill=none](5c)[below=2cm of 1c]{}; 
   \node[draw=none,fill=none](501c)[left=0.5cm of 5c]{$\Lambda_3^{(1)}$}; 
   \node[draw=none,fill=none](502c)[right=0.5cm of 5c]{}; 
   \node[draw=none,fill=none](6c)[above=0.4cm of 5c]{}; 
   \node[draw=none,fill=none](601c)[left=0.5cm of 6c]{}; 
   \node[draw=none,fill=none](602c)[right=0.5cm of 6c]{}; 
     \node[draw=none,fill=none](7c)[above=0.3cm of 6c]{}; 
   \node[draw=none,fill=none](701c)[left=0.5cm of 7c]{}; 
   \node[draw=none,fill=none](702c)[right=0.5cm of 7c]{}; 

    \node[draw=none,fill=none,align=center](92c)[below=0.1cm of 2c]{(d) `Decoupling limit' of \\ N interacting fields};
   
     \draw[-,ultra thick,red] (501c) to (502c);
      \draw[-,ultra thick,blue,dotted] (601c) to (602c);
       \draw[-,ultra thick,blue,dotted] (701c) to (702c);


    \node[draw=none,fill=none](8c)[below=3.5cm of 1c]{}; 
   \node[draw=none,fill=none](801c)[left=0.5cm of 8c]{${\Lambda_3^{(N)}}$}; 
   \node[draw=none,fill=none](802c)[right=0.5cm of 8c]{}; 
   \node[draw=none,fill=none](10c)[below=0.0cm of 8c]{}; 
   \node[draw=none,fill=none](1001c)[left=0.5cm of 10c]{}; 
   \node[draw=none,fill=none](1002c)[right=0.5cm of 10c]{}; 
   
    \node[draw=none,fill=none](11c)[below=0.6cm of 10c]{}; 
   \node[draw=none,fill=none](1101c)[left=0.2cm of 11c]{}; 
   \node[draw=none,fill=none](1102c)[right=0.2cm of 11c]{}; 
   
       \node[draw=none,fill=none](112c)[below=0.1cm of 10c]{}; 
   \node[draw=none,fill=none](11012c)[left=0.5cm of 112c]{}; 
   \node[draw=none,fill=none](11022c)[right=0.5cm of 112c]{}; 
   
     \draw[-,ultra thick,red] (801c) to (802c);
       \draw[-,ultra thick,blue,dotted] (1001c) to (1002c);
       \draw[-,ultra thick,blue,dotted] (11012c) to (11022c);
        \draw[-,ultra thick,black] (1101c) to (1102c);

    \node[draw=none,fill=none](1cc)[right=0.5cm of 1c]{}; 
     \node[draw=none,fill=none](1102cc)[right=0.05cm of 1102c]{}; 
        \node[draw=none,fill=none](1101cc)[left=0.0cm of 1101c]{$N^{-1/2} \Lambda_3^{(N)}$}; 
  
\end{tikzpicture}
\caption{Higher order interaction terms and the associated decoupling limit for theories discussed in section \ref{sec-cubic}. {\bf (a)} We show higher order interactions coming from two isolated bimetric links. Solid lines correspond to interactions coming from a bimetric interaction term with coupling constant $m_{(1)}$, whereas dotted lines come from an analogous term with $m_{(2)}$. $m_{(2)} < m_{(1)}$ here. We only explicitly show scales $\Lambda_3$ and $\Lambda_5$ here. Note that, depending on the precise theory chosen, $\Lambda_3^{(2)}$ could also lie above $\Lambda_5^{(1)}$. 
{\bf (b)} Zooming in on a `generalised' decoupling limit where the lowest order contributions from both bimetric interactions in (a) are kept. We here choose a dRGT-like theory that tunes all contributions from scales smaller than $\Lambda_3$ to zero. Hence $\Lambda_3^{(2)}$ and $\Lambda_3^{(1)}$ are the smallest scales for the two interaction terms respectively. At the cubic level both $\{1,0,2\}$ and $\{0,2,1\}$ terms contribute to these scales.
{\bf (c)} The `generalised' decoupling limit for a trimetric line connecting three sites with two bimetric links.  The thick lines now correspond to the would-be scales for the two links taken individually, i.e. the bigravity theory shown in (b). The dotted lines are coming from  $\{1,0,2\}$ contributions, where  $\Lambda_3^{(+)}$ and $\Lambda_3^{(-)}$ label the smallest and largest resulting scales respectively. In principle there is a whole tower of scales sitting in between these extremes. Depending on the relative value of $m_{(2)}$ and $m_{(1)}$, $\Lambda_3^{(+)} \sim \Lambda_3^{(-)} \sim \Lambda_3^{(2)}$, if $m_{(1)} \to \infty$ or $\Lambda_3^{(+)} = \Lambda_{3,min}^{(+)}$ and $\Lambda_3^{(-)} \to \infty$ if $m_{(1)} = m_{(2)}$. $\Lambda_{3,min}^{(+)}$ is the lowest possible scale, obtained when all links are identical.  Generically $\Lambda_{3,min}^{(+)} \le \Lambda_3^{(+)} \le \Lambda_3^{(2)}$ and  $\Lambda_3^{(2)} \le \Lambda_3^{(-)} \le \infty$.
Note that, since all the terms shown here can in principle contribute at roughly the same scale, we here choose to show a `generalised` decoupling limit with all these terms still present. $\{0,2,1\}$ terms lead to contributions at the $\Lambda_3^{(2)}$ scale if  $\Lambda_3^{(2)} = \Lambda_3^{(1)}$, but otherwise add four different additional scales in between the highest and lowest $\{1,0,2\}$ scales.
{\bf (d)} The analogous `generalised' decoupling limit for $N$ interacting spin-2 fields. Again $\{0,2,1\}$ terms still lead to contributions at the $\Lambda_3^{(i)}$ scales, if all $\Lambda_3^{(i)}$ are identical.
$\{1,0,2\}$ terms will give contributions to `permutations modes' linearly combining the original \St scalars $\pi^{(i)}$ - the dotted lines. The lowest and largest such scale will have analogous behaviour to that discussed in the trimetic case (c). In particular there will be a lowest scale `sum mode', which, if all coupling constants $m_{(i)}$ are identical, will take its minimal value $\Lambda_{3,min}^{(+)} = N^{-1/2} \Lambda_3^{(N)}$ for large $N$. In fact, as discussed below, it is the second and third term in \eqref{tri-lambda3-terms} which yield interactions contributing to this scale. Also note that this scale is identical to the one discussed in the specific context of a circle theory in \cite{deRham:2013awa}.
} \label{fig-dec}
\end{figure}
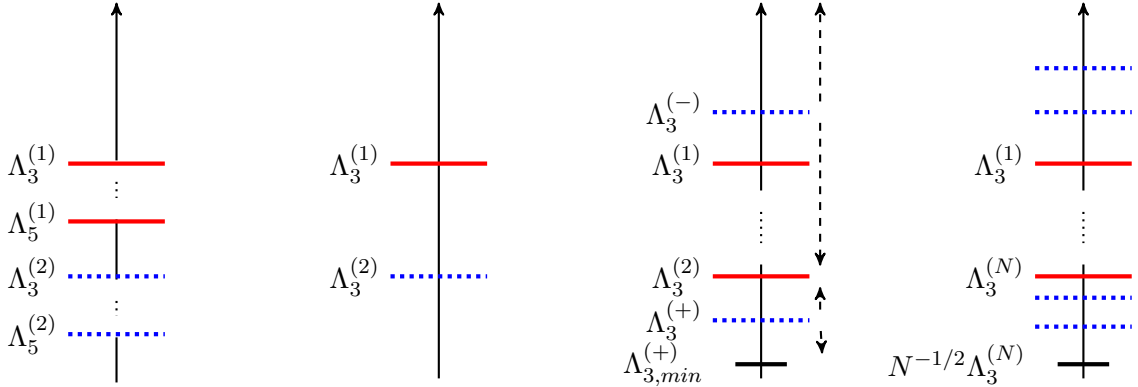

Having canonically normalised all fields and de-mixed at quadratic order, we can now consider higher order interaction terms and indulge ourselves in power counting to identify the scales suppressing these terms. In particular we will be interested in the decoupling limit, i.e. the limit where one zooms in on the interaction(s) suppressed by the smallest scale and hence most relevant at low energies. As we shall see, having multiple interacting spin-2 fields can qualitatively change this limit when compared to the analogous massive/bi-gravity cases \cite{deRham:2010ik,deRham:2010kj,Gabadadze:2013ria,Ondo:2013wka,Fasiello:2013woa}. Throughout this section the explicit examples we provide for multi-spin-2 field models will be of the `line theory' type discussed above. Analogous examples are readily constructed for branching vertices or $n$-field interaction terms, but the simpler line theory examples suffice to illustrate generic features of interacting spin-2 field theories. Note that we also do not consider cases with potentially ghost-inducing loops here - we discuss these in \cite{loops}.  
\\

{\bf Bimetric theories: }
We begin by quickly reviewing higher order interactions and the decoupling limit in massive/bi-gravity, which will serve as a reference point later on. In theories with two spin-2 fields as described by \eqref{toybimaction}, the nature of the \St trick (cf. \eqref{goldstone2}) leads to interaction terms
\be \label{Gen-int-bim}
m^2 M_{Pl}^2 h^{n_h} (\pa A)^{n_A} (\pa^2 \pi)^{n_\pi} \to m^{2-n_A-2n_\pi} M_{Pl}^{2 - n_h - n_A - n_\pi} \hat h^{n_h} (\pa \hat A)^{n_A} (\pa^2 \hat \pi)^{n_\pi}
\ee  
where $m^2 M_{Pl}^2$ is the coupling constant of the interaction term, the arrow signifies the effect of canonical normalisation, which is trivial in theories with a single bimetric interaction. $n_h,n_A,n_\pi$ denote the powers of $h_{\mn},A^\mu$ and $\pi$ respectively. Note that in dRGT massive gravity there is only one $h, A$ and $\pi$, since the second spin-2 field is non-dynamical. In a full bigravity model, however, there are two $h$, which will generically both have interaction terms of the type shown here. Since the \St expansion of the metric also introduces terms of the form $A^\mu \pa_\mu g_{\alpha \beta}$ (again cf. \eqref{goldstone2}), it is not always true that $h$ enters with no derivative, $A$ enters with one and $\pi$ with two (at least prior to a re-summing procedure - see appendix \ref{appendix-nonlocal}), so not enforcing any derivative structure we denote the canonically normalised interaction terms by
\be
M_{Pl}^{2- n_h - n_A - n_\pi} m^{2 - n_A - 2 n_\pi } \{n_h, n_A, n_\pi\} = \Lambda_{\lambda}^{4-n_h-2n_A-3n_\phi} \{n_h,n_A,n_\pi\}.
\ee
Here we have introduced the ordering parameters $\Lambda$ and $\lambda$, setting the scale of the interaction and satisfying
\be \Lambda_\lambda = (M_P m^{\lambda - 1})^{1/\lambda}, \qquad \lambda = \frac{3n_\phi + 2n_A + n_h - 4}{n_\phi + n_A + n_h - 2},
\ee
A larger $\lambda$ then corresponds to a smaller scale of suppression. For the different possible cubic interaction terms we have
\begin{align}
\nn\{0,0,3\} &\to \lambda = 5   &\{0,3,0\} &\to \lambda = 2 \\
\nn\{0,1,2\} &\to \lambda = 4   &\{2,0,1\} &\to \lambda = 1 \\
\nn\{1,0,2\} &\to \lambda = 3   &\{1,2,0\} &\to \lambda = 1 \\
\nn\{0,2,1\} &\to \lambda = 3   &\{2,1,0\} &\to \lambda = 0 \\
\{1,1,1\} &\to \lambda = 2   &\{3,0,0\} &\to \lambda = -1  \label{lambdahierarchy}
\end{align}
Note that higher order interaction terms can have comparably low suppression scales, e.g. the smallest quartic scale is set by $\{0,0,4\}$ and leads to $\lambda = 4$ and the highest order pure-scalar interactions $\{0,0,n_\pi \to \infty\}$ have $\lambda = 3$.

We can see that the pure-scalar cubic interaction term suppressed by $\Lambda_5$ generically is the least suppressed term in the theory. Phrasing it explicitly in terms of $m$, its scale is set by $(M_{Pl} m^{4})^{1/5}$. With such a term the scattering amplitude $\pi \pi \to \pi \pi$ obtained by patching together two cubic interactions scales like $(E/\Lambda_5)^{10}$ \cite{ArkaniHamed:2002sp}. The theory becomes strongly coupled at $\Lambda_5$, which is also the cutoff scale of the theory (taking into account quantum corrections may postpone unitarity violation to scales higher than this cutoff \cite{ArkaniHamed:2002sp}) . dRGT massive gravity tunes interaction terms such that all pure-scalar interaction terms become total derivatives and hence vanish. This is essentially the higher order generalisation of Fierz-Pauli tuning at quadratic order \eqref{GenFPC}. Eliminating the pure scalar interaction in this way also automatically turns terms with $\lambda = 4$ such as $\{0,1,2\}$ into total derivatives \cite{Hinterbichler:2011tt}, leaving us with a theory where the least suppressed interaction terms have a scale $\Lambda_3$  (and a corresponding $\Lambda_3$ cutoff) \cite{deRham:2010ik,deRham:2010kj,Hinterbichler:2011tt}. Again phrasing this explicitly in terms of $m$, at cubic order its scale is set by $(M_{Pl} m^{2})^{1/3}$.

The conformal transformation used to eliminate scalar-tensor mixing will turn $\{n_h ,n_A, n_\pi\}$ terms into $\{n_h-n, n_A, n_\pi + n\}$ ones. However, it is worth stressing that the factors in front of the scalar fields in the conformal transformations will be $\sim m^2$, and the canonical normalisation of the fields involves a factor of $m^{-2}$ for the scalar (as compared to tensor) modes. As a result the new interaction terms introduced by the conformal transformations retain their original scaling.  That is, the conformal transformations will give terms, after canonical normalisation,
\be
\Lambda_{\lambda}^{4-n_h-2n_A-3n_\pi} \sum_n \{n_h-n,n_A,n_\pi + n\}.
\ee
Note that the bigravity statement, that interaction terms introduced by the conformal transformations retain their original scaling, will not strictly remain true in the $N$ spin-2 field case discussed below, since the $m^2_{(i)}$ and $m^{-2}_{(j)}$ that cancel out in the bigravity case ($i=j$ here) no longer cancel there (since $i \neq j$ for all interactions in theories with $N > 2$). More precisely, the scaling will stay of the same functional form, but will yield contributions where the `original' $m^2_{(i)}$ is replaced by a different $m^2_{(j)}$ - more on this in the following sections.
\\

{\bf The decoupling limit: }
Having understood the scale of all interaction terms one can proceed to the decoupling limit. Heuristically this amounts to zooming in on the interaction suppressed by the lowest scale, while setting all other interactions to zero. This limit is particularly interesting because the physics at low energies is dominated by the decoupling limit interaction term(s). Typically this is consequently the relevant limit for computing the cosmological evolution as well as e.g. spherically symmetric solutions around massive bodies. More formally we can define the decoupling limit by identifying the lowest suppression scale $\Lambda_{dec}$\footnote{As discussed above this will generically be $\Lambda_5$, but can be tuned to a higher scale such as $\Lambda_3$.} and taking the limit
\begin{align}
M_{Pl} &\to \infty, &m_{} &\to 0, &\Lambda_{dec} &\quad \text{fixed}.
\end{align}
Note that the Galilean shift symmetry imposed on $\pi$ in the pure-scalar action prior to demixing (a consequence of the nature of the \St trick and the introduction of a $U(1)$ symmetry for $\pi$ via $A^\mu \to \tilde{A}^\mu + \pa^\mu \pi$) results in decoupling limit interactions for $\pi$ of the Galileon type \cite{Nicolis:2008in}\footnote{Note the earlier work of   \cite{Fairlie:1991qe,Horndeski:1974wa}. This already contains Galileon-type interactions, which were independently re-discovered by \cite{Nicolis:2008in}.}. This is shown explicitly in the massive and bigravity cases in \cite{deRham:2010ik,Hinterbichler:2011tt,deRham:2012kf,Fasiello:2013woa}.

What changes when we consider theories with more than two spin-2 fields? Generically there are now multiple mass scales (or equivalently: several interaction terms with different coupling constants)\footnote{In addition there are also the in principle different Planck masses $M_{Pl}$ for each of the spin-2 fields.}. As a result there are now multiple ways of sending all mass scales to zero and all Planck scales to infinity, while keeping a number of distinct `decoupling' scales $\Lambda_{dec}$ fixed. As a result there is no longer a unique decoupling limit\footnote{We thank Kurt Hinterbichler for pointing this out to us.} and instead we have a number of `decoupling limits' $DL^{(k)}$ labelled by the index $k$
\begin{align}
&DL^{(k)}:  &\forall_i \forall_j: M_{Pl}^{(i)} &\to \infty \text{ and }  m_{(j)} \to 0, &\Lambda_{dec}^{(k)} &\quad \text{fixed}.
\end{align}
In what follows we will pay special attention to the limit corresponding to the lowest such scale, i.e. the strong-coupling scale of the theory. We will call this scale $\Lambda_\lambda^{(+)}$. However, in order to understand this limit properly, we will keep a full hierarchy of interaction terms, which could all in principle contribute to this lowest scale. As we will show, which terms contribute and what their precise hierarchy is depends on the nature of the interaction terms and the coupling constants of different interactions. As a final comment, it is worth pointing out that the strong-coupling scale inferred by zooming in on the lowest scale of the theory is not necessarily the cut-off of the theory, but instead one may hope to have discovered a phenomenologically interesting (and strongly coupled) Vainshtein regime - cf. related discussions in \cite{deRham:2013awa}.
\\

{\bf Trimetric $\Lambda_5$ theories: }
We now move on to consider interactions for $N$ spin-2 fields at cubic order and higher. The analogue of \eqref{Gen-int-bim} in this case is
\be \label{lambda5gen}
m_{(c)}^2 M_{Pl}^2 h^{n_h} (\pa A)^{n_A} (\pa^2 \pi)^{n_\pi} \to m_{(c)}^{2-n_A} M_{Pl}^{2 - n_h - n_A} \hat h^{n_h} (\pa \hat A)^{n_A} (\pa^2 \pi)^{n_\pi},
\ee  
In other words, this corresponds to the interactions generated by a term with coupling constant $m_{(c)}^2 M_{Pl}^2$; $h,A,\pi$ are shorthand for $h_{(i)},A_{(j k)},\pi_{(m n)}$, where one of $\{j,k\}$ and one of $\{m,n\}$ has to be identified with $i$. All possible cross-terms allowed by this constraint will in principle be generated, depending on the structure of $\cal{S}_{\text{int}}$. Note that there is no $\hat\pi$ here, since canonical normalisation applies to the propagating modes $\chi_{(l)}$ only. Consequently we have to first express interaction terms in terms of the $\chi_{(l)}$ before identifying the scale of all interactions. The fact that the $\chi_{(l)}$ are linear combinations of $\pi$'s makes this straightforward.
It is also worth pointing out that for this reason we roughly expect the hierarchy \eqref{lambdahierarchy} to stay intact. $\pi_{(i)}$ interactions will combine to yield $\chi_{(l)}$ ones, so that $\{n_h,n_A,n_\pi\}$ combine to  $\{n_h,n_A,n_\chi\}$ ones. As we shall see the scale suppressing these terms changes, however. 

As an initial concrete example consider a trimetric line theory as in \eqref{toylineaction}. Generically the cubic interaction terms suppressed by the smallest scale (corresponding to $\Lambda_5$ in the bigravity case) will come from
\be
m_{(1)}^2 M_{Pl}^2 (\pa^2 \pi_{(1)})^3 + m_{(2)}^2 M_{Pl}^2 (\pa^2 \pi_{(2)})^3,
\ee
where $\pi_{(1)},\pi_{(2)}$ are the \St scalars associated with interaction terms with coupling constants $m_{(1)}^2 ,m_{(2)}^2$ respectively. Both interaction terms are of the same functional form (only the coupling constants differ) - note that the relative sign between the two terms depends on how the \St fields $Y$ are introduced, see section \ref{sec-link} for details.
We now wish to re-express these interactions in terms of the kinetic eigen-modes $\phi_{(1)}$ and $\phi_{(2)}$, which we recall are related to $\pi_{(1)},\pi_{(2)}$ via $\pi_{(1)} = \frac{1}{\sqrt{2}} (\phi_{(1)} + \phi_{(2)})$ and $\pi_{(2)} = \frac{1}{\sqrt{2}} (\phi_{(1)} - \phi_{(2)}) \alpha^{-2}$ (see previous section). The result is
\begin{eqnarray}
\nn &&\frac{m_{(1)}^2 M_{Pl}^2}{2\sqrt{2}} \left[ (\pa^2 \phi_{(1)})^3 + (\pa^2 \phi_{(2)})^3 + 3 (\pa^2 \phi_{(1)}) (\pa^2 \phi_{(2)})^2 + 3 (\pa^2 \phi_{(1)})^2 (\pa^2 \phi_{(2)}) \right] \\
&+& \frac{m_{(1)}^6 M_{Pl}^2}{2\sqrt{2} m_{(2)}^4} \left[  (\pa^2 \phi_{(1)})^3 - (\pa^2 \phi_{(2)})^3 + 3 (\pa^2 \phi_{(1)}) (\pa^2 \phi_{(2)})^2 - 3 (\pa^2 \phi_{(1)})^2 (\pa^2 \phi_{(2)})   \right].
\end{eqnarray}
We can now canonically normalise all the fields involved, again as discussed in the previous section, and the interaction terms become
\begin{eqnarray}
\nn &&\frac{1}{M_{Pl} m_{(1)}^4 } \left[ \frac{1}{27} (\pa^2 \hat\phi_{(1)})^3 + \frac{1}{3\sqrt{3}}(\pa^2 \hat\phi_{(2)})^3 + \frac{1}{3} (\pa^2 \hat\phi_{(1)}) (\pa^2 \hat\phi_{(2)})^2 + \frac{1}{3\sqrt{3}} (\pa^2 \hat\phi_{(1)})^2 (\pa^2 \hat\phi_{(2)}) \right] \\
\nn &+& \frac{1}{M_{Pl} m_{(2)}^4 } \left[ \frac{1}{27} (\pa^2 \hat\phi_{(1)})^3 - \frac{1}{3\sqrt{3}}(\pa^2 \hat\phi_{(2)})^3 + \frac{1}{3} (\pa^2 \hat\phi_{(1)}) (\pa^2 \hat\phi_{(2)})^2 - \frac{1}{3\sqrt{3}} (\pa^2 \hat\phi_{(1)})^2 (\pa^2 \hat\phi_{(2)}) \right], \\ \label{lambda5intnorm}
\end{eqnarray}
 where a hat denotes canonical normalisation. We now still have to diagonalise the mass matrix for $\hat\phi$ in order to express cubic order interactions in terms of the modes $\hat\chi_{(1)},\hat\chi_{(2)}$, which independently propagate at quadratic order. Note that, if $m_{(2)} = m_{(1)}$, 
 $\hat\phi_{(i)} = \hat\chi_{(i)}$ and \eqref{lambda5intnorm} already shows the cubic action for the independently propagating modes. However, in general we have to diagonalise the mass matrix
\be \label{M-L5}
 M_{\hat\phi} = m_{(1)}^2 \begin{pmatrix}
   1+\alpha^2 & \sqrt{3} (1-\alpha^2) \\
\sqrt{3} (1-\alpha^2) &   3 (1+\alpha^2)
 \end{pmatrix},
\ee
where we are ignoring overall numerical factors, since we do not care about the numerical coefficients in front of mass terms for $\hat\chi_{(1)},\hat\chi_{(2)}$ (these will vanish in the decoupling limit anyway). We now find the normalised eigenvectors of this matrix and use these to diagonalise \eqref{M-L5} and in the process express cubic interactions in terms of $\hat\chi_{(1)},\hat\chi_{(2)}$. We re-emphasize that, since we are using an orthogonal matrix to diagonalise the mass term, the resulting kinetic matrix for   $\hat\chi_{(1)},\hat\chi_{(2)}$ will also be canonically normalised and diagonal (since the one for $\hat\phi$ already was).
As a result of completing the full diagonalisation procedure, we obtain the following expression for cubic interaction
\be
\sum_{p,q=0,1,2,3}^{p+q=3} (\pa^2 \hat\chi_{(1)})^p (\pa^2 \hat\chi_{(2)})^q \left(\frac{1}{m_{(1)}^4} \sum_i c_{i,p,q} \left(\frac{m_{(2)}}{m_{(1)}}\right)^{2i}
+ \frac{1}{m_{(2)}^4} \sum_i d_{i,p,q} \left(\frac{m_{(2)}}{m_{(1)}}\right)^{2i}\right),
\ee
where $c_{i,p,q}$ and $d_{i,p,q}$ are constant coefficients. Note that it is also possible to write this in closed form as
\be \label{C-eqn}
{\frac{\zeta^3}{M_{\text{Pl}} m_{(1)}^4}} \left[C_{1} (\pa^2 \hat \chi_{(1)})^3 +  C_{2}  (\pa^2 \hat \chi_{(1)})^2 (\pa^2 \hat \chi_{(2)}) + C_{3}  (\pa^2 \hat \chi_{(1)}) (\pa^2 \hat \chi_{(2)})^2  + C_{4} (\pa^2 \hat \chi_{(2)})^3\right]
\ee
up to numerical coefficients not depending on any mass scale and where we have defined
\begin{eqnarray}
C_1 &=&  (1-\alpha^2) \left[ -\alpha^2 (3 - 4\alpha^2 + 3\alpha^4) + (1+\alpha^2)(1-\alpha^2+\alpha^4) \xi\right]  \\
C_2 &=&  \sqrt{3} \left[ (1+\alpha^2)(2-\alpha^2 + \alpha^4)(1-\alpha^2 + 2\alpha^4) - (1+3\alpha^2 + 3\alpha^6 + \alpha^8) \xi \right]   \\
C_3 &=&  (1-\alpha^2) \left[  -(4+3\alpha^2+4\alpha^4 + 3\alpha^6 +4\alpha^8) + (1+\alpha^2)(5-\alpha^2 + 5\alpha^4) \xi  \right] \\
C_4 &=& \frac{\sqrt{3}}{9} \left[ (1+\alpha^2)(14-35\alpha^2 + 46 \alpha^4 - 35 \alpha^6 + 14 \alpha^8) - (13 - 13\alpha^2 + 8\alpha^4 - 13\alpha^6 + 13\alpha^8) \xi \right]  \nn \\
\end{eqnarray}
\begin{align}
\zeta &= \frac{1}{\sqrt{\xi (2 \xi - (1+\alpha^2))}},  &\xi &= \sqrt{1-\alpha^2 + \alpha^4}.
\end{align}

The first obvious observation to make is that, even though we have demixed the scalar action at quadratic order, at cubic order the mixing persists and interaction terms for $\hat\chi_{(1)}$ and $\hat\chi_{(2)}$ should not be looked at in isolation. There are four distinct interaction scales for the cubic action now, set by the $C$'s in \eqref{C-eqn}. 
What does this mean for the decoupling limit? It is interesting to look at two asymptotic limits.
\begin{enumerate}
\item {$\bm m_{(1)} = \bm m_{(2)} = \bm m^2$: } This is the case where $\alpha = 1$ and here \eqref{lambda5intnorm} already represents the cubic interaction for the independently propagating \dof and we see that (up to numerical coefficients) the scale of interactions is set by
\be
m^4_{eff,(\pm)} = \frac{m^8}{|m^4 \pm m^4|},
\ee
where the positive sign applies to the $(\pa^2 \hat\chi_{(1)})^3,  (\pa^2 \hat\chi_{(1)}) (\pa^2 \hat\chi_{(2)})^2$ interactions, whereas the negative sign comes in for $(\pa^2 \hat\chi_{(2)})^3, (\pa^2 \hat\chi_{(1)})^2 (\pa^2 \hat\chi_{(2)})$, i.e. these modes are infinitely suppressed here. 
The suppression scales associated with the two surviving interaction terms are
\be
\Lambda_5^{(+)} = (M_{Pl} m_{eff,(+)}^{4})^{1/5}, \quad\quad \Lambda_5^{(-)} = (M_{Pl} m_{eff,(-)}^{4})^{1/5}
\ee
and the smallest suppression scale is now $\Lambda_5^{(+)}$, where interactions carrying this scale are the ones surviving the decoupling limit.\footnote{The precise numerical coefficients in \eqref{lambda5intnorm}, enforced by the specific theory under consideration, are important to determine the precise, model-dependent, hierarchy of scales in the trimetric case. Here all relevant scales typically lie very closely together. As we shall see below, however, the $N$-dependence of these scales becomes the determining factor for theories with $N$ spin-2 fields, where $N$ is large. This will allow establishing a definitive hierarchy of interaction terms (ordered by their $N$-dependence) in these cases.} Note that this scale $\Lambda_5^{(+)}$ is lower than the corresponding $\Lambda_5$ in a bigravity theory. See figure \ref{fig-dec} for an illustration. We therefore expect such theories to become strongly coupled at a scale mildly lower than the analogous massive/bi-gravity theories.

\item {$\bm m_{(1)} \gg \bm m_{(2)}$: } This is the case where $\alpha \sim 0$, i.e. one of the masses is much heavier than the other. In this case only the $m_{(2)}$ modes in \eqref{lambda5intnorm} remain relevant, and comparing with \eqref{C-eqn} we have
\begin{align}
C_{1} &= 1, &C_{2} &= \sqrt{3}, &C_{3} &= 1  &C_{4} &= \frac{1}{3\sqrt{3}}. 
\end{align}
In other words, two of the four scales have the same value and the lowest suppression scale (which all 3 scales lie very close to here) is set by 
\be
\Lambda_5^{(+)} = (M_{Pl} m_{2}^{4})^{1/5},
\ee
i.e. the physics of the theory is completely dominated by the lighter link, as expected.
\end{enumerate} 
In a general trimetric $\Lambda_5$ theory with arbitrary $m_{(1)},m_{(2)}$ the least suppressed mode will generically have a suppression scale $\Lambda_5^{(+)}$ smaller than or equal to $\Lambda_5^{(2)}$ (where we recall we have chosen $m_{(2)} \le m_{(1)}$) - cf. Figure \ref{fig-dec}. 
\\

{\bf $\Lambda_5$ theories with $N$ spin-2 fields: }
The argument shown above can be generalised to more complex models, i.e. the generic bimetric interaction and $n$-point interaction theories discussed in the previous section. There will always be (several) `sum' and `difference' modes, where the sum/difference term in the denominator of the effective masses mimics the way the $\chi_{(l)}$ are composed in terms of the $\pi_{(i)}$. Some effective sum mode $m^4_{eff,(+)}$ will always be the lightest one and $\Lambda_\lambda^{(+)}$ consequently always the decoupling limit scale. Note, however, that, just as in the trimetric example discussed above, (some of) the other effective masses associated with `difference modes' may only be marginally heavier, so that generically one should keep all modes with similar suppression scales when investigating the phenomenology of some `generalised decoupling limit'.

Let us now see how the above argument for trimetric $\Lambda_5$ theories generalises for an $N$ spin-2 field theory. Especially interesting will be how suppression scales scale with $N$, the number of spin-2 fields. In a generic theory a $\Lambda_5$ term can be written
\be
\frac{1}{\Lambda_5^5} C_{i j k}^{\mu\nu\rho\sigma\alpha\beta} \pi^{i}_{\mu\nu} \pi^{j}_{\rho\sigma} \pi^{k}_{\alpha\beta},
\ee
and when written in terms of the propagating modes $\hat{\chi}$ (using \eqref{diagonalisation transformation}) the coefficient matrix becomes 
\be \label{C-l5}
\tilde{C}_{i j k}^{\mu\nu\rho\sigma\alpha\beta} = C_{i_2 j_2 k_2}^{\mu\nu\rho\sigma\alpha\beta} \sum_{i_1,j_1,k_1} U^K_{i_2 i_1} N_{(i_1)} U^M_{i_1 i} U^K_{j_2 j_1} N_{(j_1)} U^M_{j_1 j} U^K_{k_2 
k_1} N_{(k_1)} U^M_{k_1 k},
\ee
since the normalisation matrix is diagonal.
In a `line' theory the kinetic terms will only mix scalars from adjacent links, and hence the kinetic term matrix will be tri-diagonal, and in the specific case in which the quadratic interactions take on the form found in the dRGT theory (when extended to bigravity), the smallest eigenvalues of this matrix scale as $N^{-2}$ for large $N$. In such a theory (and in fact for any theory with purely bimetric interactions) the terms in equation \eqref{C-l5} involve only fields of identical (original) flavour, i.e $C_{a b c} \propto \delta_{a b} \delta_{b c}$ (note that we do not sum over the $b$ index until substituting into \eqref{C-l5} and that, to avoid clutter, the space-time tensor structure is suppressed). Writing
\be \label{Utilde}
\tilde{U}^{KM}_{a c} = \sum_b U^K_{a b} \lambda_{(b)}^{-1/2} U^M_{b c} 
\ee
we can re-express equation \eqref{C-l5} using \eqref{Utilde} and $C_{a b c} \propto \delta_{a b} \delta_{b c}$ to obtain
\be \label{C-UT}
\tilde{C}_{i j k} = \sum_l  \tilde{U}^{KM}_{l i} \tilde{U}^{KM}_{l j} \tilde{U}^{KM}_{l k},
\ee
where we have ignored the tensor structure relating to the Greek space-time indices, which is of no importance here. 
The most quickly growing entries in the coefficient matrix, associated with the smallest eigenvalues $\lambda_{(i)}$ of the kinetic scalar matrix, scale as $\tilde{C} \sim N^\frac{5}{2}$, corresponding to a  suppression scale
\be \label{LC5}
\Lambda_C = \Lambda_5 / \sqrt{N}.
\ee

We can understand this scaling behaviour by noting that the canonical normalisation matrix $N_{a b}$ scales as $N$ for the smallest eigenvalues (the corresponding eigenmodes will be those with the most strongly $N$-suppressed scale). Furthermore, the elements of the eigenvectors of $\tilde U$ will generically scale as $\mathcal{O}(N^{-\frac{1}{2}})$. In other words, this second factor comes in because of the orthogonal nature of $U^K$ and $U^M$. Overall this results in a scaling behaviour for the least suppressed modes
\be
\tilde{C} \sim N^{3} \sum \left( \frac{1}{\sqrt{N}} \right)^3 = \frac{N^3 N}{N^{\frac{3}{2}}} = N^{\frac{5}{2}}. \label{lambda5 scaling line}
\ee
That is, the effective suppression scale is $\Lambda = \Lambda_5 / \sqrt{N}$ as we found above. Note that care must be taken with this brute-force way of obtaining the scaling. While we have confirmed that it gives the correct result here, as we shall see below, if the flavour structure of $C_{i j k}$ is more complicated, then it is possible for cancellations to occur which lead to a slower scaling with $N$ than would otherwise be expected from this argument.

An interesting observation for the $\Lambda_5$ line theory is that the same scaling can be obtained in a simplified way here. If $U^K$ is taken to be the orthogonal matrix of normalised eigenvectors of $K$, then the scaling of $\tilde{C}$ with $N$ happens to be unaffected by the (unitary) diagonalisation of the mass terms. It will be interesting to investigate whether there is some deeper, model-independent, reason why this is the case. However, taking the observation at face value, we here have
\be
\tilde{C}_{i j k} \sim | \lambda_{(i)} \lambda_{(j)} \lambda_{(k)} |^{-\frac{1}{2}} \sum_{l} U^K_{l i}U^K_{l j}U^K_{l k}, \label{lambda5 scaling}
\ee
where $\lambda$ are the eigenvalues of the (pre-mass demixing) kinetic term matrix, which we already found scale as $N^{-2}$. The same reasoning as for equation \eqref{lambda5 scaling line} then establishes the scaling $\Lambda = \Lambda_5 / \sqrt{N}$.
\\

{\bf Trimetric $\Lambda_3$ theories: } 
As seen in dRGT and Hassan-Rosen bigravity, higher order interactions can be tuned such that the lowest relevant scale is $\Lambda_3$. Suppose we now build an interacting spin-2 field theory with $N$ fields by linking sites with such ghost-free dRGT-like interaction terms and their generalisations to interacting spin-2 fields \cite{deRham:2010ik,deRham:2010kj,Hassan:2011tf,Hassan:2011zd,Hassan:2011ea, Hinterbichler:2012cn}\footnote{To explicitly include the interacting spin-2 field generalisations of \cite{Hinterbichler:2012cn} one would need to work in the vielbein picture - we will discuss this in \cite{vielbein}.}. Again we will first give an explicit example for the trimetric case in order to illustrate the types of interaction terms one obtains and then consider the scaling with $N$ for theories with a larger number of spin-2 fields. Focusing on the cubic terms, we expect interactions relevant in the decoupling limit to come from terms carrying the $\Lambda_3$ scale in Hassan-Rosen-like models
\be \label{tri-lambda3-terms}
\sim h (\pa^2 \pi)^2 + (\pa \pi)  (\pa h) \pa^2 \pi + (\pa \pi)^2 \pa^2 h  + (\pa A)^2 \pa^2 \pi.
\ee
The first three terms are all of the $\{1,0,2\}$ type, where the second and third term involving derivatives of $h$ come from higher order terms in the expansion of the metric - cf. equation \eqref{goldstone3} and also appendix \ref{appendix-nonlocal}. Note that it is consistent to ignore the final ($\{0,2,1\}$) term, since vector perturbations can classically be set to zero \cite{deRham:2010gu,Hinterbichler:2011tt}. Consequently we will first focus on the $\{1,0,2\}$ piece in isolation, before considering the final term later on. 

Again consider the trimetric line example \eqref{toylineaction}. We focus on the relevant scalar-tensor interactions involving the `central node'. This term will be affected by the conformal transformation 
\be
{\bf h}_{(0)} \to {\bf \bar h}_{(0)} + c_1 {\bm \eta} \pi_{(1)} + c_2 {\bm \eta} \pi_{(2)},
\ee
we use to eliminate scalar-tensor mixing at quadratic order in the fields. Here this transformation will now introduce a scalar mixing in the cubic and higher order interactions even before re-expressing $\pi_{(1)},\pi_{(2)}$ in terms of the kinetic eigen-modes $\phi_{(1)},\phi_{(2)}$. 
In particular, for the $\{1,0,2\}$ terms in equation \eqref{tri-lambda3-terms}, this results in interaction terms of the type
\be \label{tri-lambda3-terms2}
\sim \pi_{(i)} (\pa^2 \pi_{(j)})^2 + (\pa \pi_{(i)})  (\pa \pi_{(j)}) \pa^2 \pi_{(i)} + (\pa \pi_{(i)})^2 \pa^2 \pi_{(j)}.
\ee
We now want to show how such terms combine in terms of the kinetic eigen-modes $\phi_{(l)}$. In this section we will show this explicitly only for the cubic $h (\pa^2 \pi_{(i)})^2$ term --- it is straightforward to perform the analogous calculation for the second and third term and we will show combined results in a more compact notation in the following subsection. For the time being, however, it is useful to see the explicit way in which interaction terms combine in terms of the propagating modes. So for cubic $h (\pa^2 \pi_{(i)})^2$ interactions we schematically have
\begin{eqnarray}
&\sim & m_{(2)}^2 h_{(0)} (\pa^2 \pi_{(2)})^2 + m_{(1)}^2 h_{(0)} (\pa^2 \pi_{(1)})^2 \\
&\to & c_1 m_{(2)}^2 \pi_{(1)} (\pa^2 \pi_{(2)})^2 + c_2 m_{(2)}^2 \pi_{(2)} (\pa^2 \pi_{(2)})^2 + c_2 m_{(1)}^2 \pi_{(2)} (\pa^2 \pi_{(1)})^2 + c_1 m_{(1)}^2 \pi_{(1)} (\pa^2 \pi_{(1)})^2, \nn
\end{eqnarray}
where we only show the pure scalar interactions post-conformal transformation and we recall that $h_{(0)}$ corresponds to the `central' node in our trimetric theory. Re-expressing the interaction terms in terms of $\phi_{(1)}$ and $\phi_{(2)}$, we recall $\pi_{(1)} = \frac{1}{\sqrt{2}} (\phi_{(1)} + \phi_{(2)})$ and $\pi_{(2)} = \frac{1}{\sqrt{2}} (\phi_{(1)} - \phi_{(2)}) \alpha^{-2}$, one obtains 
\begin{eqnarray}
\nn && ( \phi_{(1)} + \phi_{(2)} ) \left[ (\pa^2 \phi_{(1)})^2 + (\pa^2 \phi_{(2)})^2  - 2 (\pa^2 \phi_{(1)}) (\pa^2 \phi_{(2)})  \right]  m_{(2)}^2 m_{(1)}^2 \alpha^{-4} \\
\nn &+& ( \phi_{(1)} - \phi_{(2)} ) \left[ (\pa^2 \phi_{(1)})^2 + (\pa^2 \phi_{(2)})^2  - 2 (\pa^2 \phi_{(1)}) (\pa^2 \phi_{(2)})  \right] m_{(2)}^4 \alpha^{-6} \\
\nn &+& ( \phi_{(1)} - \phi_{(2)} ) \left[ (\pa^2 \phi_{(1)})^2 + (\pa^2 \phi_{(2)})^2  + 2 (\pa^2 \phi_{(1)}) (\pa^2 \phi_{(2)})  \right]  m_{(1)}^2 m_{(2)}^2 \alpha^{-2} \\
 &+& ( \phi_{(1)} + \phi_{(2)} ) \left[ (\pa^2 \phi_{(1)})^2 + (\pa^2 \phi_{(2)})^2  + 2 (\pa^2 \phi_{(1)}) (\pa^2 \phi_{(2)}) \right] m_{(1)}^4,
\end{eqnarray}
up to overall numerical coefficients and where we have not canonically normalised the $\phi_{(i)}$ fields yet\footnote{Here one can also directly see that the conformal transformation generates interaction terms which do not retain their original scaling, since each $c_{(i)}$ only knows about one mass scale $m_{(j)}$ and cross-terms $i \neq j$ consequently change the mass-scale in front of interaction terms.}. The factors coming in as multipliers at the end of each line come from the pre-factor of the interaction term, the dimension of $c_{(i)}$ and the $\alpha$ factor in $\pi_{(2)} = \frac{1}{\sqrt{2}} (\phi_{(1)} - \phi_{(2)}) \alpha^{-2}$ respectively. Canonically normalising and combining terms and mass scales, this set of cubic interactions reduces to
\be
\frac{m_{(1)}^2 + m_{(2)}^2}{M_{Pl} m_{(1)}^2 m_{(2)}^2} \hat\phi_{(1)} (\pa^2 \hat\phi_{(1)})^2 + 
\frac{m_{(1)}^2 + m_{(2)}^2}{M_{Pl} m_{(1)}^2 m_{(2)}^2} \hat\phi_{(1)} (\pa^2 \hat\phi_{(2)})^2 - 
2 \frac{m_{(1)}^2 - m_{(2)}^2}{M_{Pl} m_{(1)}^2 m_{(2)}^2} \hat\phi_{(1)} (\pa^2 \hat\phi_{(1)})  (\pa^2 \hat\phi_{(2)}). \label{innerlambda3}
\ee
where we see that the surviving interaction terms for the kinetic eigenmodes have effective mass scales
\be
m_{eff,(\pm)}^2 = \frac{m_{(1)}^2 m_{(2)}^2}{|m_{(1)}^2 \pm m_{(2)}^2|}. 
\ee
We can also compute the contribution from the `outer nodes' giving rise to $h_{(1)}$ and $h_{(2)}$. There we have
\begin{eqnarray}
\sim  m_{(2)}^2 h_{(2)} (\pa^2 \pi_{(2)})^2 + m_{(1)}^2 h_{(1)} (\pa^2 \pi_{(1)})^2 
\to  c_2 m_{(2)}^2 \pi_{(2)} (\pa^2 \pi_{(2)})^2 + c_1 m_{(1)}^2 \pi_{(1)} (\pa^2 \pi_{(1)})^2,
\end{eqnarray}
Again we re-express the interaction terms in terms of $\phi_{(1)}$ and $\phi_{(2)}$ and find
\begin{eqnarray}
\nn && \frac{1}{M_{Pl} m_{(2)}^2} ( \hat\phi_{(1)} - \hat\phi_{(2)} ) \left[ (\pa^2 \hat\phi_{(1)})^2 + (\pa^2 \hat\phi_{(2)})^2  - 2 (\pa^2 \hat\phi_{(1)}) (\pa^2 \hat\phi_{(2)})  \right]  \\
 \nn &+& \frac{1}{M_{Pl} m_{(1)}^2}  ( \hat\phi_{(1)} + \hat\phi_{(2)} ) \left[ (\pa^2 \hat\phi_{(1)})^2 + (\pa^2 \hat\phi_{(2)})^2  + 2 (\pa^2 \hat\phi_{(1)}) (\pa^2 \hat\phi_{(2)}) \right] \\
\nn &=& \frac{1}{M_{Pl} m_{eff,(+)}^2} \left[ \hat\phi_{(1)} (\pa^2 \hat\phi_{(1)})^2 + 
\hat\phi_{(1)} (\pa^2 \hat\phi_{(2)})^2 + 2 \hat\phi_{(2)} (\pa^2 \hat\phi_{(1)})  (\pa^2 \hat\phi_{(2)}) \right] \\
&+& \frac{1}{M_{Pl} m_{eff,(-)}^2} \left[ \hat\phi_{(2)} (\pa^2 \hat\phi_{(1)})^2 
+  \hat\phi_{(2)} (\pa^2 \hat\phi_{(2)})^2 + 2 \hat\phi_{(1)} (\pa^2 \hat\phi_{(1)})  (\pa^2 \hat\phi_{(2)}) \right], \label{outerlambda3}
\end{eqnarray}
up to relative signs between interaction terms in the final expression, as these will depend on the relative size of $m_{(1)}^2$ and $m_{(2)}^2$, now that we have expressed everything in terms of $m_{eff,(\pm)}^2$. When combining contributions from \eqref{innerlambda3} and \eqref{outerlambda3}, we find that all interaction terms only depend on these two mass scales.

Again we now have to diagonalise the mass matrix for $\hat\phi$ in order to express cubic order interactions in terms of the modes $\hat\chi_{(1)},\hat\chi_{(2)}$, which independently propagate at quadratic order. Note that, if $m_{(2)} = m_{(1)}$, 
 $\hat\phi_{(i)} = \hat\chi_{(i)}$ and \eqref{outerlambda3} already shows the cubic action for the independently propagating modes. Just as in \eqref{M-L5} in the $\Lambda_5$ case above we have to diagonalise the mass matrix
\be \label{M-L3}
 M_{\hat\phi} =m_{(1)}^2 \begin{pmatrix}
   1+\alpha^2 & \sqrt{3} (1-\alpha^2) \\
\sqrt{3} (1-\alpha^2) &   3 (1+\alpha^2)
 \end{pmatrix} 
\ee
Everything proceeds as in the $\Lambda_5$ case and we obtain the following expression for  cubic interactions
\begin{eqnarray}
&& \sum_{p,q=0,1,2,3 ; r,s=0,1}^{p+q=3 ; r+s=1} (\hat\chi_{(1)})^r (\hat\chi_{(2)})^s (\pa^2 \hat\chi_{(1)})^p (\pa^2 \hat\chi_{(2)})^q \nn \\ &\cdot &  \left(\frac{1}{m_{(1)}^2 M_{\text{Pl}}} \sum_i c_{i,p,q,r,s} 
\left(\frac{m_{(2)}}{m_{(1)}}\right)^{2i}
+ \frac{1}{m_{(2)}^2 M_{\text{Pl}}} \sum_i d_{i,p,q,r,s} \left(\frac{m_{(2)}}{m_{(1)}}\right)^{2i}\right), \label{scheme-L3}
\end{eqnarray}
where $c_{i,p,q,r,s}$ and $d_{i,p,q,r,s}$ are again numerical coefficients. Closed form expressions as in \eqref{C-eqn} can also be derived, but the relevant scales can already be read off from \eqref{scheme-L3}. We re-emphasise that we have only been considering the $h (\pa^2 \pi)^2$ contribution to the $\{1,0,2\}$ terms in \eqref{tri-lambda3-terms} here. In general additional different interaction scales will arise from the other two  $\{1,0,2\}$ contributions in \eqref{tri-lambda3-terms}.

What about the $A^\mu$-dependent $(\pa A)^2 (\pa^2 \pi_{(i)})$ interaction we have ignored so far? As pointed out above it is consistent to set this to zero. But if it is present, what suppression scale is associated with these interactions now? In our trimetric example we have
\begin{eqnarray}
\nn &\sim & m_{(1)}^2 M_{Pl}^2 (\pa A_{(\pi_{(1)})})^2 \pa^2 \pi_{(1)} + m_{(2)}^2 M_{Pl}^2 (\pa A_{(\pi_{(2)})})^2 \pa^2 \pi_{(2)} \\
&\to &  (\pa \hat{A}_{(\pi_{(1)})})^2 \pa^2 \pi_{(1)} +  (\pa \hat{A}_{(\pi_{(2)})})^2 \pa^2 \pi_{(2)}, 
\end{eqnarray}
where we have canonically normalised the $A$'s in the second line, but now need to re-express $\pi_{(1)},\pi_{(2)}$ in terms of  $\phi_{(1)},\phi_{(2)}$ before doing the same for scalar modes as before. This leads to
\begin{eqnarray}
\nn &\sim &  (\pa \hat{A}_{(\pi_{(1)})})^2 (\pa^2 \phi_{(1)} + \pa^2 \phi_{(2)} ) +  (\pa \hat{A}_{(\pi_{(2)})})^2 (\pa^2 \phi_{(1)} - \pa^2 \phi_{(2)} ) \alpha^{-2} \\
&\sim &  \frac{1}{M_{Pl} m_{(1)}^2}  (\pa \hat{A}_{(\pi_{(1)})})^2 (\pa^2 \hat\phi_{(1)} + \pa^2 \hat\phi_{(2)} ) +  
\frac{1}{M_{Pl} m_{(2)}^2} (\pa \hat{A}_{(\pi_{(2)})})^2 (\pa^2 \hat\phi_{(1)} - \pa^2 \hat\phi_{(2)} ). \label{vector-mixing}
\end{eqnarray}
Contrary to the $h (\pa^2 \pi_{(i)})^2$ term, there is no effect analogous to the conformal transformation mixing $\pi_{(1)}$'s and $\pi_{(2)}$'s here, which means there is no mixing between $A_{\pi_{(1)}}$ and $A_{\pi_{(2)}}$ modes. As a final step we should now again re-express \eqref{vector-mixing} in terms of the modes $\hat\chi_{(1)},\hat\chi_{(2)}$. As a result we obtain
\begin{eqnarray}
&&\frac{1}{m_{(1)}^2 M_{\text{Pl}}}  (\pa \hat{A}_{(\pi_{(1)})})^2 (\pa^2 \hat\chi_{(1)}) \sum_{i=0}^{\infty} \tilde c_{i,1} 
\left(\frac{m_{(2)}}{m_{(1)}}\right)^{2i} + \frac{1}{m_{(1)}^2 M_{\text{Pl}}}  (\pa \hat{A}_{(\pi_{(1)})})^2 (\pa^2 \hat\chi_{(2)}) \sum_{i=0}^{\infty} \tilde c_{i,2} 
\left(\frac{m_{(2)}}{m_{(1)}}\right)^{2i} \nn \\ &+&
\frac{1}{m_{(2)}^2 M_{\text{Pl}}}  (\pa \hat{A}_{(\pi_{(2)})})^2 (\pa^2 \hat\chi_{(1)}) \sum_{i=0}^{\infty} \tilde d_{i,1} 
\left(\frac{m_{(2)}}{m_{(1)}}\right)^{2i} + \frac{1}{m_{(2)}^2 M_{\text{Pl}}}  (\pa \hat{A}_{(\pi_{(2)})})^2 (\pa^2 \hat\chi_{(2)}) \sum_{i=0}^{\infty} \tilde d_{i,2} 
\left(\frac{m_{(2)}}{m_{(1)}}\right)^{2i}, \nn \\ \label{scheme-vector-L3}
\end{eqnarray}
where $\tilde c$ and $\tilde d$ are numerical coeffcients {\it not} identical to the $c,d$ coefficients of \eqref{scheme-L3}. 
  
Combining our results for $h (\pa^2 \pi)^2$ and $(\pa A)^2 (\pa^2 \pi)$ terms, it becomes clear that the decoupling limit has qualitatively changed from the bimetric $\Lambda_3$ case. Instead of just having two suppression scales $\Lambda_3$ associated with $m_{(1)}^2$ and $m_{(2)}^2$ respectively, we now have several\footnote{How many scales exactly depends on the theory under consideration - see examples given below.} scales associated with $\{1,0,2\}$ interactions and four scales associated with $\{0,2,1\}$ terms.  In particular we observe that interacting spin-2 field theories generically break the degeneracy between the scale of $(\pa A)^2 (\pa^2 \pi)$ and $h (\pa^2 \pi)^2$ interactions. This is illustrated in figure \ref{fig-dec}. Again it is useful to concentrate on two asymptotic cases in order to see how this happens explicitly.
\begin{enumerate}
\item {$\bm m_{(1)} = \bm m_{(2)} = \bm m^2$: } This is the case where the modes in \eqref{outerlambda3} and \eqref{vector-mixing} have already demixed the mass term as well and consequently $\hat\phi=\hat\chi$. Up to numerical coefficients the scale of interactions for the $\{1,0,2\}$ terms is set by
\be
m^2_{eff,(\pm)} = \frac{m^4}{|m^2 \pm m^2|},
\ee
whereas the $\{0,2,1\}$ terms have an effective mass $m^2_{eff} = m^2$. As in the $\Lambda_5$ example $m^2_{eff,-}$ is infinitely massive. Consequently the relevant interaction scales are
\begin{align}
\nn \Lambda_3^{(+)} &= (M_{Pl} m_{eff,(+)}^{2})^{1/3},  
 &\Lambda_3^{(2)} &= (M_{Pl} m^{2})^{1/3},
\end{align}
corresponding to the  `sum mode' arising from the $\{1,0,2\}$ terms (the difference mode is the one with infinite mass and hence infinitely suppressed) and the $1/m^2$ suppression carried by $(\pa A)^2 (\pa^2 \pi)$ respectively. This is an explicit example of how interacting spin-2 field theories break the degeneracy between the scale of $\{1,0,2\}$ and $\{0,2,1\}$ interactions. Again the smallest suppression scale is $\Lambda_3^{(+)}$ and strictly speaking only interactions carrying this scale survive the decoupling limit. As before, however, one should be cautious that $\Lambda_3^{(+)}$ and $\Lambda_3^{(2)}$ may lie very close together, depending on the theory under consideration. As we will see in the next subsection, however, for large $N$ theories the $N$-dependence of interaction terms will become the dominant factor unambiguously establishing which suppression scales are lowest.

\item {$\bm m_{(1)} \gg \bm m_{(2)}$: } Again we also consider the case where one of the masses is much heavier than the other. The $m_{(1)}$ dependence of interaction terms then drops out and we find that $\{1,0,2\}$ terms are of the form
\begin{eqnarray}
\frac{1}{m_{(2)}^2 M_{\text{Pl}}}\sum_{p,q=0,1,2,3 ; r,s=0,1}^{p+q=3 ; r+s=1} (\hat\chi_{(1)})^r (\hat\chi_{(2)})^s (\pa^2 \hat\chi_{(1)})^p (\pa^2 \hat\chi_{(2)})^q  d_{0,p,q,r,s}, \label{scheme-L3-extreme}
\end{eqnarray}
whereas $\{0,2,1\}$ terms become
\begin{eqnarray}
\frac{1}{m_{(2)}^2 M_{\text{Pl}}}  (\pa \hat{A}_{(\pi_{(2)})})^2 (\pa^2 \hat\chi_{(1)}) \tilde d_{0,1} + \frac{1}{m_{(2)}^2 M_{\text{Pl}}}  (\pa \hat{A}_{(\pi_{(2)})})^2 (\pa^2 \hat\chi_{(2)}) \tilde d_{0,2}, \label{scheme-vector-L3-extreme}
\end{eqnarray}
The strong coupling scale is therefore now set by (up to numerical coefficients) 
\be
\Lambda_3^{(+)} = \Lambda_3^{(2)} = (M_{Pl} m_{2}^{2})^{1/3}.
\ee
Not surprisingly this is the same scale as for bigravity with $m = m_{(2)}$ - the much more massive $m_{(1)}$ modes are irrelevant to all intents and purposes here. It is worth pointing out that the extreme case considered here corresponds to the largest possible value for $\Lambda_3^{(+)}$ - typically this scale will always lie lower than any individual $\Lambda_3^{(n)}$ corresponding to a bigravity theory formed out of one of the theory's links.
\end{enumerate}

{\bf $\Lambda_3$ theories with $N$ spin-2 fields: } 
As discussed in the $\Lambda_5$ example above, one can generalise the trimetric $\Lambda_3$ argument to more general $N$ spin-2 field models of the $\Lambda_3$ type. These will also give rise to several `sum' and `difference modes', with an effective sum mode $\Lambda_3^{(+)}$ providing the lowest decoupling scale. We emphasize that these modes will always come from linear combinations of  $\{1,0,2\}$ interactions, re-iterating the qualitative change in the decoupling limit for multiple interacting spin-2 fields. 

Let us again see what happens when we consider line theories with a large number of fields. More specifically we will be particularly interested in finding out how the suppression scale varies with the number of spin-2 fields. As explained above, in the $\Lambda_3$ case there will be two qualitatively different types of contributions to the pure scalar cubic terms. They are of the form $\pi (\pa^2 \pi)^2$ and $(\pa \pi)^2 (\pa^2 \pi)$  --- cf. equation \eqref{tri-lambda3-terms2}. We can compactly write these interaction terms as
\be \label{N-l3-int}
\frac{1}{\Lambda_3^3} C_{A, i j k}^{\mn\alpha\beta} \pi^i \pi^j_{\mn} \pi^k_{\alpha\beta}, \qquad \text{and} \qquad \frac{1}{\Lambda_3^3} C_{B, i j k}^{\mn} \pi^i_{\mn} \pi^{j,\alpha} \pi^k_{\alpha}.
\ee
Specialising again to the case of a `line' theory in which the quadratic interactions are of the dRGT/Hassan-Rosen type, the flavour index structures of $C_A$ and $C_B$ are:
\be \label{C-L3N}
C^A_{i j k} \propto \left( \beta_L ( \delta_{i+1,j} - \delta_{i,j} ) + \beta_R ( \delta_{i,j} - \delta_{i-1,j} )\right) \delta_{j,k}, \quad \text{and} \quad C^B_{i j k} \propto \left( \delta_{i,j} - \delta_{i-1,j} \right) \delta_{j,k},
\ee
where $\beta_L$ and $\beta_R$ are constants, there is \emph{no} summation over repeated indices here, and the tensor structure leads to terms $\pi_i X^\mu_{(2),\mu}(\pi_j,\pi_k)$ and $(\Box \pi_i) \pi_j^\alpha \pi^k_\alpha$ respectively. We emphasize that lower case Latin indices are `flavour' indices, whereas Greek indices are space-time indices. Expressing the interactions \eqref{N-l3-int} using equations \eqref{C-l5}-\eqref{C-UT}, we find
\begin{align}
\tilde{C}^A_{i j k} &\sim \sum_{l} \left[ \beta_L (\tilde U^{KM}_{l-1,i} - \tilde U^{KM}_{l,i} ) + \beta_R (\tilde U^{KM}_{l,i} - \tilde U^{KM}_{l+1,i} )\right] \tilde U^{KM}_{l j} \tilde U^{KM}_{l k}, \\\tilde{C}^B_{i j k} &\sim \sum_{l} \left[ \tilde U^{KM}_{l,i} - \tilde U^{KM}_{l+1,i} \right] \tilde U^{KM}_{l j} \tilde U^{KM}_{l k}.
\end{align}

We now again want to find out how the effective scale of interactions scales with $N$. 
Compared to the $\Lambda_5$ case, we now have a more complicated flavour structure, so care must be taken to track the cancellations of terms this implies. A brute-force scaling argument analogous to \eqref{lambda5 scaling line} would be blind to this subtlety.
More specifically we find that the tensor structure imposes certain symmetries on the flavour structure. So for example the $A$-type terms are totally symmetric in the three fields, and the $B$-type terms are symmetric in the last two fields. With this in mind, one finds that the most quickly growing entries in the coefficient matrix scale as $\tilde{C}^A \sim N^\frac{1}{2}$ and $\tilde{C}^B \sim N^\frac{3}{2}$ for large $N$, corresponding to suppression scales 
\be
\Lambda_A = \Lambda_3 / N^{\frac{1}{6}}, \qquad \text{and} \qquad \Lambda_B = \Lambda_3 / \sqrt{N}.
\ee
We note that the $B$-type terms scale fastest, and hence will come to dominate. This also means the decoupling limit will be insensitive to te coefficients $\beta_{L,R}$, which only affect $A$-type terms - cf. \eqref{C-L3N}. In analogy to \eqref{LC5}, the strong coupling scale of the theory will therefore be set by the B-type terms and we have 
\be
\Lambda_C = \Lambda_3 / \sqrt{N}.
\ee  
Note that this scaling is the same as was found for the strong coupling scale in dimensional deconstruction models \cite{deRham:2013awa}. Those models deal with `loop', rather than `line', theories, but we see that in the large-$N$ limit they scale in the same way.
In this context it is perhaps also worth mentioning the mass eigenvalue prescription used by \cite{Deffayet:2005yn,Deffayet:2003zk,ArkaniHamed:2003vb,Schwartz:2003vj,deRham:2013awa} in order to deal with what we have dubbed `scalar mixing'. In the specific context of a circle theory this Fourier based method is a highly effective way of not only obtaining the propagating modes, but also the scaling of the interaction terms with $N$.

Finally, it is again interesting to note that, at least for the line theories under consideration, the $N$-dependence for the `decoupling' interaction terms is not affected by the mass-diagonalisation, just as we found in the $\Lambda_5$ case. Ignoring the mass-diagonalisation (essentially setting $U^M_{ ab} = \delta_{a b}$), the analogous expression to \eqref{lambda5 scaling} is now:
\begin{align}
\tilde{C}^A_{i j k} &\sim | \lambda_{(i)} \lambda_{(j)} \lambda_{(k)} |^{-\frac{1}{2}} \sum_{l} \left[ \beta_L (U^K_{l-1,i} - U^K_{l,i} ) + \beta_R (U^K_{l,i} - U_{l+1,i} )\right] U^K_{l j}U^K_{l k}, \\\tilde{C}^B_{i j k} &\sim | \lambda_{(i)} \lambda_{(j)} \lambda_{(k)} |^{-\frac{1}{2}} \sum_{l} \left[ U^K_{l,i} - U^K_{l+1,i} \right] U^K_{l j}U^K_{l k}.
\end{align}

\section{Example I: EBI (Eddington-Born-Infeld) Bigravity} \label{sec-EBI}

\begin{figure}[tp]
\centering
\begin{tikzpicture}[-,>=stealth',shorten >=0pt,auto,node distance=2cm,
  thick,main node/.style={circle,fill=blue!10,draw,font=\sffamily\large\bfseries},arrow line/.style={thick,-},barrow line/.style={thick,->},no node/.style={plain},rect node/.style={rectangle,fill=blue!10,draw,font=\sffamily\large\bfseries},red node/.style={rectangle,fill=red!10,draw,font=\sffamily\large\bfseries},green node/.style={circle,fill=green!20,draw,font=\sffamily\large\bfseries},yellow node/.style={rectangle,fill=yellow!20,draw,font=\sffamily\large\bfseries}]

 \node[main node](100){};
  \node[main node] (101) [right=3cm of 100] {};
 
  \node[main node,] (2)  [right=2cm of 101]{};
   \node[main node] (1) [right=3cm of 2] {};
   
    \node[main node,fill=black!100,scale=0.7] (5)  [right=3cm of 1]{};
  \node[main node] (7) [below left=1cm of 5] {};
   \node[main node] (8) [below right=1cm of 5] {};
   
     \node[main node](80)[above left=1.0cm of 5]{};
       \node[main node](81)[above right=1.0cm of 5]{};
        \node[draw=none,fill=none](82)[above=1.5cm of 5]{};
         \node[draw=none,fill=none](83)[right=1.2cm of 100]{};
         \node[draw=none,fill=none](84)[right=1.2cm of 2]{};

  \path[every node/.style={font=\sffamily\small}]
  (100) edge node {$g^{\mu\nu} f_{\mu\nu} = Tr[g^{-1} f] $} (101)
    (2) edge node {$e_m \left( \sqrt{g^{-1} f} \right)$} (1)
      (7) edge node [above left] {} (5)
       (8) edge node [above right] {} (5);


\draw[-,] (5) to (80);
\draw[-,] (5) to (81);
    
    \node[draw=none,fill=none](90)[above=1.2cm of 5]{$g^{\mu\nu}_{(0)} g_{\nu \alpha}^{(1)} g^{\alpha \beta}_{(2)} g_{\beta \mu}^{(3)}$};

   \node[draw=none,fill=none](92)[below of=83]{(a) EBI bigravity};
   
   \node[draw=none,fill=none](93)[below of=84]{(b) Hassan-Rosen bigravity};
   
   \node[draw=none,fill=none](94)[below of=5]{(c) 4 spin-2 ``EBI''};
\end{tikzpicture}
\caption{Examples I-III shown in sections \ref{sec-EBI}-\ref{sec-EBI4}. From left to right: (a) Eddington-Born-Infeld (EBI) bigravity as discussed in section \ref{sec-EBI}. (b)  Hassan-Rosen bigravity with a dRGT-like interaction term as discussed in section \ref{sec-dRGT}. (c) A multi-gravity extension of EBI bigravity as discussed in section \ref{sec-EBI4}. We here consider a 4 spin-2 field theory with a single EBI-like interaction term tracing over the product of all four spin-2 fields.}
\end{figure}
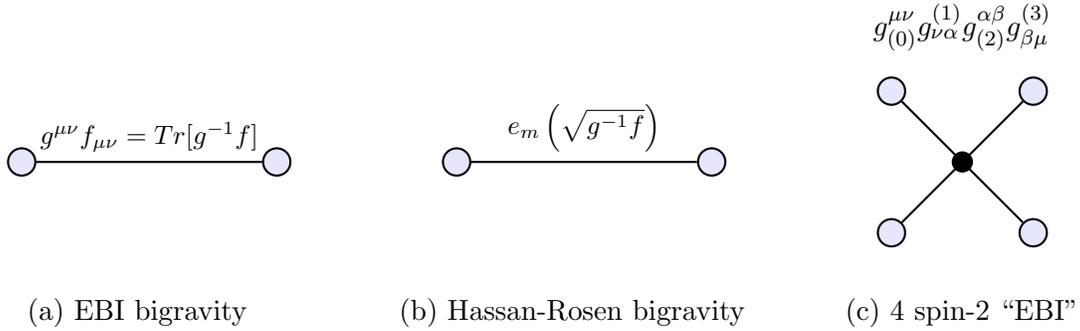

Throughout the previous sections we have provided a number of examples to illustrate the features of interacting spin-2 field models. In this and the following two sections we wish to complete this array of examples by explicitly showing how the \St analysis can be carried out for three such theories and e.g. how it can be used to demonstrate the presence or absence of ghosts (at least in a given limit/up to a given order in interaction terms).

We begin by considering the EBI bigravity theory presented in \cite{Banados:2008rm} in its explicit bigravity formulation \cite{Banados:2008fi}
\be \label{EBI-action}
{\cal S}_{\text{EBI}} = M^2_{\text{Pl}} \int d^4 x \left( \sqrt{-g}R[g] + \sqrt{-f}R[f]  + \frac{m^2}{2} \sqrt{-g} g^{\mn} f_{\mn}\right).
\ee
Note that different versions of EBI theories exist \cite{Banados:2010ix} based on work by \cite{Deser:1998rj,Vollick:2003qp,Vollick:2005gc}, where effectively one of the two metrics becomes an auxiliary field, so we are left with a single dynamical spin-2 field. Also see \cite{Scargill:2012kg,Cho:2012vg,Banados:2010ix,Avelino:2012ue,Cho:2013pea,Pani:2011mg} for related phenomenological studies in this version of the theory, 
\comment{[\jn{Phenomenology for which version?]}}
e.g. motivated by the fact that in EBI-type theories cosmological solutions sourced by a perfect fluid can avoid the initial big bang singularity and naturally produce a period of exponential expansion akin to inflation. Though see \cite{Pani:2012qd} for possible problems with the description of compact objects. In what follows we will only consider the bigravity version \eqref{EBI-action}.
\footnote{Note that the EBI bigravity interaction term, $\mathrm{tr} [g^{-1}f]$, is reminiscent of one of the known ghost free interaction terms from the dRGT model and Hassan-Rosen bigravity we discuss in the next section, $e_2(\sqrt{g^{-1}f}) = \frac{1}{2}( (\mathrm{tr} [\sqrt{g^{-1}f}])^2 - \mathrm{tr} [g^{-1}f] )$, only lacking the first trace of the square root.}
\\

{\bf The generalised Fierz-Pauli condition:}
In its bimetric formulation the Eddington-Born-Infeld interaction term can be written as
\be
{\cal L}_{\text{int}}^{\text{EBI}} = \sqrt{-g} g^{\mn}f_{\mn}.
\ee
As discussed in sections \ref{sec-link} and \ref{sec-gold}, we restore the gauge-invariance of the theory by introducing a single \St field $Y^\mu = x^\mu + \tilde{A}^\mu \to x^\mu + {A}^\mu + \pa^\mu \pi$ for the $f_{\mu\nu}$ metric, pulling it onto the `site' of the $g_{\mu\nu}$. We will ignore the \St vector field ${A}$, and focus on the \St scalar $\pi$. We also expand each metric about a common, flat background,
\be
g^{\mn} \to \eta^{\mn} + h^{\mn}, \qquad f_{\mn} \to \eta_{\mn} + l_{\mn}.
\ee
Up to terms quadratic in the fields, the result is\footnote{Note that we have not re-summed the non-local terms due to $l_{\mn}(x + \partial \pi)$, as described in appendix \ref{appendix-nonlocal}, and hence there is a $\pi^\mu \partial_\mu l^\nu_\nu$ term, however this is of no concern here as we shall see that a ghost already appears at quadratic order in the pure scalar sector.}
\bea
{\cal L}_{\text{int}}^{\text{EBI}}  &=& 4 + 2 \pi^\mu_\mu + \pi^{\mn}\pi_{\mn}\nonumber \\
&& -\: h^\mu_\mu + l^\mu_\mu + h^{\mn}h_{\mn} + h^{\mn}l_{\mn} - h^\mu_\mu l^\nu_\nu \nonumber \\
&& +\: 2 h^{\mn}\pi_{\mn} + 2 l_{\mn}\pi^{\mn} - h^\mu_\mu \pi^\nu_\nu + \pi^\mu \partial_\mu l^\nu_\nu. \label{EBI_pre_conformal_trans}
\eea
The pure-scalar part of this action (still prior to demixing) is  
\be \label{EBI-purescalarpre}
{\cal L}_{\text{int, pre-demixing}}^{\text{EBI, pure-scalar}}  = 4 + 2 \pi_\alpha^{\alpha} + \pi^{\mu\nu} \pi_{\mu\nu}
\ee
This action needs to satisfy the generalised Fierz-Pauli condition \eqref{GenFPC} to avoid ghost-like degrees of freedom. But in fact comparison with \eqref{GenFPA} shows that $a=1$ and $b=0$ here, so we have $a + b = 1 \neq 0$ and the generalised Fierz-Pauli condition is violated. Since the dangerous quadratic self-interactions for $\pi$ consequently do not vanish up to total derivatives, a ghost-like degree of freedom enters the theory. This is also shown in table \ref{table1}.
\\

{\bf The full \St analysis:}
Returning to \eqref{EBI_pre_conformal_trans}, we now wish to demix scalar and tensor modes, in the process introducing a standard kinetic term for $\pi$. As described in section \ref{sec-gold}, this is achieved by the conformal transformations 
\be
h_{\mn} \to \bar{h}_{\mn} + c_g \pi \eta_{\mn}, \qquad l_{\mn} \to \bar{l}_{\mn} + c_f \pi \eta_{\mn}.
\ee
The new terms generated when these transformations are applied to \eqref{EBI_pre_conformal_trans} are
\be
2(c_f - c_g) \pi \pi^\mu_\mu + 4c_f \pi^\mu \pi_\mu + 4(c_f - c_g) \pi + 4c_g (c_g - c_f) \pi^2 + (2c_g - c_f) \bar h^\mu_\mu \pi - c_g \bar l^\mu_\mu \pi,
\ee
while the linearised kinetic terms give (cf. \eqref{Rlinconf})
\be
{\cal L}_{\text{EH}}^{\text{lin},g} + {\cal L}_{\text{EH}}^{\text{lin},f} \to \bar {\cal L}_{\text{EH}}^{\text{lin},g} + \bar {\cal L}_{\text{EH}}^{\text{lin},f} + 3(c_g^2 + c_f^2) \pi \pi^\mu_\mu + 2 (c_g h_{\mn} + c_f l_{\mn}) (\eta^{\mn} \pi^\rho_\rho - \pi^{\mn}).
\ee
Thus post conformal transformation the EBI Lagrangian is, to quadratic order and neglecting the \St vector $ A^\mu$,
\bea
\mathcal{L}^{\text{EBI}} &=& M_\mathrm{Pl}^2 \Bigg\{ \left[3(c_g^2 + c_f^2) - m^2 (c_g + c_f) \right] \pi \pi^\mu_\mu  + \bar h_{\mn} \hat{\mathcal{E}}^{\mn \alpha\beta} \bar h_{\alpha\beta} + \bar l_{\mn} \hat{\mathcal{E}}^{\mn \alpha\beta} \bar l_{\alpha\beta}  \\
&&+\:m^2 \left[ 2 (c_f-c_g) \pi - 2 c_g (c_f - c_g) \pi^2 + 2 + \pi^\mu_\mu + \frac{1}{2} \pi^{\mn}\pi_{\mn} \right] \nonumber \\
&& +\:\frac{m^2}{2} \left( \bar l^\mu_\mu - \bar h^\mu_\mu + \bar h^{\mn} \bar h_{\mn} + \bar h^{\mn} \bar l_{\mn} - \bar h^\mu_\mu \bar l^\nu_\nu - c_g \bar l^\mu_\mu \pi + (2 c_g - c_f) \bar h^\mu_\mu \pi \right) \nonumber \\
&&+\:(m^2 - 2 c_f) \bar h_{\mn} \pi^{\mn} + (m^2 - 2 c_g) \bar l^{\mn} \pi_{\mn} + \left(2 c_f - \frac{m^2}{2}\right) \bar h^\mu_\mu \pi^\nu_\nu + \left(2 c_g - \frac{m^2}{2}\right) \bar l^\mu_\mu \pi^\nu_\nu \Bigg\}. \nonumber
\eea
The scalar-tensor kinetic mixing can be partially removed by choosing $c_g = \frac{m^2}{2} = c_f$, giving
\bea
\mathcal{L}^{\text{EBI}} &=& M_\mathrm{Pl}^2 \Bigg\{ \frac{m^4}{2} \pi \pi^\mu_\mu + 2 + m^2 \pi^\mu_\mu + \frac{m^2}{2} \pi^{\mn}\pi_{\mn}  +  \bar h_{\mn} \hat{\mathcal{E}}^{\mn \alpha\beta} \bar h_{\alpha\beta} + \bar l_{\mn} \hat{\mathcal{E}}^{\mn \alpha\beta} \bar l_{\alpha\beta} \\
&& +\:\frac{m^2}{2} \left( \bar l^\mu_\mu - \bar h^\mu_\mu + \bar h^{\mn} \bar h_{\mn} + \bar h^{\mn} \bar l_{\mn} - \bar h^\mu_\mu \bar l^\nu_\nu \right) - \frac{m^4}{4} \left( \bar h^\mu_\mu \pi + \bar l^\mu_\mu \pi \right) + \frac{m^2}{2}\left( \bar h^\mu_\mu \pi^\nu_\nu + \bar l^\mu_\mu \pi^\nu_\nu \right) \Bigg\}. \nonumber 
\eea
As discussed in section \ref{sec-gold} the scalar-tensor mixing has only partially been removed by the conformal transformations and there are still residual mixing terms involving $\bar h^\alpha_\alpha$ and $\bar l^\alpha_\alpha$. These are removed by adding appropriate gauge-fixing terms enforcing the traceless nature of $\bar h$ and $\bar l$, leaving us with
\bea \label{EBI-L-final}
\mathcal{L}^{\text{EBI}} &=& M_\mathrm{Pl}^2 \Bigg\{ \frac{m^4}{2} \pi \pi^\mu_\mu + 2 + m^2 \pi^\mu_\mu + \frac{m^2}{2} \pi^{\mn}\pi_{\mn} \\
&& +\: \bar h_{\mn} \hat{\mathcal{E}}^{\mn \alpha\beta} \bar h_{\alpha\beta} + \bar l_{\mn} \hat{\mathcal{E}}^{\mn \alpha\beta} \bar l_{\alpha\beta} + \frac{m^2}{2} \left( \bar h^{\mn} \bar h_{\mn} + \bar h^{\mn} \bar l_{\mn} \right) \Bigg\}. \nonumber 
\eea

%
\begin{table}[t]
\centering
\renewcommand{\arraystretch}{1.4}
    \begin{tabular}{ | l | l | c | p{5cm} |}
    \hline
    (n,m) & EBI scalar terms & Galilean invariants & Total derivative combinations \\ \thickhline
     \rowcolor[gray]{0.95} (0,0) & $1$ & -- &  1 \\ \hline
     (0,1) & -- & $\pi$ &  -- \\ \hline
    \rowcolor[gray]{0.95} (2,1) & $\pi_{\mu}^\mu$ & -- &  $\pi_{\mu}^\mu$ \\ \hline
 (2,2) & $ m^4 \pi \pi^\mu_\mu  $ & $\pi_\mu \pi^\mu$ & $\pi_\mu \pi^\mu - \pi \pi^\mu_\mu$ \\ \hline
 \rowcolor[rgb]{1,0.6,0.6} (4,2) & $m^2 \pi^\mu_\alpha \pi_\mu^\alpha$ & -- & $\pi^\mu_\alpha \pi_\mu^\alpha - \pi_\mu^\mu \pi_\nu^\nu$ \\ \hline
    \end{tabular}
\caption[EBI scalar interaction term]{EBI bigravity scalar interaction terms from \eqref{EBI-L-final} up to numerical prefactors and grouped by $(n,m)$, where $n$ denotes the number of derivatives $\pa$ and $m$ denotes the order in the field $\pi$ for the given term. We list all EBI bigravity terms in the second column and go up to order $(n,m) = (4,2)$ - sufficient to see the appearance of ghosts that result from a violation of the generalised Fierz-Pauli tuning \eqref{GenFPC}. Note that $\pi$ has no potential terms $\{0,1\}$. Gray (light shaded) contributions vanish up to total derivatives, while white contributions give Galilean invariants, i.e. ghost-free terms. Red (dark shaded) contributions denote terms which result in the equation of motion for $\pi$ being higher than second order and hence give rise to Ostrogradski ghosts.}
\label{table1}
\end{table}

Extending the argument of section \ref{sec-gold}, it is now straightforward to see why a failure of the generalised Fierz-Pauli tuning \eqref{GenFPC} pre-demixing cannot be cured by taking into account scalar-tensor mixing terms. We denote the order of a given scalar interaction term by $(n,m)$, where $n$ counts the number of derivatives $\pa$ and $m$ the order in the field $\pi$. Importantly different $(n,m)$ cannot be related by integration by parts, so they cannot cancel up to total derivatives. Any $(n,m)$ terms that produce dangerous ghost-like interactions must therefore vanish by themselves. Now consider the $\pi^\alpha_\alpha \pi^\beta_\beta$ and $\pi_{\alpha \beta} \pi^{\alpha \beta}$ terms relevant for the generalised Fierz-Pauli condition \eqref{GenFPC}. These are $\{4,2\}$ terms, as expected, since these interactions are generated via the \St trick prior to any demixing procedure. Mixing terms $h (\pa^2 \pi)^n$ will always generate $\{ 2n, n+1 \}$ scalar terms, so can never cancel potentially dangerous $\{2n,n\}$ scalar interactions already present prior to demixing. This is why the dangerous $\pi^{\mn}\pi_{\mn}$ interaction present in \eqref{EBI-L-final} was already visible in \eqref{EBI-purescalarpre}. 
\comment{Table \ref{table1} summarises the , showing that at 4th order in derivatives and second order in the Goldstone field $\pi$ a ghost appears for the EBI interaction term\footnote{Further ghost-like interaction terms will occur at higher order as well, but there exact form is not of interest here, given that no higher order terms can cure the low order instability shown for $(4,2)$.}}

It is interesting to note that the ordinary kinetic term for $\pi$ in \eqref{EBI-L-final} already has the wrong sign. In addition, the higher-derivative terms in \eqref{EBI-L-final} are non-degenerate, so given that they are not of the Galilean invariant form (cf. table \ref{table1}) it is straightforward to appeal to Ostrogradski's theorem to infer the presence of ghosts. To see this explicitly and also in order to determine the mass scale of the ghost-like \dof, we consider the non-total-derivative scalar piece from  \eqref{EBI-L-final} and write 
\bea
\nn \int d^4x \frac{M_\mathrm{Pl}^2 m^2}{2} \left( m^2 \pi \pi^\mu_\mu + \pi^{\mn} \pi_{\mn} \right) &=& \int d^4x \left( \frac{1}{2} \hat \pi \Box \hat \pi + \sigma \Box \hat \pi - \frac{1}{2} m^2 \sigma^2 \right) \\
&=& \int d^4x \frac{1}{2} \left( \hat\pi \Box \hat\pi - \sigma \Box \sigma - m^2 \sigma^2 \right),
\eea
where we have canonically normalised the scalar field $\pi \to m^{-2} M_{\text{Pl}}^{-1}\hat\pi$ and introduced an auxiliary field $\sigma$ in the first line. In the second line we then perform a field re-definition $\hat\pi \to \hat\pi - \sigma$ to demix the fields. Thus we see that there is in fact just one ghost in the theory which is massless and the theory already displays ghost-like behaviour before any cubic or higher order interactions are taken into account.
One further comment may be of interest. If the $\pi$ field did have a mass term, then the field re-definition used above would generate a $\sim m^2 \pi \sigma$ coupling, i.e. a mixing in the potential $V(\pi,\sigma)$. This is a quirk of the theory in $D=4$ with the Planck masses of the two metrics equal. 
\\

\section{Example II: Ghost-free Bigravity } \label{sec-dRGT}

Ghost-free massive gravity theories have recently been constructed, first for a single dynamical spin-2 field \cite{deRham:2010ik,deRham:2010kj,Hassan:2011hr} and then an extension for two dynamical and interacting spin-2 fields \cite{Hassan:2011tf,Hassan:2011zd,Hassan:2011ea}. The particular form of the interaction term $V(g^{-1} f)$ is what ensures ghost-freedom in these theories. Focusing on the second case, Hassan-Rosen bigravity, the action for in $D$ space-time dimensions\footnote{Note that contrary to the rest of this paper, in this section we generalise both away from $D = 4$ (cf.  \cite{Hinterbichler:2012cn}) and to non-equal Planck masses for the spin-2 fields. The reader will note that this does not change the procedure in any fundamental way.} is
\be
\label{dRGT-act} 
S=\int d^Dx \bigg[ M_g^2 \sqrt{- g}\,R[g] + M_f^2 \sqrt{- f}\,R[f] + m^2 M_{\text{eff}}^2 \sqrt{- g}\,\sum_{n=0}^{D} \beta_n\,
  e_n(\sqrt{g^{-1} f})\bigg] ,
\ee
where $M_{\text{eff}}^{-2} = M_g^{-2} + M_f^{-2}$.
Here the matrix square root is defined via $\sqrt{g^{-1}f} \sqrt{g^{-1}f} = g^{\mu \lambda}f_{\lambda \nu}$ and $e_k(M)$ denote the elementary symmetric polynomials of the
eigenvalues $\lambda_n$ for a matrix $M$.  In four dimensions $\sqrt{g^{-1}f}$ is a $4
\times 4$ matrix and we can write the elementary polynomials as 
\bea
e_n(M) &=& \delta^{\alpha_1 \ldots \alpha_n}_{[\beta_1 \ldots \beta_n]} M_{\alpha_1}^{\beta_1} \ldots M_{\alpha_n}^{\beta_n},
\eea
where we have defined 
\be
\delta^{\alpha_1 \ldots \alpha_n}_{[\beta_1 \ldots \beta_n]}  \equiv \frac{1}{n!}  \varepsilon^{\alpha_1 \ldots \alpha_n}
\varepsilon_{\beta_1 \ldots \beta_n}.
\ee
Using the notation $[M]=\text{Tr} [M]$ we can explicitly write out the five relevant elementary symmetric polynomials in four dimensions
\begin{align}
e_0(M)&= 1,  &e_1(M)&= [M], \nn \\  
\nn e_2(M)&= \tfrac{1}{2}([M]^2-[M^2]), &e_3(M)&= \tfrac{1}{6}([M]^3-3[M][M^2]+2[M^3]),
\end{align}
\vspace{-0.5cm}
\be
e_4(M) = \tfrac{1}{24}([M]^4-6[M]^2[M^2]+3[M^2]^2 +8[M][M^3]-6[M^4]).
\ee
In particular note that $e_D(M) = \text{det} (M)$, $e_k(M)= 0$ for $k>D$, 
where $e_D (M)$ is the characteristic polynomial for $M$, i.e. its determinant. There are $D$ free parameters $\beta_n$, two of which correspond to cosmological constants ($\beta_0$ and $\beta_D$) and there is degeneracy between scaling the $\beta_n$ and $m$.  
Here we now wish to apply the \St machinery to this theory as a check for the `analysis recipe' developed throughout this paper to be compared with the results of \cite{deRham:2010ik,deRham:2010kj,Hassan:2011hr,Fasiello:2013woa}.
\\

{\bf The \St analysis: } 
We now restore gauge invariance for \eqref{dRGT-act}, where in the language of section \ref{sec-multiSt} we take \emph{approach I}. In other words, we only introduce a single \St field $Y$ mapping $f$ to the site of $g$ or vice versa. Note that what seems like a trivial and arbitrary choice in this context actually lies beneath the duality presented in \cite{deRham:2013hsa}. The two different choices of link field $Y_{f g}$ or $Y_{g f}$ will lead to two different galileon theories for the \St theories in the decoupling limit, which are nevertheless physically equivalent, i.e. dual to one another. With an interaction $e_m(\sqrt{g^{-1} f})$, we here choose to map $f$ to the site of $g$, i.e. work with the \St fields $Y_{f g}$\footnote{Due to the fact that $\sqrt{-g} e_m(\sqrt{g^{-1} f}) = \sqrt{-f} e_{D-m}(\sqrt{f^{-1} g})$, the equivalent Lagrangian with the \St fields introduced for $g$ instead of $f$ can be obtained from what follows by exchanging $h$ and $l$ and replacing $m$ by $D-m$.}. 
Each metric is expanded about Minkowski as
\be
g_{\mn} = \eta_{\mn} + h_{\mn}, \qquad f_{\mn} = \eta_{\mn} + l_{\mn},
\ee
\comment{, the deviations from which of $g$ and $f$ respectively are denoted by $h$ and $l$.}where, contrary to the previous example, we have defined perturbations for $\bf{g}$ and not $\bf{g}^{-1}$. We have chosen this here since if one is building theory graphs with more links, such as in section \ref{sec-can}, then it can be helpful to treat all fields in the same way. Since the action is linear in the $e_i$ we can consider each separately and confirm that, to a given order, no ghost is present. 

\begin{enumerate}
\item {\it The pure scalar part prior to demixing: } 
For an $e_m$ interaction it can be shown that the pure scalar part is
\be
\cL_\pi = m^2 M_{\text{eff}}^2 \sum_{n=0}^{m} \frac{1}{n! (m-n)!} \frac{(D-n)!}{(D-m)!} \cL^{\mathrm{TD}}_{(n)}(\pi),
\ee
where $\cL^{\mathrm{TD}}_n(\pi)$ indicates the $n^\text{th}$ order total derivative combination of second derivatives of $\pi$ - for more details see appendix \ref{appendix-TD}. Thus the Lagrangian is purely total derivative and vanishes (up to boundary terms). Note that this is not only true at quadratic order in the field $\pi$, but at all higher orders too. Phrased in terms of $\{n_h,n_A,n_\pi\}$ as defined in section \ref{sec-cubic} this means that all pure scalar interactions from $\{0,0,3\}$ to $\{0,0,n_\pi \to \infty\}$ associated with suppression scales $\Lambda_5$ to $\Lambda_3$ are eliminated up to total derivatives. Since this tuning also results in $A$-dependent $\Lambda_4$ suppressed terms vanishing, this raises the cutoff of the theory to $\Lambda_3$.

\item {\it Scalar-tensor mixing: }
The mixing with $h$ takes the form (ignoring terms with more than one $h$ since these will have higher suppression scales and hence vanish in the decoupling limit)
\be
\cL_{h\pi} = m^2 M_{\text{eff}}^2\, h_{\mn} \frac{1}{2} \sum_{n=0}^m \frac{1}{n! (m-n)!} \frac{(D-n-1)!}{(D-m-1)!} X^{\mn}_{(n)}(\pi),
\ee
where $X_{(n)}(\pi)$ are the conserved tensors formed from second derivatives of $\pi$ as defined in \eqref{X-def}. If the $f$ metric is fixed to be flat as in dRGT massive gravity, there are no further scalar-tensor mixing terms. However, as $f$ is dynamical there will be further terms involving the mixing of $l$ with $\pi$.
Such terms take the form 
\be \label{dRGT-second-mixing}
\cL_{l\pi} = m^2 M_{\text{eff}}^2\, l_{\mn}(Y) \frac{1}{2} \sum_{n=0}^m \frac{1}{n! (m-n)!} \frac{(D-n-1)!}{(D-m)!} ((D-n) \eta^{\mn} \cL^{\mathrm{TD}}_{(n)} - (D-m) X^{\mn}_{(n)}(\pi)),
\ee
where $l_{\mn}(Y) = l_{\mn}(x + A)$, i.e. we have re-summed the expansion of the metric around $x^\mu$ into the form $l_{\mn}(Y)$ - see appendix \ref{appendix-nonlocal}.

At lowest order in the scalar-tensor mixing we can simplify these re-summed expressions to an even simpler form. The mixing \eqref{dRGT-second-mixing} then becomes
\begin{eqnarray}
\nn {\cal L}_{l \pi} &=& m^2 M_{\text{eff}}^2 \left\{ l_{\mn}(x) \left[ (\ldots) \eta^{\mn} {\cal L}_{(1)}^{\text{TD}} - (\ldots) X_{(1)}^{\mn}(\pi) \right] + \pi^\alpha \pa_\alpha l_{\mn} \left[ (\ldots) \eta^{\mn} {\cal L}_{(0)}^{\text{TD}} - (\ldots) X_{(0)}^{\mn}(\pi) \right] \right\}\\
&\to & m^2 M_{\text{eff}}^2 \left\{ 0 \times \bar l^{\alpha}_{\alpha} {\cal L}_{(1)}^{\text{TD}} - (D-2) c \bar l_{\mn} X_{(1)}^{\mn}(\pi) - (D-2) c c_l \pi X_{\mu,(1)}^{\mu}(\pi) \right\},
\end{eqnarray}
%
where $\to$ denotes the effect of the conformal transformation $l \to \bar l + c_l \pi$, and $c$ is given by \eqref{cdRGT}. The $\bar l^\alpha_\alpha$ term actually vanishes by itself (but otherwise would be removed by gauge-fixing), $\bar l_{\mn} X_{(n)}^{\mn}(\pi)$ cancels against terms generated from the Ricci scalar $R[l]$ via the conformal transformation (see section \ref{sec-gold}) and $\pi X_{\mu,(1)}^{\mu}(\pi)$ gives a contribution to a kinetic term for $\pi$.

\item {\it Demixing and ghost-free scalar interactions:}
From the above we see that transformations 
\begin{align}
h_{\mn} &\to \bar h_{\mn} - c \frac{m^2 M_{\text{eff}}^2}{M_g^2} \eta_{\mn} \pi, 
&l_{\mn} &\to  \bar l_{\mn} + c \frac{m^2 M_{\text{eff}}^2}{M_f^2} \eta_{\mn} \pi 
\label{doubleconf}
\end{align}
will eliminate scalar-tensor mixing at quadratic order in the fields, where we have defined the coefficient $c$ to be 
\be \label{cdRGT}
c \equiv \frac{1}{2 (m-1)!} \frac{(D-3)!}{(D-m-1)!}.
\ee
 This means we obtain a scalar kinetic term \eqref{dRGT-kin} 
\be
- \frac{1}{2} m^4 M^2_{\text{eff}} (D-1)(D-2) c^2 \pi\Box\pi
\ee 
and the structure of the interaction term ensures there are no further dangerous interactions at quadratic order in $\pi$, i.e. Fierz-Pauli tuning is upheld. The fact that this tuning generalises to eliminate all pure-scalar higher order interactions ensures that the decoupling scale of the theory is $\Lambda_3$.  One could now write out higher-order interaction terms in the decoupling limit to some given order in the $\pi$ and show they are ghost-free up to this given order along the lines sketched out in \cite{deRham:2010ik}. However, the infinite derivative expansion of $l_{\mn}(Y)$ means that for this purpose it is more economical to analyse interactions in a re-summed format, cf. \cite{Fasiello:2013woa}.

\item {\it Re-summing interactions: }
The infinite derivative expansion of $l_{\mn}(Y)$ in the scalar-tensor mixing involving $l$ creates a difficulty in analysing the full decoupling limit of the theory, rather than the same limit to some finite order in $\pi$. Following \cite{Fasiello:2013woa} it is therefore advisable to re-sum $l_{\mn}(Y)$  After a little algebra and using the techniques discussed in appendix \ref{appendix-nonlocal}, the mixing terms \eqref{dRGT-second-mixing} become
\be
\cL_{l\pi} = \frac{m^2 M_{\text{eff}}^2}{2(m-1)!} l_{\mn}(Y) \partial_{\alpha}Y^{\nu} X^{Y \alpha \mu}_{(m-1)}.
\ee
In particular, in terms of the field $\phi$ related to $\pi$ through $x + \partial \phi = (x + \partial \pi)^{-1}$, we find
\be \label{resummed-mixing}
\cL_{l \phi} = m^2 M_{\text{eff}}^2\, l_{\mn}(x) \frac{1}{2} \sum_{n=0}^{D-m} \frac{1}{n! (m-1)!} \frac{(D-n-1)!}{(D-m-n)!} X^{\mn}_{(n)}(\phi).
\ee
$\pi$ and $\phi$ are essentially the \St scalars corresponding to choosing a \St field $Y$ mapping from the $f$-site to the $g$-site and vice versa respectively. In other words, by \Sting the scalar-tensor mixing involving $l$ with $Y_{(g f)}$ and that involving $h$ with $Y_{(g f)}$, we are essentially employing a hybrid approach, since we use two \St fields to \St a single interaction term, although they are of course not independent, but related via $Y_{(g f)}= Y^{-1}_{(f g)}$. Having expressed the scalar-tensor mixing in this hybrid format, the conformal transformations demixing at quadratic order in the fields are now
\begin{align} 
h_{\mn} &\to  \bar h_{\mn} - c \frac{m^2 M_{\text{eff}}^2}{M_g^2} \eta_{\mn} \pi,
&l_{\mn} &\to  \bar l_{\mn} - c \frac{m^2 M_{\text{eff}}^2}{M_f^2} \eta_{\mn} \phi,
\label{doubleconfdem}
\end{align}
where $c$ is defined as in \eqref{cdRGT}.  This will remove the scalar tensor mixing (at lowest order) and give scalar kinetic terms
\be \label{dRGT-kin}
-\frac{1}{2} m^4 M_{eff}^4 (D-1)(D-2) c^2(M_g^{-2} \pi \Box \pi + M_f^{-2} \phi \Box \phi).
\ee
Note that we can express $\phi$ in terms of an infinite series in $\pi$ (see appendix \ref{appendix-nonlocal}) to find that, again at lowest order, $\phi = -\pi$. This explains why both conformal transformations \eqref{doubleconf} and \eqref{doubleconfdem} successfully demix scalar and tensor terms at lowest order.

Having employed the above re-summation procedure, investigating higher order mixing terms becomes more straightforward.
The canonical transformations \eqref{doubleconfdem} will introduce terms of the form
\be
\sim \sum_n^{m} \pi X^\mu_{\mu,(n)} (\pi) + \sum_n^{D-m} \phi X^\mu_{\mu,(n)} (\phi),
\ee
where
\be
\pi X^{\mu}_{(n) \mu}(\pi) = \pi \partial_{\mu_1}(\delta^{\mu \mu_1 \hdots \mu_n}_{[\mu \nu_1 \hdots \nu_n]} \pi^{\nu_1} \pi_{\mu_2}^{\nu_2} \hdots \pi_{\mu_n}^{\nu_n}) = (D-n) \pi \partial_{\mu_1}(\delta^{\mu_1 \hdots \mu_n}_{[\nu_1 \hdots \nu_n]} \pi^{\nu_1} \pi_{\mu_2}^{\nu_2} \hdots \pi_{\mu_n}^{\nu_n}),
\ee
and analogously for $\phi X^\mu_\mu (\phi)$. As discussed in appendix \ref{appendix-TD}, upon integration by parts this becomes $\pi_\mu \pi_\nu X^{\mn}_{(n-1)}(\pi)$, i.e. the ($n+1$)-th Galileon term, and so these will not introduce any new ghosts. For a full treatment of the Hassan-Rosen bigravity decoupling limit we refer to \cite{Fasiello:2013woa}. As far as this paper is concerned, there is an important point of principle here, however.  The effective field theory description outlined here is very powerful in understanding what the propagating degrees of freedom of a general $N$ interacting spin-2 field theory are, how to identify the associated energy and cutoff scales etc. However, if we want to make statements about interaction terms to all orders, the inherent non-locality introduced by the \St expansion of a spin-2 field $h(x+A)$ means it becomes necessary to perform a re-summation as outlined here and discussed further in appendix \ref{appendix-nonlocal}.
\end{enumerate}

\section{Example III: A quartic EBI-like interaction} \label{sec-EBI4}

The previous two examples are explicitly bimetric, so we would like to finish the example section(s) of this paper with a higher order example. By using the arguments presented throughout this paper we here wish to show that $N$-metric generalisations of the bimetric EBI models considered in section \ref{sec-EBI} also always contain ghost-like instabilities at quadratic order in the Goldstone fields $\pi$. As such consider an EBI-like theory with four dynamical spin-2 fields, each equipped with its own Ricci scalar, and an interaction term 
\be
{\cal S}_{int} = \int d^4 x \sqrt{g_{(0)}} g^{\mu\nu}_{(0)} g_{\nu \alpha}^{(1)} g^{\alpha \beta}_{(2)} g_{\beta \mu}^{(3)}.
\ee
In order to restore gauge-invariance we introduce three \St fields along the lines discussed in sections \ref{sec-link} and \ref{sec-gold}, which we label $\pi_{(j)}$. We can demix tensor and scalar contributions via
\be
{\bf h}_{(i)} \to {\bf \bar{h}}_{(i)} + \sum_j^{j=1,2,3} c_{(j)} \pi_{(ji)} {\bm \eta} \qquad \forall i.
\ee
where ${\bf h}_{(i)} = {\bf g}_{(i)}  - {\bm \eta}$. Note that, when we use the shorthand $\pi_{(3)}$, we really mean $\pi_{(3,0)}$, but in this specific example the extra index is redundant, since we have mapped all spin-2 fields to the 0-site. 
Now we recall that, even before demixing and as shown in equation \eqref{GenFPA}, the interaction term needs to satisfy a generalised Fierz-Pauli condition at quadratic order to avoid ghost-like degrees of freedom. The pure-scalar action prior to demixing is  
\begin{eqnarray}
\nn {\cal L}_{\text{int, pre-mixing}}^{\text{pure-scalar}} &=& \eta^{\mu\nu} \left( \delta_\mu^\beta + \pi_\mu^{\beta,(1)}  \right)  \left( \delta_\alpha^\gamma + \pi_\alpha^{\gamma,(1)}  \right) \eta_{\beta \gamma}
\left( \delta^\alpha_\lambda + \pi^\alpha_{\lambda,(2)}  \right)  \left( \delta^\sigma_\rho + \pi^\sigma_{\rho,(2)}  \right) \\
&\cdot & \eta^{\lambda \rho}
\left( \delta_\sigma^\chi + \pi_\sigma^{\chi,(3)}  \right)  \left( \delta_\nu^\kappa + \pi_\nu^{\kappa,(3)}  \right) \eta_{\chi \kappa}. \label{4spin2St}
\end{eqnarray}
At quadratic order in the fields this reduces to
\be
{\cal L}_{\text{int, pre-mixing}}^{\text{pure-scalar}}  = 4 + 2\sum_i \pi_\alpha^{\alpha,(i)} + \sum_i \pi^{\mu\nu}_{(i)} \pi_{\mu\nu}^{(i)} + 4 \sum_{i,j}^{i \neq j} \pi^{\mu\nu}_{(i)} \pi_{\mu\nu}^{(j)} + {\cal O}(\pi^3).
\ee
This maximally violates the generalised Fierz-Pauli condition at quadratic order, i.e. none of the dangerous self-interaction and cross-terms vanish up to total derivatives (cf. equation \eqref{GenFPA})
\be
a_{(ij)} + b_{(ij)} \neq 0   \quad\quad \forall i,j.
\ee
Consequently ghost-like degrees of freedom enter at quadratic order in the fields and it is straightforward to see how this generalises to an arbitrary $N$ spin-2 field generalisation of EBI bigravity. The pattern of equation \eqref{4spin2St} will be reproduced and all possible dangerous terms will be present, since there is no means for these to cancel given the interaction term.

\section{Summary and conclusions} \label{sec-conc}

Throughout this paper we have generalised and extended the `Effective field theory for massive gravitons' developed by \cite{ArkaniHamed:2002sp} to an `Effective field theory for interacting spin-2 fields' in light of recent progress in massive gravity - in particular the constructions of dRGT gravity \cite{deRham:2010ik,deRham:2010kj}, Hassan-Rosen bigravity \cite{Hassan:2011tf,Hassan:2011zd,Hassan:2011ea} and Hinterbichler-Rosen multi-gravity  \cite{Hinterbichler:2012cn}. The key results of this paper are as follows
\begin{itemize}
\item We discussed how the \St trick can be used to explicitly restore gauge invariance in interacting spin-2 field theories. While in bigravity there are two different ways to do so, which already lead to very interesting duality theorems \cite{deRham:2013hsa}, in multi-gravity there is a multitude of possible approaches. We investigated these and showed that they are also ultimately all dual to one another. In particular we established a minimal approach that maximally simplifies interaction terms and makes the propagating degrees of freedom explicit.
 
\item We investigated generalised scalar-tensor mixing in multi-spin-2 field theories and how to remove this, bringing out various subtleties not present in massive and bi-gravity. Especially intriguing is a generalised `Fierz-Pauli' condition one can derive for the pure-scalar interactions at quadratic order, even before any demixing has taken place.
Gauge fixing issues relevant in eliminating scalar-tensor mixing at lowest order in this context are also discussed.

\item Another phenomenon absent in massive and bi-gravity, but generic in multi-gravity, is that of scalar mixing at quadratic order in the fields. We investigated this mixing and showed how it changes some of the propagating degrees of freedom of the theory into linear combinations of the \St scalars.

\item Scalar mixing at quadratic order has important consequences for the structure of higher order interactions, in particular the decoupling limit, as well. We presented results showing that this limit is qualitatively changed in interacting spin-2 field theories. More specifically they generically possess a strong coupling scale that is lower than that in analogous bigravity models (cf. related results in \cite{deRham:2013awa}). Also only some of the interactions surviving in the decoupling limit in massive and bi-gravity contribute to this new lowest scale in multi-gravity theories (i.e. multi-gravity models break the degeneracy between different terms contributing to the same scale\footnote{Note that there is an important caveat here: Depending on the exact nature of the theory, this degeneracy may be strongly or weakly broken. If the breaking is only weak, then all previously relevant interactions will have scales clustered together very closely. In other words, strictly discarding all terms but those suppressed by the lowest scale may not produce the correct low-energy physics, since model-dependently this may amount to discarding terms that contribute at roughly the same scale as those which are kept.}), changing the physics of the decoupling limit. 
 
\item Finally we applied some of the machinery developed here to a few concrete model examples. Of particular interest is our proof that EBI (Eddington-Born-Infeld) bigravity has a ghost-like instability, as well as generalisations of this theory to higher-order interactions. We also showed how Hassan-Rosen bigravity fits into our formalism.
\end{itemize}  
Theories of interacting spin-2 fields stand to promise several new and intriguing features absent in theories with up to two spin-2 fields. Various tantalizing prospects are on the horizon: 
Higher-dimensional theories of gravity where the extra dimension(s) are discretised are intrinsically linked to particular interacting spin-2 field theories where the theory graph forms one big loop with nearest site interactions \cite{ArkaniHamed:2001ca,ArkaniHamed:2002sp,ArkaniHamed:2003vb,Schwartz:2003vj,deRham:2013awa}. We will investigate this and the general phenomenology of loops further in \cite{loops}. In this context one may also wonder whether the large $N$ limit of multi-gravity theories possesses interesting generic features on top of the ones discovered so far. Perhaps new ghost-free interaction terms appear in the context of several fields, that are absent whenever just one dynamical spin-2 field is present. Another interesting topic for further investigation will be to establish whether consistent multi-gravity theories permit the existence of superluminalities - cf. discussions in \cite{Burrage:2011cr,Gruzinov:2011sq,Deser:2012qx,Deser:2013eua}.
Finally one should stress that both the `Effective field theory of massive gravitons' as well as its extension presented here exclusively deal with constructing and analysing `potential' interaction terms, i.e. terms at zeroth order in derivatives. Investigating whether derivative interaction terms other than Ricci scalars are admissible along the lines of \cite{Hinterbichler:2013eza,Kimura:2013ika,Folkerts:2011ev} may lead to exciting new avenues in constructing theories of one-to-N spin-2 fields.  We hope that the toolbox developed throughout this paper will prove useful in tackling some of these outstanding issues in the future.
\\

\noindent {\bf Acknowledgements: } We would like to thank Claudia de Rham, Kurt Hinterbichler, Macarena Lagos, Angnis Schmidt-May and Andrew Tolley for very useful discussions, correspondence and comments on a draft of this paper. We also thank Sigurd K. N\ae ss and Nemanja Kaloper for very helpful discussions at various stages throughout this work. JN, JHCS and PGF were supported by the STFC, BIPAC and the Oxford Martin School.

\appendix

\section{Appendix I: Total derivatives and ghost-free interaction terms}
\label{appendix-TD}


dRGT and Hassan-Rosen bigravity theories, that raise the cutoff to $\Lambda_3$, do so by turning the pure-scalar interactions into total derivative combinations. At $n$-th order in the Goldstone fields $\pi$ a unique total derivative can be formed out of $\pi_{\alpha\beta}$ \cite{Nicolis:2008in}
\be \label{TD}
{\cal L}^{\text{TD}}_{(n)} = \delta^{\alpha_1 \ldots \alpha_n}_{[\beta_1 \ldots \beta_n]} \pi^{\beta_1}_{\alpha_1} \ldots \pi^{\beta_n}_{\alpha_n},
\ee
where, in $D$ dimensions, we have defined a tensor $\delta^{\alpha_1 \ldots \alpha_n}_{[\beta_1 \ldots \beta_n]}$ separately anti-symmetric in its indices $\alpha_1 \ldots \alpha_n$ and $\beta_1 \ldots \beta_n$ in terms of the totally antisymmetric tensor $\varepsilon$ via
\be \label{delta-def}
\delta^{\alpha_1 \ldots \alpha_n}_{[\beta_1 \ldots \beta_n]}  \equiv \frac{1}{(D-n)!}\varepsilon^{\alpha_1 \ldots \alpha_n \lambda_1 \ldots \lambda_{D-n}}
\varepsilon_{\beta_1 \ldots \beta_n \lambda_1 \ldots \lambda_{D-n}}.
\ee
Note that the antisymmetry enforced by this expression means that ${\cal L}^{\text{TD}}_{(n)} (\pi^{(i)},\ldots,\pi^{(i)} )$ is a total derivative as well as ${\cal L}^{\text{TD}}_{(n)} (\pi^{(i)},\ldots,\pi^{(j)} )$, i.e. each $\pi$ appearing in \eqref{TD} can be any of the Goldstone fields in the theory. It is easy to prove that ${\cal L}^{\text{TD}}_{(n)} $ is a total derivative by integrating by parts once and finding
\be
- \delta^{\alpha_1 \ldots \alpha_n}_{[\beta_1 \ldots \beta_n]} \pi_{\alpha_1} \pa^{\beta_1} \left(\pi^{\beta_2}_{\alpha_2} \ldots \pi^{\beta_n}_{\alpha_n}\right) = 0,
\ee
which vanishes, since upon expansion the $\pa^{\beta_1} \left(\pi^{\beta_2}_{\alpha_2} \ldots \pi^{\beta_n}_{\alpha_n}\right)$ term yields contributions that are symmetric under exchange of $\beta_1$ and one other $\beta_i$, whereas the $\delta^{\alpha_1 \ldots \alpha_n}_{[\beta_1 \ldots \beta_n]}$ tensor enforces anti-symmetry in the same two indices.

Now it was noticed by \cite{Deffayet:2011gz} that the flat-space Galilean invariant contributions to the Lagrangian, which do not have Ostrogradski ghosts by design, can be written in one of three forms, which are all related by total derivatives
\bea
{\cal L}^{\text{Gal,1}}_{(n+2)} &=& \delta^{\alpha_1 \ldots \alpha_{n+1}}_{[\beta_1 \ldots \beta_{n+1}]} \pi^{\beta_1}_{\alpha_1} \ldots \pi^{\beta_n}_{\alpha_n} \pi_{\alpha_{n+1}} \pi^{\beta_{n+1}}, \\
{\cal L}^{\text{Gal,2}}_{(n+2)} &=& \delta^{\alpha_1 \ldots \alpha_{n}}_{[\beta_1 \ldots \beta_{n}]} \pi^{\beta_2}_{\alpha_2} \ldots \pi^{\beta_n}_{\alpha_n} \pi_{\alpha_{1}} \pi_{\gamma} \pi^{\beta_{1} \gamma}, \\
{\cal L}^{\text{Gal,3}}_{(n+2)} &=& \delta^{\alpha_1 \ldots \alpha_{n}}_{[\beta_1 \ldots \beta_{n}]} \pi^{\beta_1}_{\alpha_1} \ldots \pi^{\beta_n}_{\alpha_n} \pi_{\gamma} \pi^{\gamma}.
\eea
Here $(n+2)$ counts the order of the term in the $\pi^{(i)}$.  ${\cal L}^{\text{Gal}}_{(1)} = \pi$ and ${\cal L}^{\text{Gal}}_{(2)} = \pi_\mu \pi^\mu$. These expressions become useful when analysing the properties of the scalar field Lagrangian arising via scalar-tensor mixing in theories where the scalar Lagrangian pre-mixing is a total derivative.  

If the pure scalar Lagrangian pre-mixing with the metric is a total derivative, then we have via \eqref{TD} that
\be
{\cal L}_{\text{pre-mixing}} =\sum_n c_n \delta^{\alpha_1 \ldots \alpha_n}_{[\beta_1 \ldots \beta_n]} \pi^{\beta_1}_{\alpha_1} \ldots \pi^{\beta_n}_{\alpha_n}.
\ee 
where the $c_n$ are some constant coefficients. But because of the \St expansion of the metric this will only be the first term in the series $\frac{1}{4} \eta^{\mn}{\cal L}_{\text{pre-mixing}} ( \eta_{\mn} + h_{\mn} + \ldots    )$. Upon conformally transforming the metric perturbation $h_{\mu\nu}^{(i)} \to \bar h_{\mu\nu}^{(i)} + a \pi_{(i)}\eta_{\mu\nu}$, at lowest order in the mixing we get a new contribution to the scalar field Lagrangian of the form
\be
{\cal L}_{\text{mixing,1}} =\pi \delta^{\alpha_1 \ldots \alpha_n}_{[\beta_1 \ldots \beta_n]} \pi^{\beta_1}_{\alpha_1} \ldots \pi^{\beta_n}_{\alpha_n},
\ee
which, upon integrating by parts, becomes
\be
{\cal L}_{\text{mixing,1}} = \pi \delta^{\alpha_1 \ldots \alpha_n}_{[\beta_1 \ldots \beta_n]} \pi^{\beta_1}_{\alpha_1} \ldots \pi^{\beta_n}_{\alpha_n} \to - \pi_\mu \pi^\mu \delta^{\alpha_2 \ldots \alpha_n}_{[\beta_2 \ldots \beta_n]} \pi^{\beta_2}_{\alpha_2} \ldots \pi^{\beta_n}_{\alpha_n} \propto {\cal L}^{\text{Gal,3}}_{(n+1)}. 
\ee
Note that this is also trivially true at next order in the mixing\footnote{Terms like these come from taking into account the \St expansion of a metric $g(x+A) = g(x) + A \pa g(x) + \ldots$. Mixing with the second term turns into `mixing' with $\pi^\mu \pi_\mu$ post-conformal transformation.}
\be
{\cal L}_{\text{mixing,2}} = \pi_\mu \pi^\mu \delta^{\alpha_1 \ldots \alpha_1}_{[\beta_2 \ldots \beta_n]} \pi^{\beta_1}_{\alpha_1} \ldots \pi^{\beta_n}_{\alpha_n} \propto {\cal L}^{\text{Gal,3}}_{(n+2)}. 
\ee
At higher orders in the mixing this is no longer guaranteed and depends on the form of the interaction term (again highlighting the importance of the square root in the ghost-free symmetric polynomial construction). 

The above argument can also be phrased in terms of the $X_{(n)}^{\mn}$, which we recall we defined in \eqref{X-def} via
\be 
X_{(n)}^{\mn}(\pi_1,\hdots,\pi_n) = \delta^{\mu \mu_1 \hdots \mu_n}_{[\nu \nu_1 \hdots \nu_n]} \pi_{\mu_1, (1)}^{\nu_1} \hdots \pi_{\mu_n,(n)}^{\nu_n}.
\ee
If the general scalar-tensor mixing pre-conformal transformation is $h_{\mn} X_{(n)}^{\mn}(\pi)$, then upon sending $h_{\mu\nu} \to \bar h_{\mu\nu} + a \pi \eta_{\mu\nu}$, we obtain a new scalar contribution
\be
\pi X^{\mu}_{\mu,(n)}(\pi) = \pi \delta^{\mu \mu_1 \hdots \mu_n}_{[\mu \nu_1 \hdots \nu_n]} \pi_{\mu_1}^{\nu_1} \hdots \pi_{\mu_n}^{\nu_n} = n! \pi \delta^{\mu_1 \hdots \mu_n}_{[\nu_1 \hdots \nu_n]} \pi_{\mu_1}^{\nu_1} \hdots \pi_{\mu_n}^{\nu_n} \propto {\cal L}^{\text{Gal,3}}_{(n+1)}.
\ee
So again this mixing leads to ghost-free scalar interaction terms at all orders $n$. It is straightforward to generalise this to conformal transformation involving more than one scalar \dof in interacting theories with mixing cross-terms involving $X_{(n)}^{\mn}(\pi_1,\hdots,\pi_n)$, where not all the $\pi_{(i)}$ are identical. Also note that we can express $X_{(n)}^{\mn}$ in terms of ${\cal L}^{\text{TD}}_{(n)}$ via \cite{deRham:2010kj,Hinterbichler:2011tt}
\be
X_{(n)}^{\mn}(\pi) = \frac{1}{n+1} \frac{\delta}{\delta \pi_{\mn}} {\cal L}^{\text{TD}}_{(n+1)}(\pi).
\ee

\section{Appendix II: Non-local scalar-tensor interactions and demixing}
\label{appendix-nonlocal}

We discussed scalar tensor mixing terms at quadratic order in the fields in section \ref{sec-gold} and we discussed the structure of higher order mixing terms in section \ref{sec-cubic}. One interesting issue that appears for higher order interaction terms is that of non-locality. Generic scalar-tensor interactions generated via \Sting interaction terms will be of the form
\be
p(\pi)^{\mn} (h_{\mn} + \pi^\alpha \partial_\alpha h_{\mn} + \frac{1}{2} \pi^\alpha \pi^\beta \partial_{\alpha \beta} h_{\mn} + \hdots),
\ee
where $p(\pi)^{\mn}$ is some function of $\pi^{\mn}$. This appears to be non-local due to the infinite number of derivatives acting on the tensor $h$. Na\"ively one may think that this automatically leads to higher order equations of motion, and hence ghosts via Ostrogradski's theorem. 
However, the non-local interaction Lagrangian generated in this fashion is degenerate, since there is no highest order derivative term, and one consequently cannot straightforwardly apply Ostrogradski's theorem here. It is therefore worthwhile to re-sum terms in the expansion in order to produce a local theory to which Ostrogradski's theorem can be applied (cf. \cite{Woodard:2006nt,Chen:2012au,Fasiello:2013woa} and our discussion of this point in section \ref{sec-dRGT}). 

The non-locality stems from the expansion of the original tensor field once the \St fields have been introduced. We recall that the introduction of \St fields essentially amounted to a co-ordinate transformation, where $Y^\mu = x^\mu + A^\mu$ and
\be
h_{\mn}(Y) = h_{\mn}(x + \partial \pi) = h_{\mn} + \pi^\alpha \partial_\alpha h_{\mn} + \frac{1}{2} \pi^\alpha \pi^\beta \partial_{\alpha \beta} h_{\mn} + \hdots,
\ee
where we have dropped contributions from the \St vector $A^\mu$. Thus, if we re-express the rest of the scalar-tensor action in terms of coordinates $Y^\mu(x)$, instead of $x^\mu(x)$, we can remove the non-locality.
Let us first analyse the way in which this works for a general $N$-metric coupling, in which \St fields are introduced to bring everything to site 1:
\be
\mathcal{S}_\mathrm{int} = \int d^Dx\, f(g_{(1)},g_{(2)},\dots,g_{(N)}) \to \int d^Dx\, f(g_{(1)},g_{(2)} \circ Y_2,\dots,g_{(N)} \circ Y_N).
\ee
In a theory where the least suppressed interaction terms scale as $\Lambda_3$, if we now go to the decoupling limit, the interactions between the tensors and the \St fields take the form\footnote{Of course to get to the decoupling limit one would have to unpack $Y$ into vector and scalar fields as described above, however after doing that, we can always repackage them back into $Y$.}
\be
\int d^Dx\, \left( h^{(1)}_{\mn}(x) f_1^{\mn} (\partial Y_2(x),\dots,\partial Y_N(x)) + \sum_{i=2}^N h^{(i)}_{\mn}(Y_i(x)) f_i^{\mn} (\partial Y_2(x),\dots,\partial Y_N(x)) \right).
\ee
The first term is straightforward to deal with since all the fields explicitly depend on $x$, so let us focus on one term of the remainder, which can be rewritten
\be
\int d^Dx\, h^{(i)}_{\mn}(Y_i(x)) f_i^{\mn} (\partial Y_2(x),\dots,\partial Y_N(x)) = \int d^DY_i\, h^{(i)}_{\mn}(Y_i) F_i^{\mn}(Y_i,x), \label{non-local-interaction}
\ee
where in the new interaction term we have
\be
F_i^{\mn}(Y_i,x) = \left| \frac{\partial x}{\partial Y_i} \right| f_i^{\mn} \left( \frac{\partial Y_2}{\partial Y_i} \frac{\partial Y_i}{\partial x},\dots,1 \frac{\partial Y_i}{\partial x},\dots,\frac{\partial Y_N}{\partial Y_i} \frac{\partial Y_i}{\partial x} \right). \label{non-local-interaction-new-interaction}
\ee
Now if the functional form of $f_i$ is such that the determinental prefactor on the RHS of \ref{non-local-interaction-new-interaction} can eliminate the $\frac{\partial Y_i}{\partial x}$'s appearing within $f_i$, then the explicit dependence on $x$ is eliminated and we have $F_i^{\mn} = F_i^{\mn}(Y_i)$. This means we can then rename the integration variable on the RHS of \ref{non-local-interaction} $Y_i \to x$ and add it back to the rest of the action.

We do not here analyse what forms of $f_i$ have the above property, but demonstrate the precise way in which this works in a specific context relevant to the ghost-free bigravity theory discussed in section \ref{sec-dRGT}. Similar to the $X_{(n)}$ tensors mentioned above, we may define
\be
X^{Y \mu}_{(n) \nu} = \delta^{\mu \mu_1 \ldots \mu_n}_{[\nu \nu_1 \ldots \nu_n]} \prod_{i=1}^{n} \partial_{\mu_i}Y^{\nu_i}.
\ee
Working in $D$ dimensions we can re-express this as
\begin{eqnarray}
\nn X^{Y \mu}_{(n) \nu} &=& \left| \frac{\partial Y}{\partial x} \right| \frac{n!}{(D-n-1)!} \frac{\partial x^\alpha}{\partial Y^\nu} \delta^{\mu \mu_1 \ldots \mu_{D-n-1}}_{[\alpha \nu_1 \ldots \nu_{D-n-1}]} \prod_{i=1}^{D-n-1} \frac{\partial x^{\nu_i}}{\partial Y^{\mu_i}} \\ &=& \left| \frac{\partial Y}{\partial x} \right| \frac{n!}{(D-n-1)!} \frac{\partial x^\alpha}{\partial Y^\nu} X^{Y^{-1} \mu}_{(D-n-1) \alpha},
\end{eqnarray}
where we have defined $X^{Y^{-1} \mu}_{(n) \nu}$ in the process. Thus a bimetric interaction with $f^\mu_\nu = \partial_\nu Y^\alpha X^{Y \mu}_{(n) \alpha}$ has $F^\mu_\nu \propto X^{Y^{-1} \mu}_{(D-n-1) \nu}$, which is a function purely of $Y$.

Upon renaming $Y \to x$, one has $\frac{\partial x}{\partial Y} \to \frac{\partial Y^{-1}}{\partial x}$. Neglecting the \St vector $A^\mu$, $Y$ satisfies $\partial_{[\mu} Y_{\nu]} = 0$, and thus $Y^{-1}$ will also satisfy this relation; since $Y$ depends on just one scalar d.o.f., $Y^{-1}$ will also. These two facts imply that we may write $\partial_\mu Y^{-1}_{\nu} = \eta_{\mn} + \partial_{\mn}\phi$, where $\phi$ is related to $\pi$ through some non-local transformation, which we specify below. Then we have
\be
X^{Y^{-1} \mn}_{(n)} = \sum_{i=0}^{n} \binom{n}{i} \frac{(D-i-1)!}{(D-n-1)!} X^{\mn}_{(i)}(\phi),
\ee
and hence we can express the scalar-tensor mixing in terms of $\phi$ as
\be
\cL_{h\phi} = h_{\mn}^{(i)} \sum_n X^{\mn,(i)}_{(n)}(\phi),
\ee
which is precisely of the form discussed in section \ref{sec-gold} that can straightforwardly be removed at linear order via a linearised conformal transformation.

For completeness we present below the form of the above-mentioned transformation between $\phi$ and $\pi$, which is derived using the fact that $Y^{-1}(Y(x)) = x$. Defining the differential operators
\be
D_{(n)} = \frac{1}{n!} \pi^{\mu_1} \dots \pi^{\mu_n} \partial_{\mu_1 \dots \mu_n} \qquad \text{and} \qquad \bar{D}_{(n)} = D_{(n)} - \frac{1}{2}D_{(n-1)}D_{(1)}
\ee
we have
\begin{align}
\nn \phi &= - \pi + \sum_{n=2}^{\infty} \left[ \bar{D}_{(n-1)}\pi + \sum_{k=1}^{n-3} \left\{ (-1)^k \sum_{i_1=k+1}^{n-2} D_{(n-1-i_1)} \dots \sum_{i_k=2}^{i_{k-1}-1} D_{(i_{k-1}-i_k)} \bar{D}_{(i_k)}\pi \right\} \right] \\
&= -\pi + \frac{1}{2}\pi^\mu \pi_\mu - \frac{1}{2} \pi^\mu \pi^\nu \pi_{\mn} + \frac{1}{2} \pi^\mu \pi_\nu \pi^\nu_\lambda \pi^\lambda_\mu + \frac{1}{6} \pi^\mu \pi^\nu \pi^\lambda \pi_{\mn \lambda} + \dots .
\end{align}

\section{Appendix III: No kinetic terms and a general curved background}
\label{appendix-curved}

In section \ref{sec-gold} we discussed that the \St scalars $\pi$ only acquire a kinetic term via mixing with tensor modes. As pointed out by \cite{ArkaniHamed:2002sp}, while this is true around some flat background, this is not the case in a general curved background. Focussing on the self-interactions for each scalar for the time being (effectively setting $a_{(ij)} = b_{(ij)} = 0$, whenever $i \neq j$), the quadratic order action for a Goldstone scalar $\pi$ will be of the following form 
\be\label{curvedS}
{\cal S}_{int,2}^{curved} \to \int d^4 x \sqrt{g} \left(  \nabla_\mu\nabla_\nu \pi \nabla^\mu \nabla^\nu \pi - \nabla_\mu \nabla^\mu \pi \nabla_\nu \nabla^\nu \pi   \right)  \to \int d^4 x \sqrt{g} \pa_\mu \pi \left[\nabla^\mu,\nabla^\nu \right] \pa_\nu \pi, 
\ee
as long as the equivalent of Fierz-Pauli tuning takes place ($a_{(ii)} + b_{(ii)} = 0$) and where we have suppressed all site-label indices $(ik)$. In a curved background the field therefore already has a kinetic term generated by curvature, even before mixing with tensor modes. Mixing will still generate a contribution to the effective kinetic term, but is no longer the sole contributor. If one now also allows cross-interactions between different Goldstone bosons, as long as the generalised Fierz-Pauli tuning is satisfied ($a_{(ij)} + b_{(ij)} = 0$) the curvature generated terms will be of the form\footnote{If one was to require the absence of cross-terms for $i \neq j$, this would be equivalent to setting $a_{(ij)} = b_{(ij)} = 0$.} 
\be
\int d^4 x \sqrt{g} \pi_\mu^{(ik)} \pi_\nu^{(jk)} R^{\mu\nu}.
\ee
Note that the derivatives on the Goldstone $\pi$ in action \eqref{curvedS} come from the \St co-ordinate transformation and from introducing the fake $U(1)$. As far as the fake $U(1)$ is concerned, the background is irrelevant because the derivative acts on a scalar. The second derivative comes from the \St coordinate transformation
\be
g_{\mu\nu}^{}(x) \to \pa_\mu Y^\alpha  \pa_\nu Y^\beta  g_{\alpha\beta}^{}(Y(x)).
\ee
We can understand how this is `covariantised' by going to the full expansion of a St\"uckelberged metric around a curved background
\begin{align}
\nn \pa_\mu Y^\alpha  \pa_\nu Y^\beta  g_{\alpha\beta}(Y(x)) &= (\delta^\alpha_\mu + \pa_\mu A^\alpha) (\delta^\beta_\nu + \pa_\nu A^\beta) (g_{\alpha \beta} + A^\sigma \pa_\sigma g_{\alpha \beta} + \hdots) \\
\nn &= g_{\mn} + \pa_\mu A^\alpha g_{\alpha \nu} + \pa_\nu A^\beta g_{\mu \beta} + A^\sigma \pa_\sigma g_{\mn} + \hdots \\
\nn &= g_{\mn} + \cL_A g_{\mn} + \hdots \\
&= g_{\mn} + \nabla_\mu A_\nu + \nabla_\nu A_\mu + \hdots,
\end{align}
where $\cL$ indicates the Lie derivative, and the ellipsis indicates terms containing higher powers of $A$.

\bibliographystyle{JHEP}
\bibliography{paper}

\end{document}